\documentclass[useAMS,usenatbib,usegraphicx,usenatbib]{mn2e}

\usepackage{ifthen}

\def\rainbow{true}

\newcommand{\subx}{_{\mathrm{x}}}
\newcommand{\kpc}{{\mathrm{kpc}}}
\newcommand{\pc}{{\mathrm{pc}}}
\newcommand{\gtapprox}{\raisebox{-0.5ex}{$\,\stackrel{>}{\scriptstyle
\sim}\,$}}
\newcommand{\ltapprox}{\raisebox{-0.5ex}{$\,\stackrel{<}{\scriptstyle
\sim}\,$}}

\title[Interactions of Jets with Inhomogeneous Cloudy Media]{Interactions of Jets with Inhomogeneous Cloudy Media}
\author[C.~J.~Saxton et al.]{
Curtis J. Saxton$^{1,2,3}$,
Geoffrey V. Bicknell$^{3,4}$,
Ralph S. Sutherland$^3$
\&
Stuart Midgley$^5$
\\
$^{1}${
Mullard Space Science Laboratory, University College London,
Holmbury St Mary, Dorking, Surrey RH5~6NT, UK
}\\
$^{2}${
Max-Planck-Institut f\"{u}r Radioastronomie,
Auf dem H\"{u}gel 69, D-53121 Bonn, Germany
}\\
$^{3}${
Research School of Astronomy \& Astrophysics,
Mt~Stromlo Observatory, Australian National University,
Weston ACT 2611, Australia
}\\
$^{4}${
Department of Physics \& Theoretical Physics, Faculty of Science,
Australian National University ACT 0200, Australia
}\\
$^{5}${
ANU Supercomputer Facility,
Australian National University ACT 0200, Australia
}
}
\begin{document}

\date{Accepted ----. Received ----; in original form ----}

\pagerange{\pageref{firstpage}--\pageref{lastpage}} \pubyear{2005}

\maketitle

\label{firstpage}

\begin{abstract}
We present two--dimensional slab--jet simulations of jets
in inhomogeneous media consisting of a tenuous hot medium
populated with a small filling factor by warm, dense clouds.
The simulations are relevant to the structure and dynamics of sources such as
Gigahertz Peak Spectrum and Compact Steep Spectrum radio galaxies,
High Redshift Radio Galaxies and radio galaxies in cooling flows.
The jets are disrupted to a degree depending upon
the filling factor of the clouds.
With a small filling factor,
the jet retains some forward momentum but also forms
a halo or bubble around the source.
At larger filling factors channels are formed in
the cloud distribution through which the jet plasma flows
and a hierarchical structure consisting of nested lobes
and an outer enclosing bubble results.
We suggest that the CSS quasar 3C48 is an example of a low filling factor jet
-- interstellar medium interaction whilst M87
may be an example of the higher filling factor type of interaction.
Jet disruption occurs primarily as a result of Kelvin--Helmholtz instabilities
driven by turbulence in the radio cocoon
not through direct jet--cloud interactions,
although there are some examples of these.
In all radio galaxies whose morphology may be the result of jet interactions
with an inhomogeneous interstellar medium
we expect that the dense clouds will be optically observable
as a result of radiative shocks driven by the pressure of the radio cocoon.
We also expect that the radio galaxies will possess
faint haloes of radio emitting material
well beyond the observable jet structure.
\end{abstract}


\section{INTRODUCTION}
\label{s:intro}

Previous models of jets propagating through the interstellar medium
of radio galaxies have almost always assumed a uniform,
usually hot ($T \sim 10^7 \> \mathrm{K}$)
interstellar medium.
These models have contributed substantially to
our understanding of the propagation of jets
in evolved radio galaxies wherein
much of the dense matter that could have previously obstructed the jets
has been cleared away.
However, it is now clear that in many \emph{young} radio galaxies,
i.e. Gigahertz Peak Spectrum (GPS) and Compact Steep Spectrum (CSS) sources,
substantial interaction between the expanding radio plasma
and dense interstellar clouds occurs.
This interaction has a number of effects:
Two of the most important are that it significantly distorts
the radio morphology and it also gives rise to shock excited line emission.
Strong observational evidence for the latter effect
has been obtained from combined Hubble Space Telescope (HST)
and radio observations of a number of CSS sources
\citep{devries1999a,labiano03a,odea2003a}
showing a strong alignment effect between optical and radio emission
and clear evidence for disturbance of the optically emitting gas
by the expanding radio lobe.
On the other hand,
the dense gas cannot have a large filling factor
in the case of GPS and CSS sources since
the expansion speed of the radio plasma is usually quite high:
$\sim 0.1 - 0.4~c$
\citep{conway2002,murgia1999a,murgia1999b,murgia2002a,murgia2003a}.
Thus the picture that is conveyed
by the combined radio and optical observations
is of radio plasma propagating through
a predominantly hot and tenuous interstellar medium
in which dense clouds are embedded with a low filling factor.
Nevertheless, as we show in this paper,
even a low filling factor of dense gas
can have a pronounced effect on the evolution of the radio lobes.

High redshift radio galaxies provide another example in
which radio plasma -- cloud interactions are clearly important.
For example, in the high redshift radio galaxy 4C41.17,
interactions between the jets and large clouds
in the Lyman-$\alpha$ emitting halo of the parent galaxy lead to
shock-excited line emission and to accelerated rates of star formation
\citep{bicknell98a}.
Returning to the low-redshift Universe,
radio plasma -- clumpy ISM interactions
may be responsible for the heating of the ISM
that is necessary to counteract the otherwise large mass inflow rates
that are inferred from X-ray observations of cooling flows.

Thus, there is ample physical motivation for detailed investigations
of the manner in which jets
and the related radio lobes interact with realistic interstellar media.
One of the most significant drivers of
this research is that all of the scenarios discussed above
represent examples of AGN-galaxy feedback which,
it has been argued
\citep[e.g.][]{silk1998a,kawakatu03a},
is a significant determinant of
the ultimate structure of galaxies and in particular,
the relation between the mass of the central black hole
and the mass of the galaxy bulge
\citep{magorrian98a,tremaine02a}.

In this first paper in this series we explore the nature of the
interaction between jets and their lobes through a series of
two--dimensional slab jet simulations.
The advantages of this approach
is that two dimensional simulations offer
the prospect of high resolution
and an initial rapid investigation of parameter space,
which capture some of the features of
a more realistic three dimensional simulations.
Fully three dimensional simulations will be presented in future papers.

The choice of coordinate system is non-trivial
when representing supersonic jets in two spatial dimensions.
Many aspects of a jet burrowing through a uniform external medium
have been adequately described in the past by axisymmetric simulations.
However an axisymmetric
simulation is less appropriate
for physical systems where the jet is expected to be bent or deflected.
When colliding directly with a dense cloud,
an axisymmetric jet must either burrow through the cloud or cease its advance.
However, a two-dimensional slab-jet can be deflected,
although both jet and cloud are essentially
infinite in the direction perpendicular to the computational grid.
Our present work is concerned with systems where jet-cloud collisions
and indirect interactions dominate.
We therefore opt for a Cartesian grid and simulate transient slab jets.

\section{CHOICE OF JET AND ISM PARAMETERS}

\subsection{Jet energy flux, cloud and ISM density}

The choice of jet parameters is based on
a self-similar model for lobe expansion
(\citet{bicknell97a},
based upon the model by \citet{begelman96a}).
In this model, the lobe is driven by a relativistic jet
which expands in such a way that $\zeta$ the ratio of hot spot to lobe pressure
is constant.
A value $\zeta \sim 2$ was inferred by Begelman
from simulations by \citet{lind89}.
Let $\rho_0$ be the atmospheric mass density at a fiducial
distance $R_0$ from the core, let
$\delta$ be
the index of the power variation of atmospheric density with
radius ($\rho = \rho_0 (R/R_0)^{-\delta}$)
and let the jet energy flux be $F_E$.
The temperature of the external medium is unimportant in this model
since it is the inertia of the background medium
that governs the expansion of the lobe.
The advance speed, $v_\mathrm{b}$
of the head of the bowshock as a function of
distance $R$ from the core is given by:
\begin{eqnarray}
\frac {v_{\rm b}}{c}
&=& \frac {3}{\left[ 18 (8-\delta)\pi\right]^{\frac{1}{3}}} \,
\zeta^{\frac{1}{6}} \,
\left( \frac {F_E}{\rho_0 R_0^2} \right)^{1/3} \, \left( \frac {R}{R_0} \right)^{(\delta-2)/3} \nonumber \\
&\approx&
0.073 \, \left( \frac {F_{E,46}}{n(100 \> \rm pc)} \right)^{\frac {1}{3}} \,
\left( \frac {R}{100 \> \rm pc} \right)^{-\frac{1}{6}}
\label{e:vb}
\end{eqnarray}
where the jet energy flux is
$10^{46} F_{E,46} \> \rm erg \> s^{-1}$, $n(100 \> \rm pc)$
is the number density at 100~pc and the numerical values are for
$\zeta=2$ and $\delta=1.5$.
Typical expansion speeds of CSS and GPS sources
are of order $(0.05 - 0.4)\> c$
with the largest velocities measured for the most compact sources
\citep{conway2002,murgia1999a}.
\citet{conway02a}
has reviewed estimates of number densities
$\sim 1 \> \rm cm^{-3}$
for three GPS sources,
and so a fiducial value of
$n(100 \> \rm pc) = 1~\mathrm{cm}^{-3}$ is reasonable.
Compatibility of the expansion
speeds estimated from equation~(\ref{e:vb})
and the observations indicates jet energy fluxes in the range of
$10^{46-47} \> \rm erg \> s^{-1}$.
Jets of this power are also consistent with a conventional ratio
$\sim 10^{-12} - 10^{-11} \> \rm Hz^{-1}$
of 1.4~GHz monochromatic radio power to jet energy flux
and a median power at 5~GHz $\sim 10^{27.5} \> \rm W \> Hz^{-1}$.
Nevertheless,
in other sources which morphologically resemble CSS radio galaxies,
such as the less radio powerful Luminous Infrared Galaxies
studied by \citet{drake2003},
it is apparent that the jet powers are somewhat less.
Thus, in our simulations, we have adopted jet
powers of $10^{45}$ and $10^{46} \> \rm erg \> s^{-1}$.

Hubble Space Telescope observations
\citep{odea2002,devries1997,devries1999a}
show abundant evidence for dense clouds
of emission-line gas associated with CSS radio galaxies and quasars.
For a temperature
$10^4 T_{\rm c,4} \> \rm K$ of the dense clouds,
a number density $n\subx \sim 1 \> \rm cm^{-3}$
and  temperature
$10^7 T_{\rm x,7}\> \rm K$ for the hot interstellar medium,
the number density in the clouds is
$n_{\rm c}\approx 10^3 \> n\subx T_{\rm x,7} /T_{\rm c,4} \> \rm cm^{-3}$,
assuming that the initial, undisturbed clouds
are in pressure equilibrium with their surroundings.
If we adopt such a value for the density of the interstellar medium
then equation~(\ref{e:vb})
implies that the velocity of advance would be $\sim 0.02 \> c$
--- much lower than what is observed.
Hence, as discussed in \S~\ref{s:intro},
the dense clouds must have a small filling factor.
Accordingly, in the simulations described below,
we create clouds of approximately the inferred
density and investigate the effect of various filling factors.

With the inferred radio source and cloud geometry,
the line emission orginates from dense,
relatively cool clouds that are shocked
as they are overrun by the expanding radio source.
A model of this type was proposed by
\citet{devries1999a}
and elaborated by \citet{bicknell2003a}.
In both treatments,
the excess pressure of the advancing radio lobe
drives shocks into the dense clouds.
In the simulations described below,
this is seen to occur but some direct jet-cloud collisions are also apparent.
For radio lobe driven shocks (the main source of shocks),
the shock velocity is:
\begin{equation}
v_{\rm sh} \approx 950 \, \left( \frac {v_{\rm b}}{0.1 \> c} \right) \,
\left( \frac {n_{\rm c}/n_{\rm x}}{10^3} \right)^{\frac{1}{2}}
\> \rm km \> s^{-1}
\end{equation}
where $n_{\rm c}$ and $n_{\rm x}$
are the cloud and ISM number densities respectively.
Shocks in
the velocity range of $300-950 \> \rm km \> s^{-1}$
are produced for bow-shock advance speeds
$\sim 0.1 \> c$
and density ratios
$n_{\rm c}/n_{\rm x}\sim 10^4 - 10^3$.
This is in accord with the
observed velocity dispersions and velocity offsets
inferred from HST spectra of a sample of CSS sources
\citep{odea2002}.

Radiative shocks do not develop instantaneously;
the time required for a shock to become fully
radiative is effectively the time, $t_4$,
required for post-shock gas to cool to $\sim 10^4~\mathrm{K}$.
In order that there be a display of optical emission lines,
$t_4$ must be less than a fraction $f \sim 1$
of the dynamical time associated with the expansion of the lobe.
In estimating $t_4$ we utilize a cooling function,
which estimates the non--equilibrium cooling following a high velocity shock,
and which is derived from the results of the one--dimensional MAPPINGS~III code
\citep{sutherland93c}.
This cooling finction is also used in our simulations described below.
The condition on $t_4$
becomes a condition on the bow-shock velocity
\begin{equation}
\left( \frac {v_b}{c} \right) < 0.07 \> f^{0.2} \,
\left(\frac{n_{\rm c}}{10^4 \> \rm cm^{-3}} \right)^{0.2} \,
\left( \frac {n_{\rm c}/n_{\rm x}}{10^4} \right)^{0.4}  \,
\left( \frac {R}{\rm kpc} \right)^{0.2}
\ ,
\label{e:vb_limit}
\end{equation}
\citep{bicknell2003a}.
This upper limit on $v_{\rm b}$
is insensitive to the various parameters,
in particular, $f$.
If condition~(\ref{e:vb_limit})
is not satisfied near the head of the radio lobe
it may be satisfied at the sides
where the bow-shock expansion speed is about a factor of 2 lower.
In this case one would expect to see emission from the sides
but not the apex of the radio source.
This condition gives a limiting bow--shock velocity $v_{\rm b}\sim 0.1 c$.
Therefore, it would be interesting to see whether the more rapidly
expanding sources have a lower ratio of emission line flux to radio power,
and whether the emission line flux
is distributed in a different manner to that in lower velocity sources.
These conditions may also be related to the causes of
the trailing optical emission in CSS sources
\citep{devries1999a}.
There are two possibilities:
(a) The apex of the bow-shock is non-radiative (if $v_{\rm b}$ exceeds the
above limit) or
(b) The number of dense clouds per unit volume
decreases at a few kpc from the core,
i.e. beyond the normal extent of the narrow line region.
We are inclined to favour option (b) but (a)
should also be kept in mind.

The simulations capture
these characteristics that depend on
the time of the effective onset of radiative shocks in the clouds.
We see a range of shock conditions
varying from almost pure adiabatic shocks to strongly radiative.
When the cooling time of gas becomes shorter than the Courant time
for the simulation,
the integration of the energy equation
ensures that the radiative shock occupies
approximately one pixel in which the gas temperature is equal to
the predefined equilibrium temperature ($10^4 \> \rm K$ in these simulations).

\subsection{\bf Overpressured jets --- Mach number and density ratio}
\label{s:jet_pars}

In order to carry out realistic 2D simulations,
we wish to establish jets with powers per unit transverse length
that are relevant to the types of radio galaxies mentioned in the introduction,
in combination with other reasonable values of parameters such as
jet width and density.
If the number density of the interstellar medium is
$\sim 1 \> \rm cm^{-3}$
on 100~pc scales,
then the jet must be overpressured.
We first show this in the context of a sub-relativistic jet model
and then support the proposition with a relativistic jet model.

Let a nonrelativistic jet have a density
$\eta$ times that of the external medium,
an internal Mach number $M$,
a specific heat ratio $\gamma$,
a pressure $\xi$ times the external ISM pressure
and a cross-sectional area $A_\mathrm{j}$.
Further, let $\rho\subx$ and $v\subx$ be the density and isothermal
sound speed of the ISM.
Then the jet mass, momentum and energy fluxes are,
collecting the dimensional quantities on the left:
\begin{eqnarray}
\dot{M}&=&
(\rho\subx \, v\subx \, A_\mathrm{j})\ \
M\sqrt{\gamma\xi\eta}
\label{mdot}
\\
\Pi&=&
(\rho\subx \, v\subx^2 \, A_\mathrm{j})\ \
\xi \, (1+\gamma M^2) \>
\label{e:thrust}
\\
F_E&=&
(\rho\subx \, v\subx^3 \, A_\mathrm{j})\ \
M\left({\frac {1}{\gamma-1} + M^2}\right)
\sqrt{{\gamma^3\xi^3}\over\eta}
\label{e:power}
\end{eqnarray}

Equation~(\ref{e:power}) clearly shows that
the energy flux can be made arbitrarily
large by decreasing the density ratio $\eta$.
However, there is a natural lower limit on $\eta$
-- the value at which an intially relativistic jet becomes transonic.
The critical value is defined by the condition
$\rho_{\rm j} c^2 / 4 p_{\rm j} = 1$
\citep{bicknell1994a},
implying that
$\eta_{\rm crit} = 4 \xi (v\subx/c)^2
\approx 5.9 \times 10^{-6} \xi T_{\rm x,7}$.
With the overpressure factor,
$\xi =1$, $\eta = 10^{-5}$, a jet diameter of 10~pc,
$T\subx = 10^7 \> \rm K$ and $M=20$, the energy flux
is only $\sim 10^{44} \> \rm erg \> s^{-1}$.
One requires an overpressure factor of about 100 to
achieve $10^{46} \> \rm erg \> s^{-1}$.

This result is easily verified
for a relativistic jet whose energy flux is given by
\begin{equation}
F_E =
(\rho\subx v\subx^2 c A_{\mathrm j})\
4\,\xi\beta_{\rm jet} \,\Gamma_{\rm jet}^2 \>
\left[1 + \frac {\Gamma_{\rm jet}-1}{\Gamma_{\rm jet}} \chi \right] \nonumber
\end{equation}
where $\chi = \rho_{\rm j} c^2 /4 p_{\rm j}$,
and $\rho_j$ and $p_j$ are the jet rest--mass density and pressure respectively.
If $\xi = 1$, $\Gamma=10$,
a jet diameter of 10~pc,
$T\subx = 10^7 \mathrm{K}$
and $\chi = 0$
(the latter being appropriate for a jet dominated by relativistic plasma),
then the energy flux is  only $1.2 \times 10^{43} \> \rm erg \> s^{-1}$.
Hence, for a relativistic jet, we require $\xi \sim 10^{2-3}$
in order to achieve the required jet power.

Is an overpressured jet physical?
The overpressure ratio refers to the pressure with respect to the
undisturbed interstellar medium.
However, the lobe itself is over-pressured
as a result of the advancing bow shock
so that it is possible for the jet to be in approximate equilibrium
with the ambient lobe.
In a real galaxy the over pressured lobe may not extend back to the core,
in which case an overpressured jet
would behave initially like a classical overpressured, underexpanded jet.

Therefore, in our simulations, we have adopted
low values of the density ratio, $\eta \sim 10^{-3}$,
Mach numbers, $M$ of the order of 10
and a high overpressure ratio in order to obtain
the necessarily high jet energy fluxes
(or their equivalent in 2D).
Such a low value of $\eta$ has the additional effect
of producing superluminal velocities ($\beta \sim 5-6$)
in some simulations,
highlighting the shortcomings  of simulating powerful jets
with a non--relativistic code.
However, the essential point is that
it is mainly the jet power that is physically important 
in jet--cloud interactions.

One could adopt the strategy of adopting non-relativistic parameters
that correspond more closely to those of relativistic jets,
for example by equating the velocity and pressure
and matching either the power or thrust
(e.g. see \citet{komissarov96a} and \citet{rosen99a}).
We have not done this explicitly here since there is no transformation
that perfectly maps a non-relativistic flow into a relativistic equivalent.
Clearly,
this initial investigation of parameter space needs to be followed by
more comprehensive simulations utilising a fully relativistic code.

\subsection{Prescription for cloud generation}
\label{s.clouds}

In our simulations we prescribe
a uniform density for the hot interstellar medium.
Within this medium we establish an initial field of dense clouds
that mimics the distribution of clouds in the interstellar medium,
with a random distribution of positions, shapes
and internal density structures.
We consider that this is more desirable than specifying,
say circular clouds, albeit with random centres;
in that case the subsequent evolution may be governed by the initial symmetry.
An advantage of the approach that we describe here
is that the distribution of the sizes, shapes
and directions of propagation of radiative shocks is
more realistic than that obtained from initially highly symmetric clouds.

We prescribe the density and distribution of clouds using a Fourier approach.
In the first step we
establish a grid of complex numbers in Fourier space
and assign a random phase to each point.
The largest wave number is compatible with
the spatial resolution of the simulations.
An amplitude filter is applied,
$\propto |\mathbf{k}|^{-5/3}$,
in order to mimic the power spectrum of structures formed by turbulence.
Taking the Fourier transform of the wavenumber data
produces an array of density fluctuations in real space.
These fluctuations, $\sigma(\mathbf{r})$,
are normalised so that their mean is $0$
and their variance is $1$.
The final number density of gas is then defined by the relation:
\begin{equation}
n=\left\{{
\begin{array}{ll}
n_\mathrm{min}+n_\sigma (\sigma-\sigma_\mathrm{min})
&\mbox{wherever $\sigma\ge\sigma_\mathrm{min}$}
\\
n\subx&\mbox{elsewhere}
\end{array}
}\right.
\ ,
\end{equation}
where $n\subx$ is the number density of hot inter-cloud gas,
$n_\mathrm{min}$
is the minimum number density of gas in a dense cloud,
$n_\sigma$ is a scaling factor
and $\sigma_\mathrm{min}$ is a threshold value
determining whether a particular cell is part of a cloud
or else hot inter-cloud medium.
In practice we choose the mean density of cloud gas,
$\bar{n}$
and set the value of $n_\sigma$ implicitly.

We apply upper and lower limits to the density distribution in
$\mathbf{k}$-space.
Amplitudes are zeroed for $k<k_\mathrm{min}$,
where $k_{\rm min}$ corresponds to the Jeans wave number.
This prevents  the insertion of clouds that would have collapsed to form stars
and also avoids the generation of a cloud field dominated by
a single large cloud-mass.
Amplitudes are also zeroed for $k> k_\mathrm{max}$
where $k_\mathrm{max}$ corresponds to 3 or 4 pixels of spatial resolution.
Fine features, at the 3-4 pixel level,
can introduce excessive and artificial sensitivity
to effects on the level of individual pixels.

Finally, radial limits are applied to the cloud field as a whole.
An outer limit, $r_\mathrm{max}$,
(typically $\sim 0.5 - 5~\kpc$)
limits the extent  of the cloud field.
The inner radius, $r_\mathrm{min}$,
is defined by the Roche limit in the vicinity of a black hole:
\begin{equation}
r_\mathrm{min} =\left({
{3\over{4\pi}}  {{M_\mathrm{bh}}\over{\rho_\mathrm{c}}}
}\right)^{1/3}
\end{equation}
If the black hole mass $M_\mathrm{bh} = 10^9 M_\odot$
and the cloud density
$\rho_\mathrm{c} \approx 15 M_\odot \> \mathrm{pc}^{-3}$
then $r_\mathrm{min} \approx 250 \> \mathrm{pc}$.

\begin{figure*}\begin{minipage}{180mm}
\begin{center}
\ifthenelse{\isundefined{\rainbow}}{
\includegraphics{f01.eps}
}{
\includegraphics{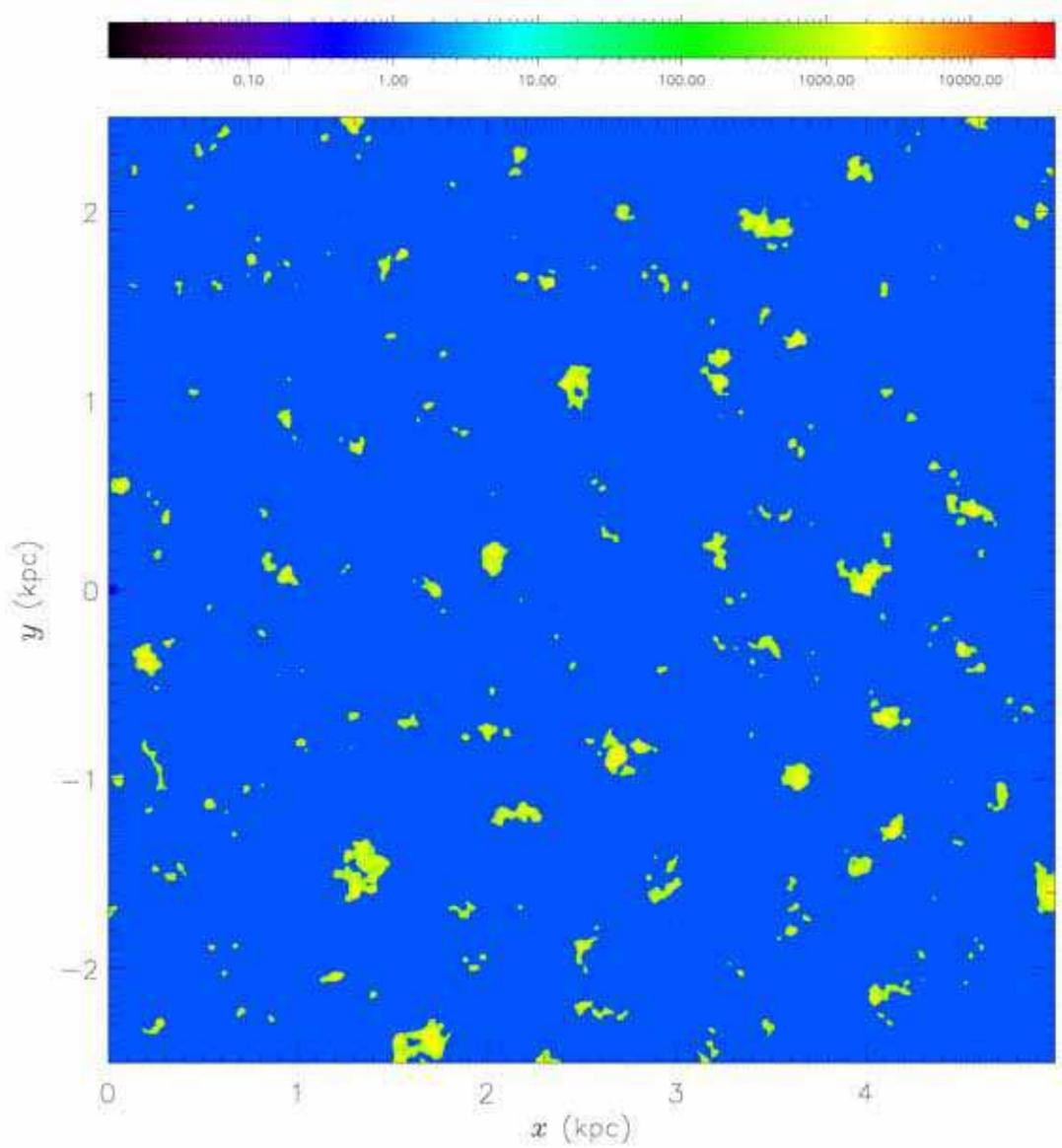}
}
\caption{
Logarithmically scaled map of density relative to the external medium,
$\rho/\rho_\mathrm{x}$, in the initial cloud field
of the simulation {\tt A1}.
The dark gray region represents the intercloud medium
and the lighter gray regions represent cloud cells.
Each cloud has an internal density structure
defined by the Fourier construction described in the text
(\S\ref{s.clouds}).
The region shown extends over
$5\ \kpc \times 5\ \kpc$
($1600 \times 1600$ cells),
with the jet origin at the middle of the left edge.
}
\label{fig.clouds.initial.1}
\end{center}
\end{minipage}\end{figure*}

\begin{figure}
\begin{center}
$\begin{array}{ccc}
\ifthenelse{\isundefined{\rainbow}}{
\includegraphics[width=7cm]{f02a.eps}
\\ \includegraphics[width=7cm]{f02b.eps}
\\ \includegraphics[width=7cm]{f02c.eps}
}{
\includegraphics[width=7cm]{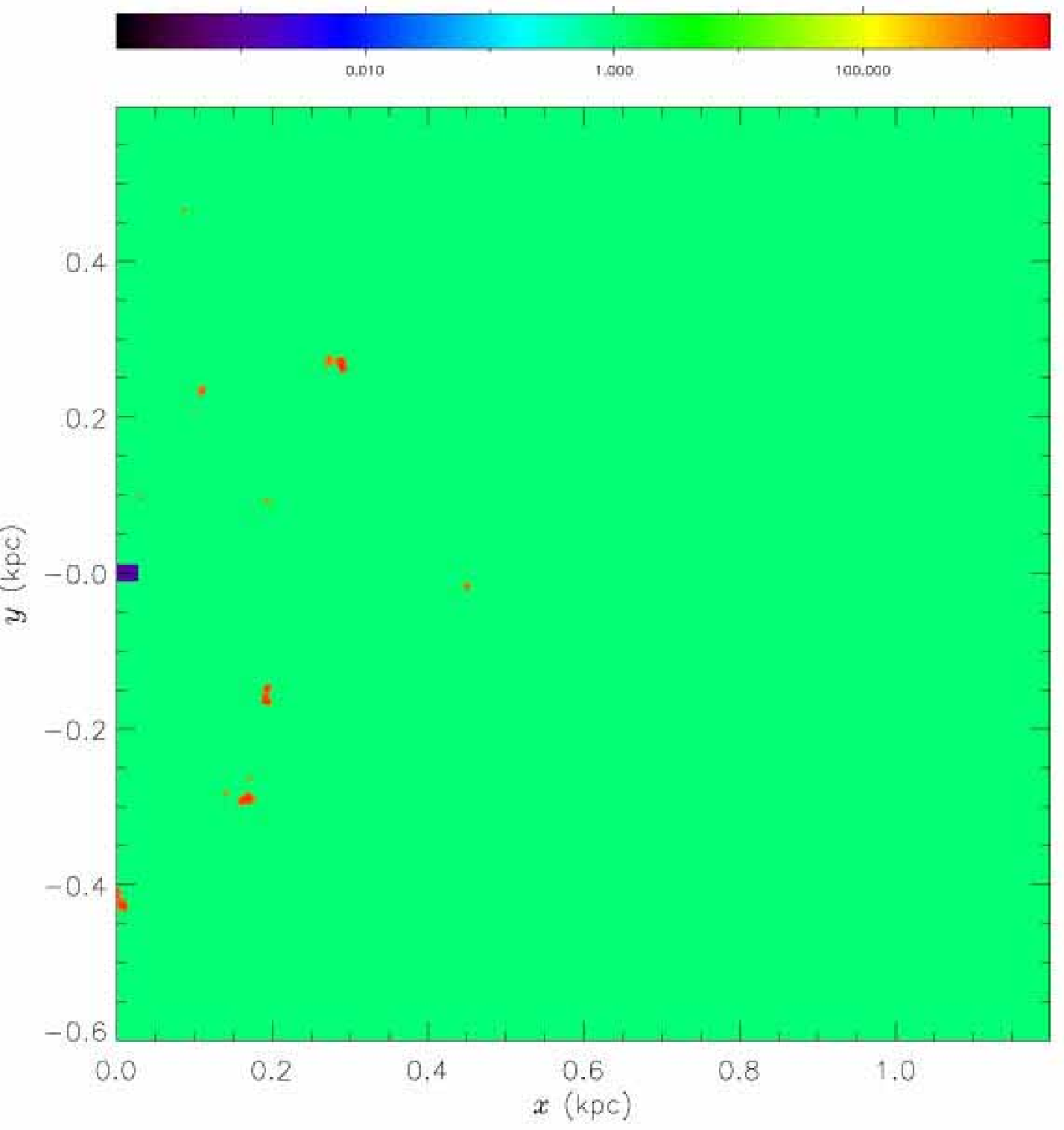}
\\ \includegraphics[width=7cm]{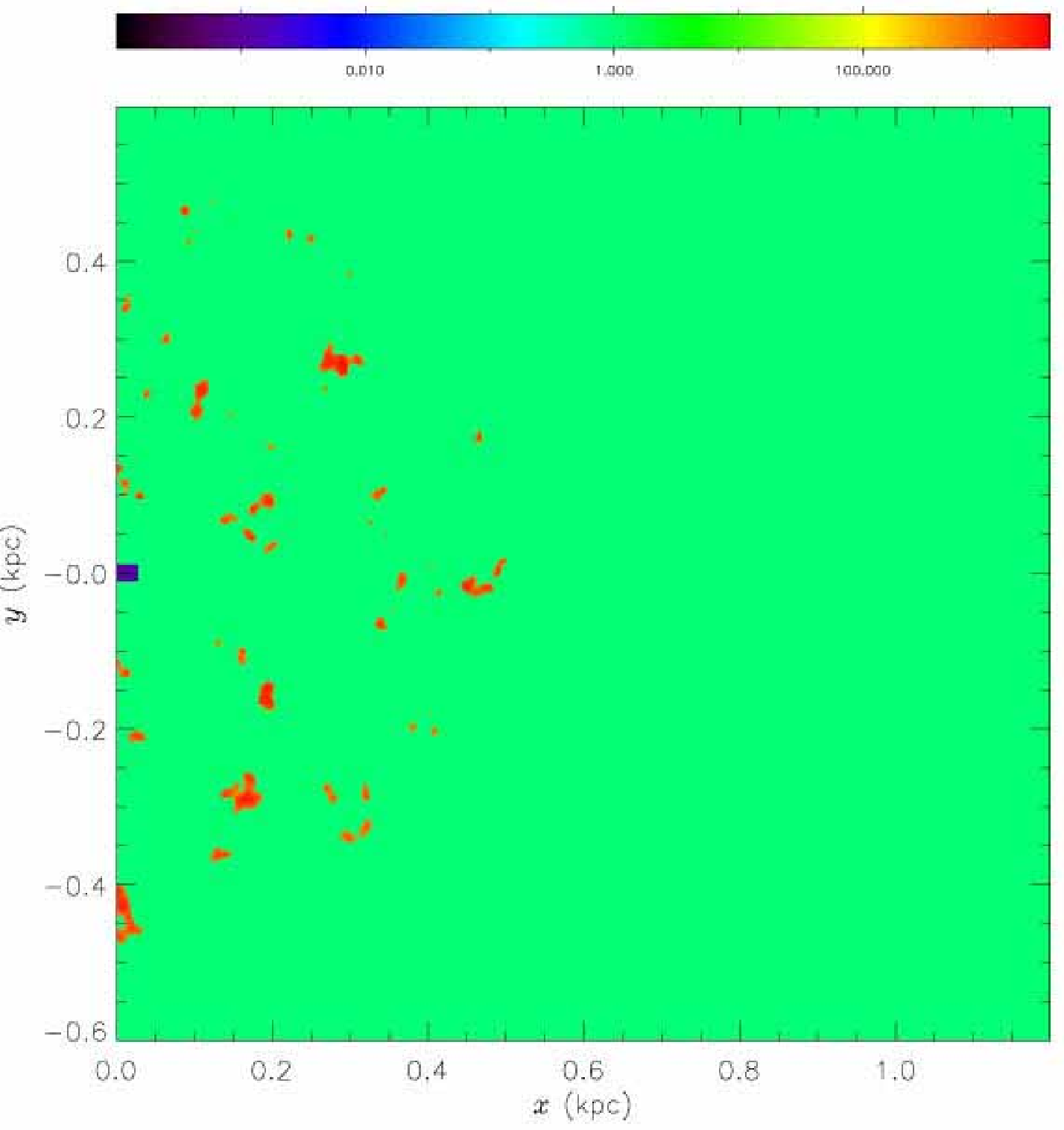}
\\ \includegraphics[width=7cm]{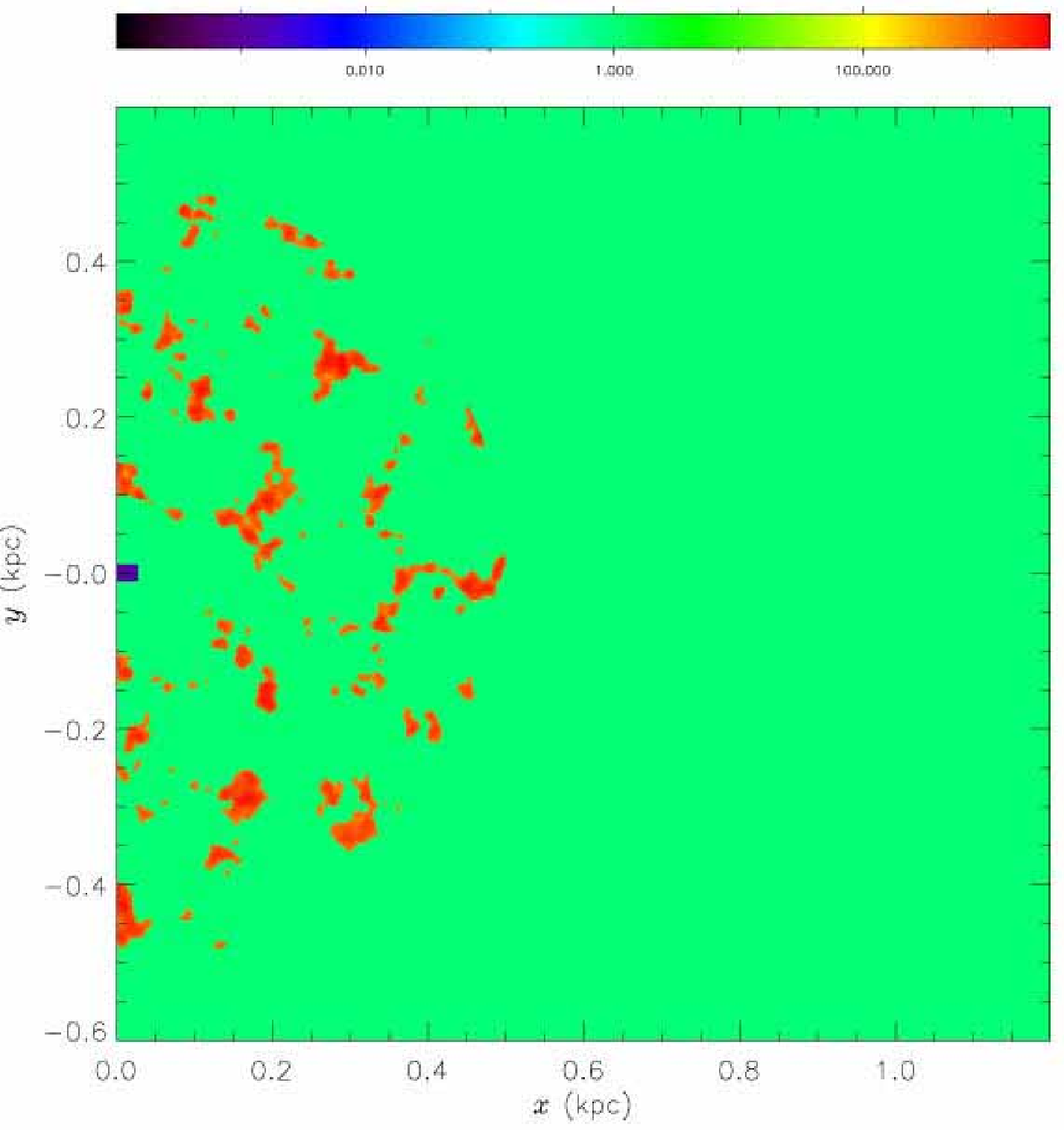}
}
\end{array}$
\caption{
Density maps of the initial cloud fields of the simulations
{\tt B1}, {\tt B2} and {\tt B3}
(from top to bottom respectively).
These cloud fields were generated from the same fractal distribution
but use a different threshold $\sigma_{\mathrm min}$
to define the cloud/intercloud boundaries.
The entire grid covers
$4 \> \mathrm{kpc} \times 4 \> \mathrm{kpc}$
($2000 \times 2000$ cells),
but the square subregion shown above is $1.2\ \kpc$ on a side.
The jet origin is at the middle of the left edge of each panel.
}
\label{fig.clouds.initial.2}
\end{center}
\end{figure}

\begin{figure*}\begin{minipage}{180mm}
\begin{center}
$\begin{array}{cc}
\includegraphics[width=7.5cm]{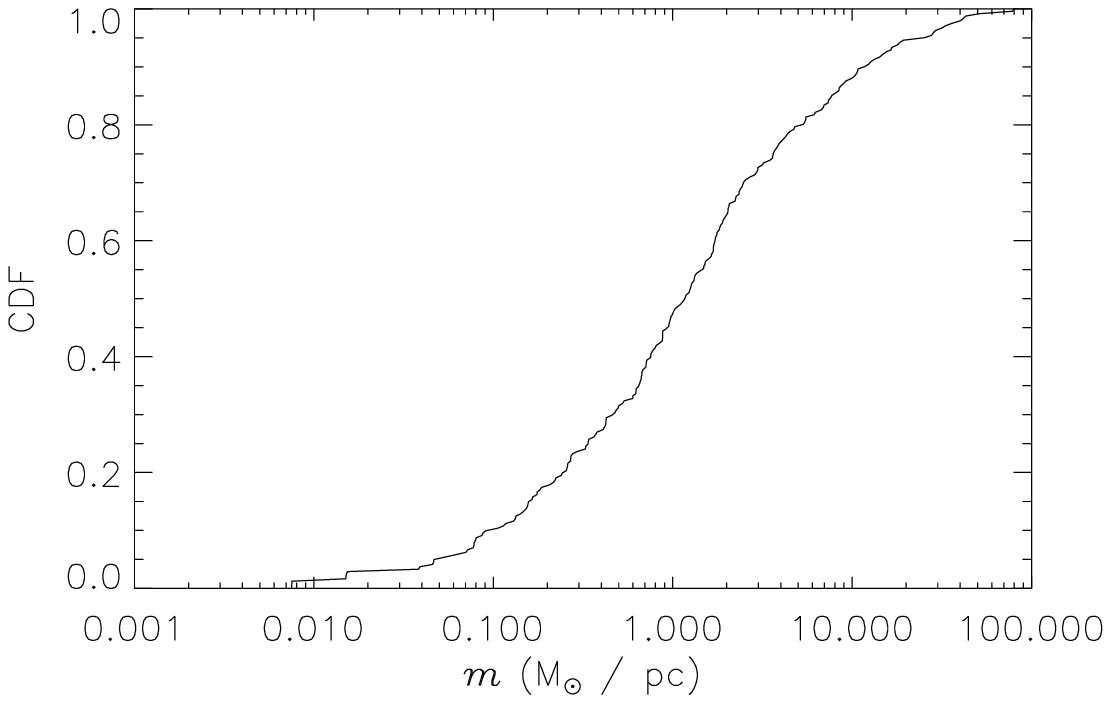}
&\includegraphics[width=7.5cm]{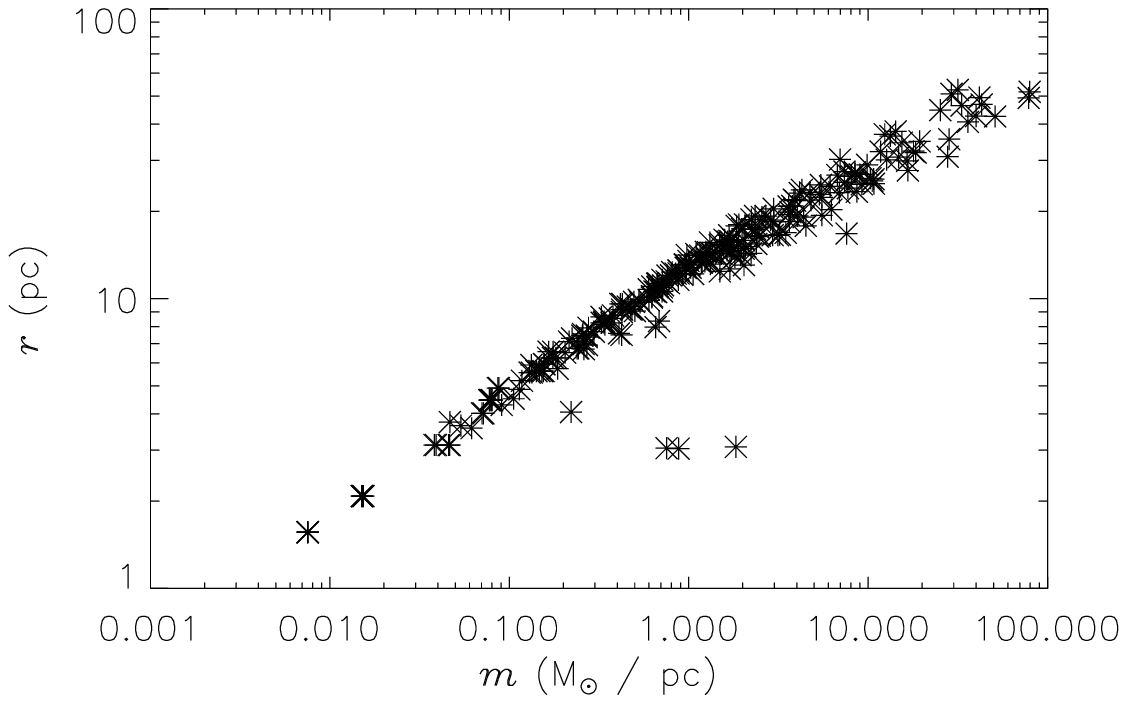}
\end{array}$
\caption{
Initial characteristics of the cloud populations
in model {\tt A1}
($\sigma_\mathrm{min}=1.9$).
The left column shows
a cumulative distribution function of the individual cloud masses per parsec.
The right column presents
a scatter plot of the characteristic depth versus cloud mass per parsec.
}
\label{fig.cloudcdf_s02}
\end{center}
\end{minipage}\end{figure*}

\begin{figure*}\begin{minipage}{180mm}
\begin{center}
$\begin{array}{cc}
\includegraphics[width=7.5cm]{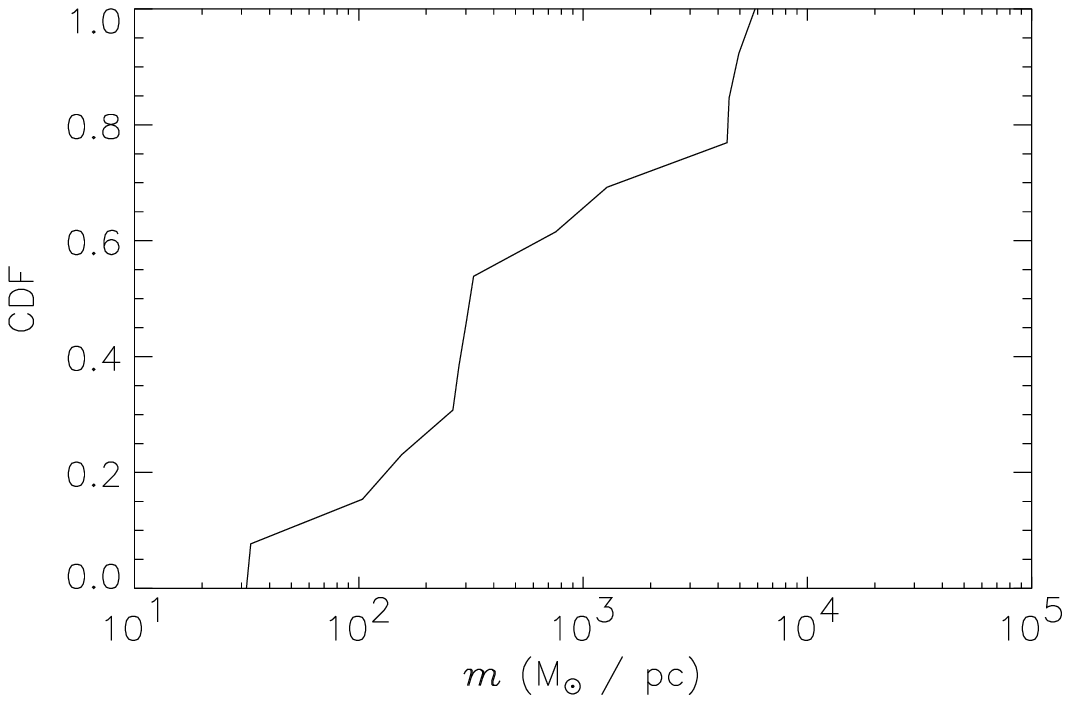}
&\includegraphics[width=7.5cm]{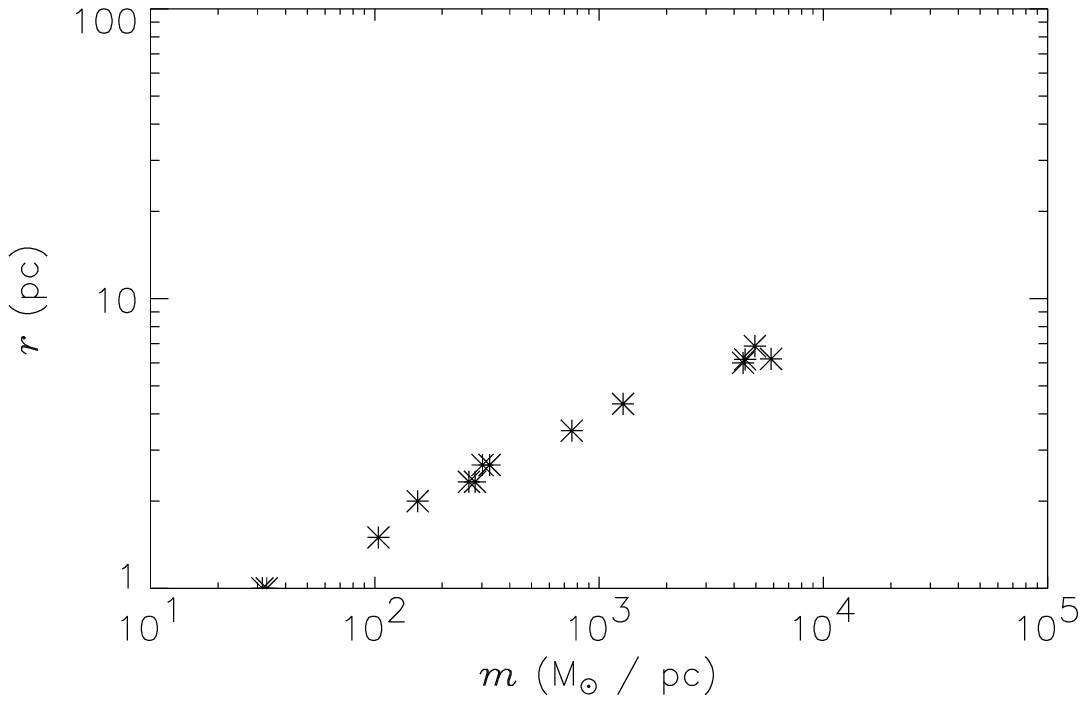}
\\
\includegraphics[width=7.5cm]{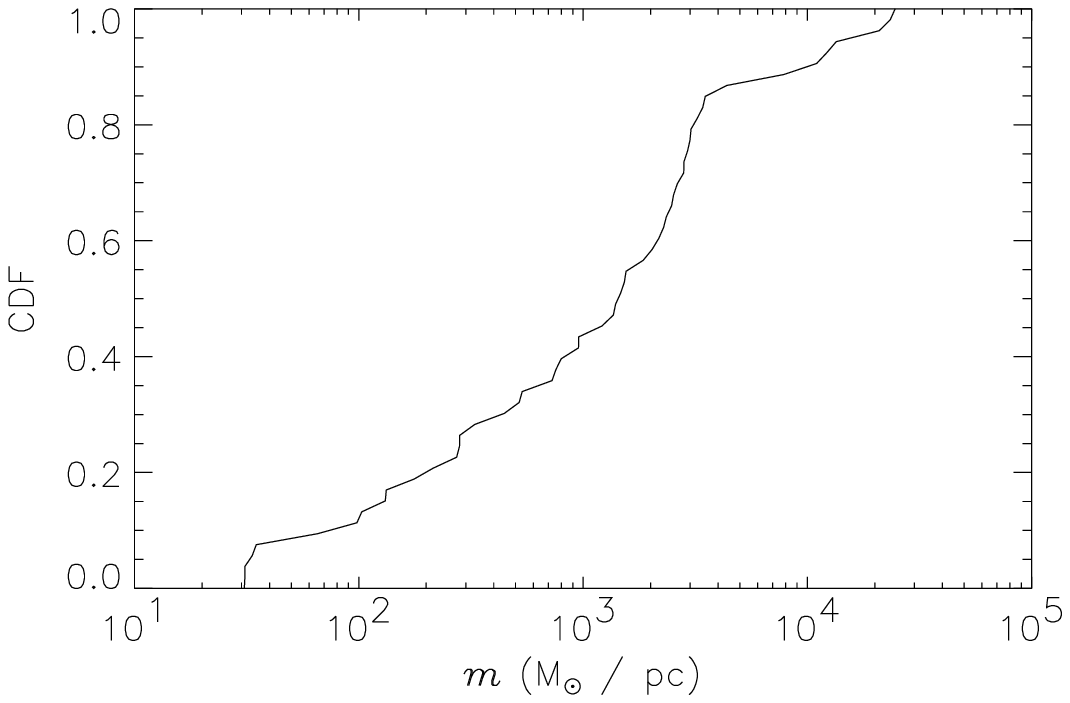}
&\includegraphics[width=7.5cm]{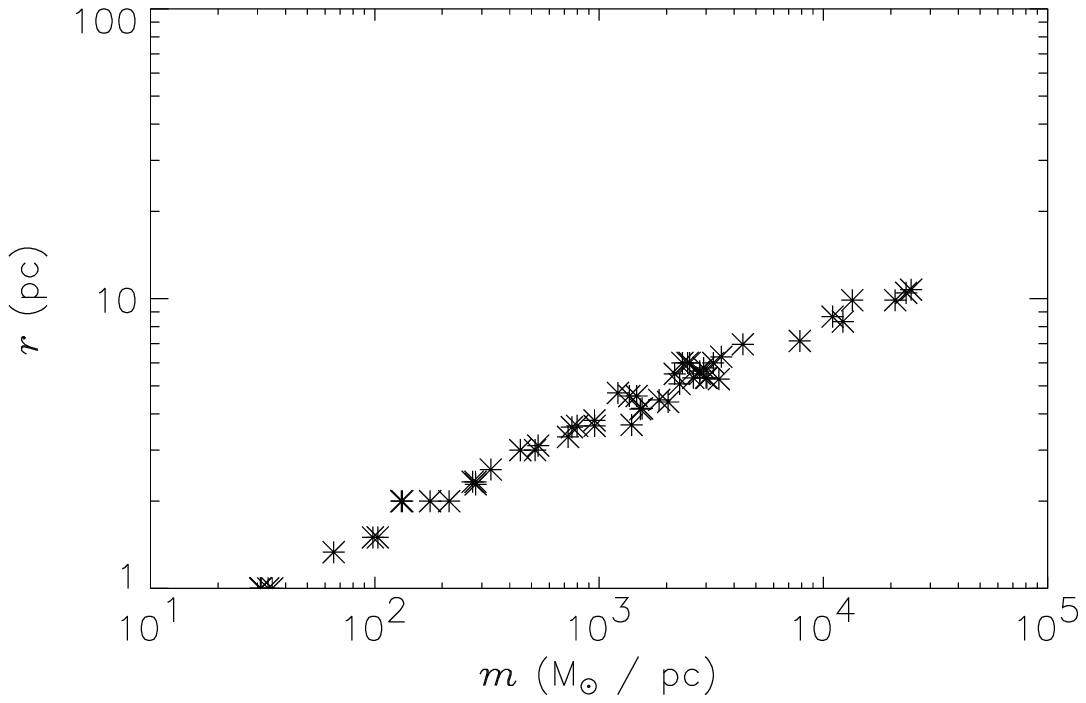}
\\
\includegraphics[width=7.5cm]{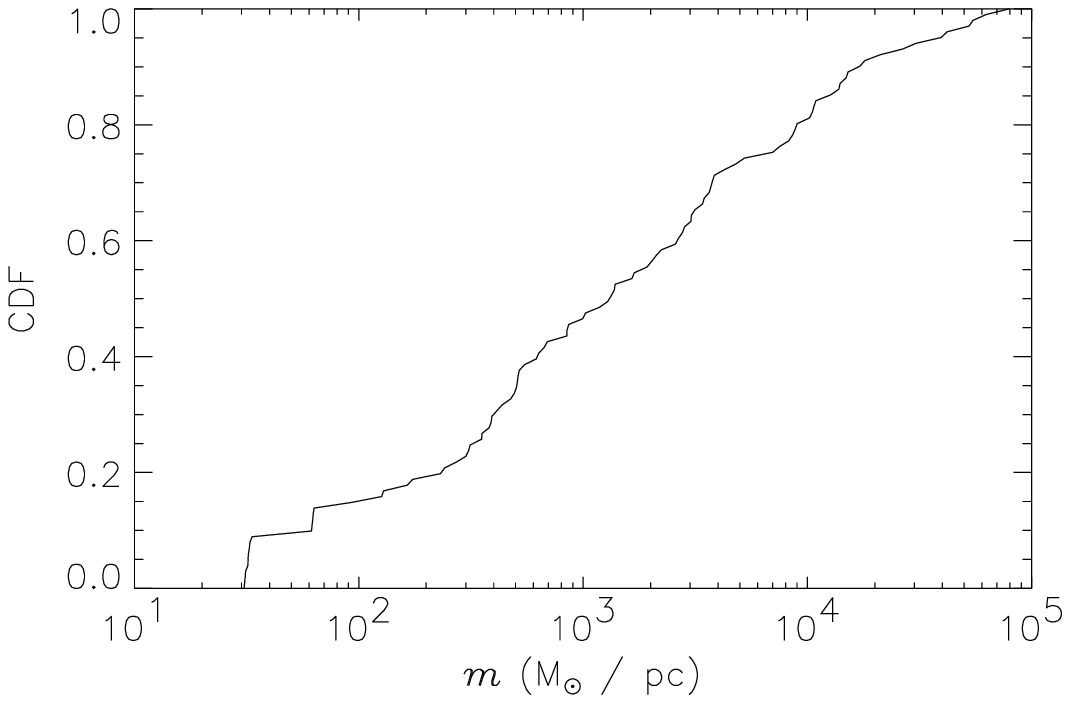}
&\includegraphics[width=7.5cm]{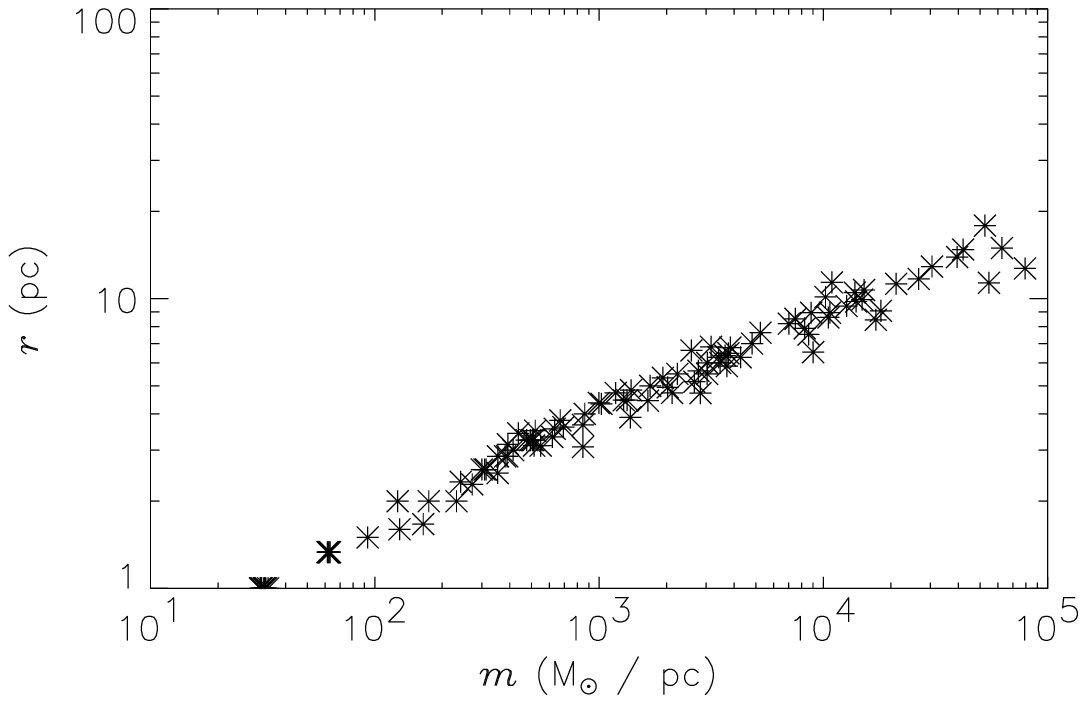}
\\
\end{array}$
\end{center}
\caption{
Cumulative distribution function of cloud masses (left panel)
and mass vs thickness relation (right panel)
as in
Figure~\ref{fig.cloudcdf_s02}
but for the models
{\tt B1}
($\sigma_\mathrm{min}=2.6$),
{\tt B2}
($\sigma_\mathrm{min}=1.9$)
and
{\tt B3}
($\sigma_\mathrm{min}=1.2$),
from top to bottom respectively.
}
\label{fig.cloudcdf_e}
\end{minipage}\end{figure*}

Table~\ref{table.cloudfield}
presents a summary of the properties
of the clouds and intercloud medium,
and the extent of cloud coverage
in our simulations
(which are each described in specific detail in \S\ref{s.simulations}).
Figures~\ref{fig.clouds.initial.1} and \ref{fig.clouds.initial.2}
represent the initial distributions of cloud and intercloud gas
in three of our simulations.
Figures~\ref{fig.cloudcdf_s02} and \ref{fig.cloudcdf_e}
show how the masses of individual clouds are distributed
in the initial conditions of four of our simulations.
As the cumulative distribution functions show,
most clouds have masses of less than a few thousand solar masses.
The scatter plots show the characteristic depth of clouds:
in mean terms the distance from an interior point to the surface,
$r=DI/S$
where $D$ is the number of spatial dimensions,
$I$ is the interior (area of a 2D cloud)
and $S$ is the surface (perimeter of a 2D cloud).
A small $r$ means that most of the cloud mass is well exposed
to surface effects,
such as ablation and the penetration of shocks from outside.
If $r$ is large than the interior regions are typically deeper
from the locally nearest surface,
and correspondingly better insulated from effects at the surface.

Large clouds, near the Jeans limit
(e.g. thickness $2r \sim 105\ \pc$ in the {\tt B} series)
are rare in every simulation.
The simulations with a greater cut-off $\sigma_{\mathrm min}$,
and hence greater filling factor of clouds,
also feature individual clouds of greater mass
than the simulations with lower $\sigma_{\mathrm min}$.

\subsection{Code}

In the simulations reported here, we have used a locally modified version of
a Piecewise Parabolic Method
\citep{colella1984} code,
known as {\tt VH-1},
published on the web by the computational astrophysics
groups at the Universities of Virginia and South Carolina\footnote{See
{\tt http://wonka.physics.ncsu.edu/pub/VH-1/}.
See also \citet{blondin93a}
for the extension of the PPM method to curvilinear coordinates.}.
Our modifications of the VH-1 code include restructuring of
the code for efficient computation
on the computers at the Australian National University
(formerly a Fujitsu VP300; now a Compaq Alphaserver SC),
addition of code for dynamically important cooling
and other code for coping with a numerical shock instability
\citep{sutherland2003a}.
The code has already been used in several adiabatic applications
(e.g. \citealt{saxton2002a}a, \citealt{saxton2002b}b)
and most recently in two-dimensional simulations of fully radiative wall shocks
\citep{sutherland2003b}.
In the latter paper, cooling was dynamically important and the code for
preventing numerically unstable shocks
was implemented and proved to be important in elminating
physically unrealistic features.
We refer to this augmented code as {\em ppmlr}.

We have recently ported the {\em ppmlr} code
to the Message Passing Interface (MPI) environment
(see Appendix~\ref{app.mpi}).
This port enables us to take advantage of
the Compaq Alphasever parallel processor architecture.
The MPI port is described in the appendix.
The simulations described here typically involve 16 processors.
Future three-dimensional simulations
will regularly use 256 processors
and in exceptional circumstances, up to 512 processors.

\begin{table*}
\centering
\begin{minipage}{140mm}
\caption{
Parameters and properties of the inter-cloud medium
and initial cloud fields
in the respective simulations.
The X-ray emitting inter-cloud medium has temperature
$T_\mathrm{x}$ and particle number density $n_\mathrm{x}$.
Clouds have a minimum density $n_\mathrm{c,min}$,
mean density $\bar{n}_\mathrm{c}$.
The cloud generation threshold is $\sigma_\mathrm{min}$
and the clouds occupy a total area $A_\mathrm{c}$,
corresponding to a filling factor of $f_\mathrm{c}$.
The total exposed cloud perimeter is $P_\mathrm{c}$.
}
\label{table.cloudfield}
\begin{center}
$
\begin{array}{ccccccccc}
\hline
\hline
\mbox{Model}
&\begin{array}{c}T_\mathrm{x}\\(\mathrm{K})\end{array}
&\begin{array}{c}n_\mathrm{x}\\(\mathrm{cm}^{-3})\end{array}
&n_\mathrm{c,min}/n_\mathrm{x}
&\bar{n}_\mathrm{c}/n_\mathrm{x}
&\sigma_\mathrm{min}
&\begin{array}{c}A_\mathrm{c}\\(\mathrm{pc}^2)\end{array}
&f_\mathrm{c}
&\begin{array}{c}P_\mathrm{c}\\(\mathrm{pc})\end{array}
\\
\hline
{\tt A1}
&10^7& 10^{-2} & 300 & 500
& 1.9 & 7.28 \times 10^5 &2.92\times10^{-2}&5.83 \times 10^4
\\
{\tt B1}
& 10^7 &1 & 500 & 1000
& 2.6 & 1.55 \times 10^3 &4.11\times10^{-3}&6.16 \times 10^2
\\
{\tt B2}
& 10^7 &1 & 500 & 1000
& 1.9 &1.15 \times 10^4 & 3.04\times10^{-2}&3.51 \times 10^3
\\
{\tt B3}
& 10^7 &1 & 500 & 1000
& 1.2 & 4.43 \times 10^4 & 1.18\times10^{-1}& 9.74 \times 10^4
\\
{\tt B3a}
& 10^7 &1 & 500 & 1000
& 1.2 & 4.43 \times 10^4 & 1.18\times10^{-1}& 9.74 \times 10^4
\\
\hline
\end{array}
$
\end{center}
\end{minipage}
\end{table*}

\begin{table*}
\centering
\begin{minipage}{180mm}
\caption{
Parameters and properties of the jets in the respective simulations.
$M$ = Mach number,
$\eta$ = ratio of jet density to intercloud medium density,
$\xi$
is the ratio of the initial jet pressure
to the pressure of the external medium,
$r_\mathrm{j}$
is one half the extent of the slab jet (the slab jet ``radius'').
An equivalent three-dimensional jet with circular cross-section
has mass flux $\dot{M}$,
thrust $\Pi$
and power $F_E$.
}
\label{table.jet}
{
\footnotesize
\begin{center}$
\begin{array}{cccccccccccc}
\hline
\hline
\mbox{Model}&M&\eta&\xi
&\begin{array}{c}r_\mathrm{j}\\(\mathrm{pc})\end{array}
&\begin{array}{c}\dot{M}\\(M_\odot / \mathrm{yr})\end{array}
&\begin{array}{c}\Pi\\(10^{33} \mathrm{dyn})\end{array}
&\begin{array}{c}F_E\\(\mathrm{erg} / \mathrm{s})\end{array}
&{{
\dot{M}
}\over{
A_\mathrm{j}\rho_\mathrm{x}c_\mathrm{x}
}}
&{{\Pi }\over{ A_\mathrm{j}\rho_\mathrm{x}c_\mathrm{x}^2 }}
&{{F_E }\over{ A_\mathrm{j}\rho_\mathrm{x}c_\mathrm{x}^3 }}
\\
\hline
{\tt A1}& 12.7&0.384~~~& 947
& 25 & 3.5\times10^{-2} &5.43~\,&10^{45}
&313&2.10 \times 10^4&2.10\times10^8
\\
{\tt B1}& 12.7& 1.16 \times 10^{-3}& 100
&10 & 1.0\times10^{-2} &9.17~\,&10^{46}
&5.58&2.22 \times 10^3&1.31\times10^8
\\
{\tt B2}& 12.7& 1.16 \times 10^{-3}& 100
&10 & 1.0\times10^{-2} &9.17~\,&10^{46}
&5.58&2.22\times 10^3&1.31\times10^8
\\
{\tt B3}& 12.7& 1.16 \times 10^{-3}& 100
&10 & 1.0\times10^{-2} &9.17~\,&10^{46}
&5.58&2.22\times 10^3&1.31\times10^8
\\
{\tt B3a}& 12.7& 1.16 \times 10^{-3}& 100
&10 & 1.0\times10^{-2} &9.17~\,&10^{46}
&5.58&2.22 \times 10^3&1.31\times10^8
\\
\hline
\end{array}
$\end{center}
}
\end{minipage}
\end{table*}

\section{SIMULATIONS}
\label{s.simulations}

In the following subsections
we describe a number of simulations that exhibit a wide range of phenomena
when we vary fundamental parameters such as the jet energy flux
($F_E$),
the jet density ratio ($\eta$)
and the filling factor of clouds ($f_\mathrm{c}$)
obstructing the path of the jet.
The jet parameters in the simulations
are presented in Table~\ref{table.jet},
and the characteristics of the clouds and intercloud media
are given in Table~\ref{table.cloudfield}.
In each simulation
the X-ray emitting inter-cloud medium has temperature
$T_\mathrm{x}=10^7\ \mathrm{K}$,
and the corresponding isothermal sound speed
is our unit of velocity in the simulations,
$v_0=v\subx=365\ \mathrm{km}\ \mathrm{s}^{-1}$.
Our unit for time measurement is
$t_0=v_0 / 1\ \pc=2.68\times10^3\ \mathrm{yr}$.

The jet is defined by a group of cells on the left boundary
where the density, pressure and velocity are kept constant.
Elsewhere on the left boundary a reflection condition is applied,
representing the effect of bilateral symmetry
of a galaxy with equally powerful jet and counter-jet.
The boundary cells on other sides of the grid
are subject to an open boundary condition:
in each cycle of the hydrodynamic computation,
the boundary cell contents are copied into the adjacent ghost cells.
This one-way boundary allows hydrodynamic waves to propagate off the grid.

In a journal paper, one is necessarily limited to
showing illustrative snapshots from simulations.
Movies of the simulations described in the following section,
which show the detailed dynamical evolution,
may be obtained from
{\tt http://macnab.anu.edu.au/radiojets/}.

\subsection{Model {\tt A1}: high density, overpressured jet}

This simulation exhibits both expected and unexpected features
of jets interacting
with a distribution of clouds in an active nucleus.
The jet is initially unimpeded by clouds and
generally behaves like a jet in a uniform medium showing the usual bow-shock,
terminal jet-shock and backflow.
Then two small clouds are overtaken by the bow-shock
and the overpressure in the cocoon
drives radiative shocks into them.
Subsequently the jet backflow is distorted by the reverse shock
emanating from the upper of the two clouds.
The interaction with the cloud field becomes more
dramatic when the apex of the bow shock strikes one of the larger clouds
just below the jet axis.
Again this drives a strong radiative shock into the cloud
and the larger reverse shock deflects both
the jet and backflow sideways.
The bow-shock then intersects the larger of the two clouds and this
produces an additional reverse shock that deflects the jet further sideways.
The jet continues onwards
past these two clouds whlist the dense clouds in the wake of the bow-shock
(i.e. immersed within the jet cocoon)
are gradually disrupted by the gas surrounding them.
A large amount of turbulence is
generated within the jet cocoon by the reverse shocks
from the bow-shock impacted clouds and the jet
flow is clearly affected but not disrupted.
At late times many of the clouds immersed within the
cocoon show trails resulting from
the ablation of gas within the turbulent cocoon.
However, the details of these features will be uncertain
until we are able to resolve them fully and
subdue the effects of numerical mass diffusion.

Notable features of this interaction are:
(I) The jet does not directly touch any of the clouds.
The radiative shocks within the clouds are all produced
as a result of the overpressure behind the bow-shock.
(II) The importance of the reverse shocks within the cocoon
in deflecting the jet and making the cocoon more turbulent.
(III) The survival of the jet despite these interactions
(aided by heaviness of the jet in this simulation,
$\eta=0.3835$).

\begin{figure*}\begin{minipage}{180mm}
\begin{center}
$\begin{array}{cc}
\ifthenelse{\isundefined{\rainbow}}{
\includegraphics[width=9cm]{f05a.eps}
&\includegraphics[width=9cm]{f05b.eps}
}{
\includegraphics[width=9cm]{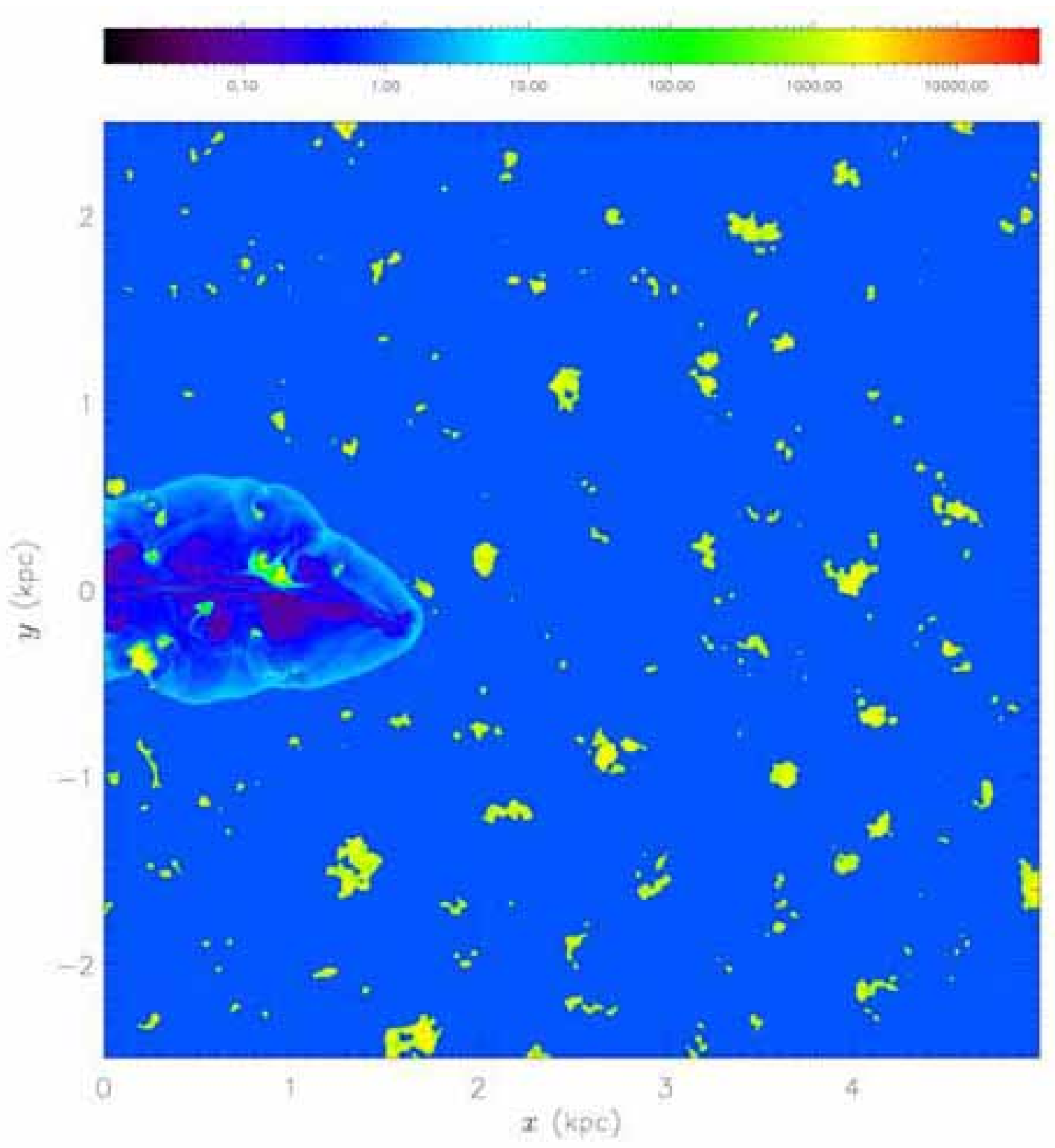}
&\includegraphics[width=9cm]{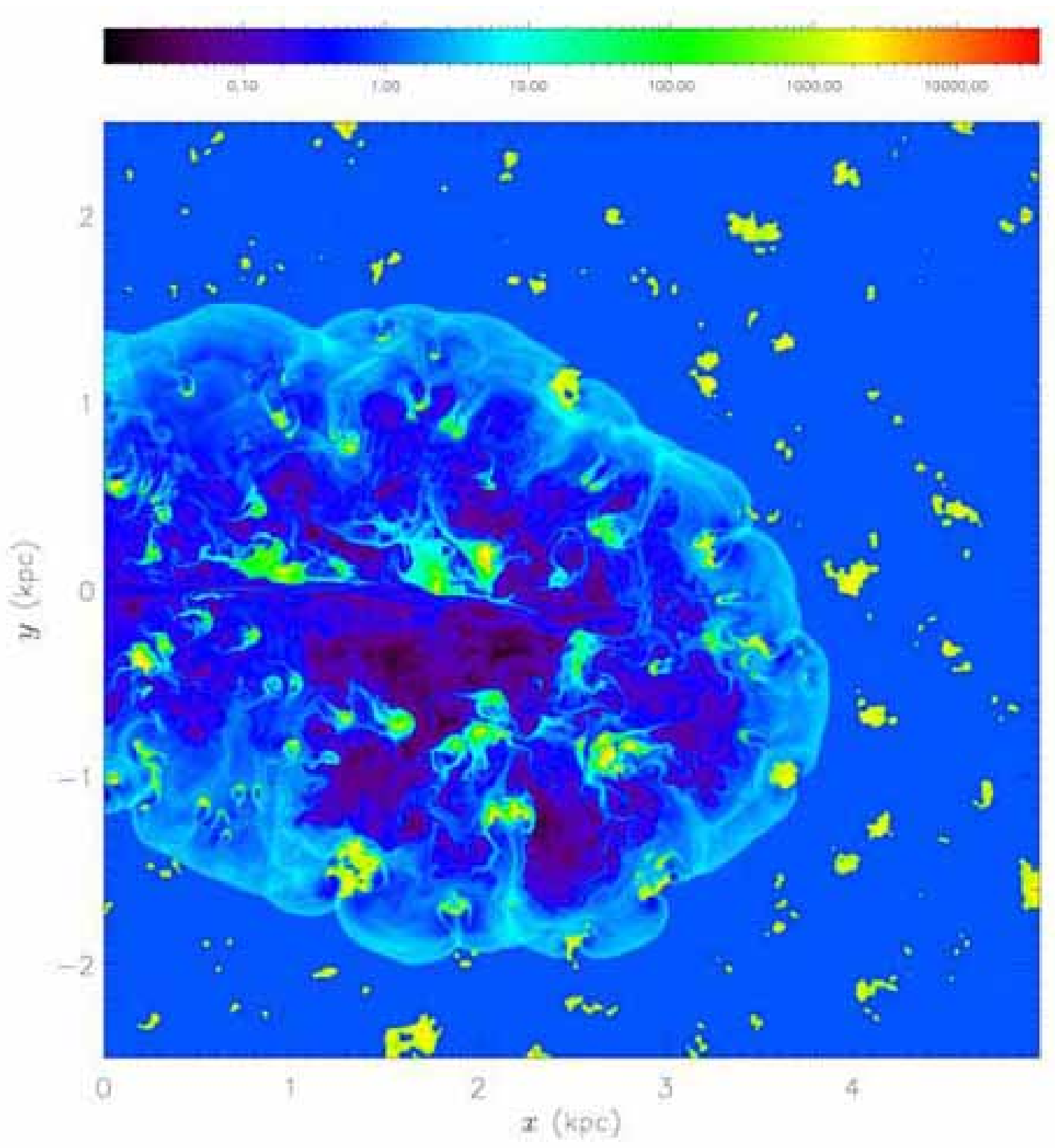}
}
\end{array}$
\caption{
Logarithmically scaled map of $\rho/\rho_\mathrm{x}$
in the model {\tt A1} at time $t=5t_0$ (left panel)
and $19t_0$ (right panel).
The intercloud medium has
number density $n_\mathrm{x}=0.01\ \mathrm{cm}^{-3}$.
}
\label{fig.slab02b.1000.dens}
\label{fig.slab02b.3800.dens}
\end{center}
\end{minipage}\end{figure*}

\begin{figure*}\begin{minipage}{180mm}
\begin{center}
$
\begin{array}{cc}
\ifthenelse{\isundefined{\rainbow}}{
\includegraphics[width=9cm]{f06a.eps}
&\includegraphics[width=9cm]{f06bv.eps}
\\
\includegraphics[width=9cm]{f06c.eps}
&\includegraphics[width=9cm]{f06dv.eps}
}{
\includegraphics[width=9cm]{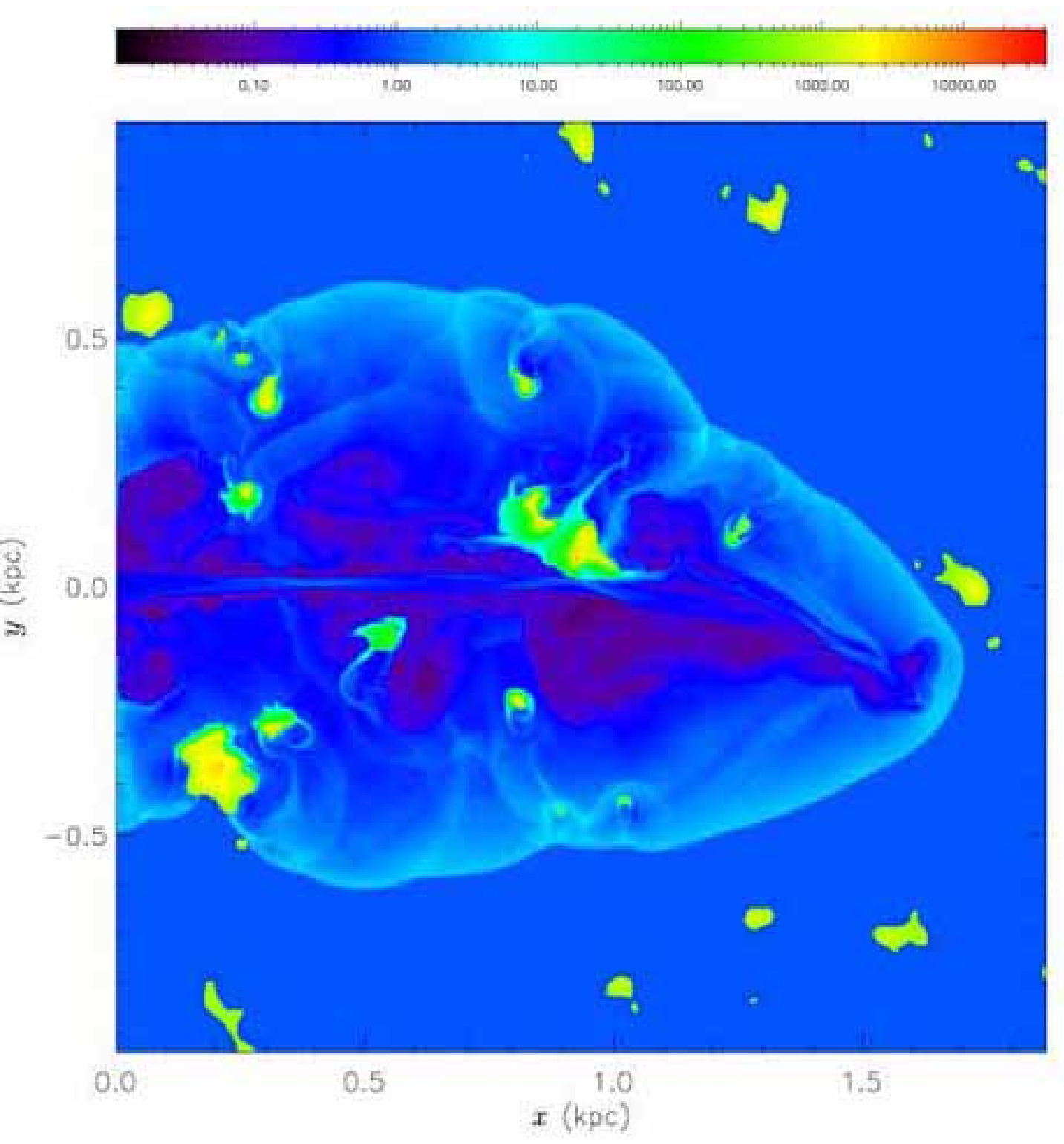}
&\includegraphics[width=9cm]{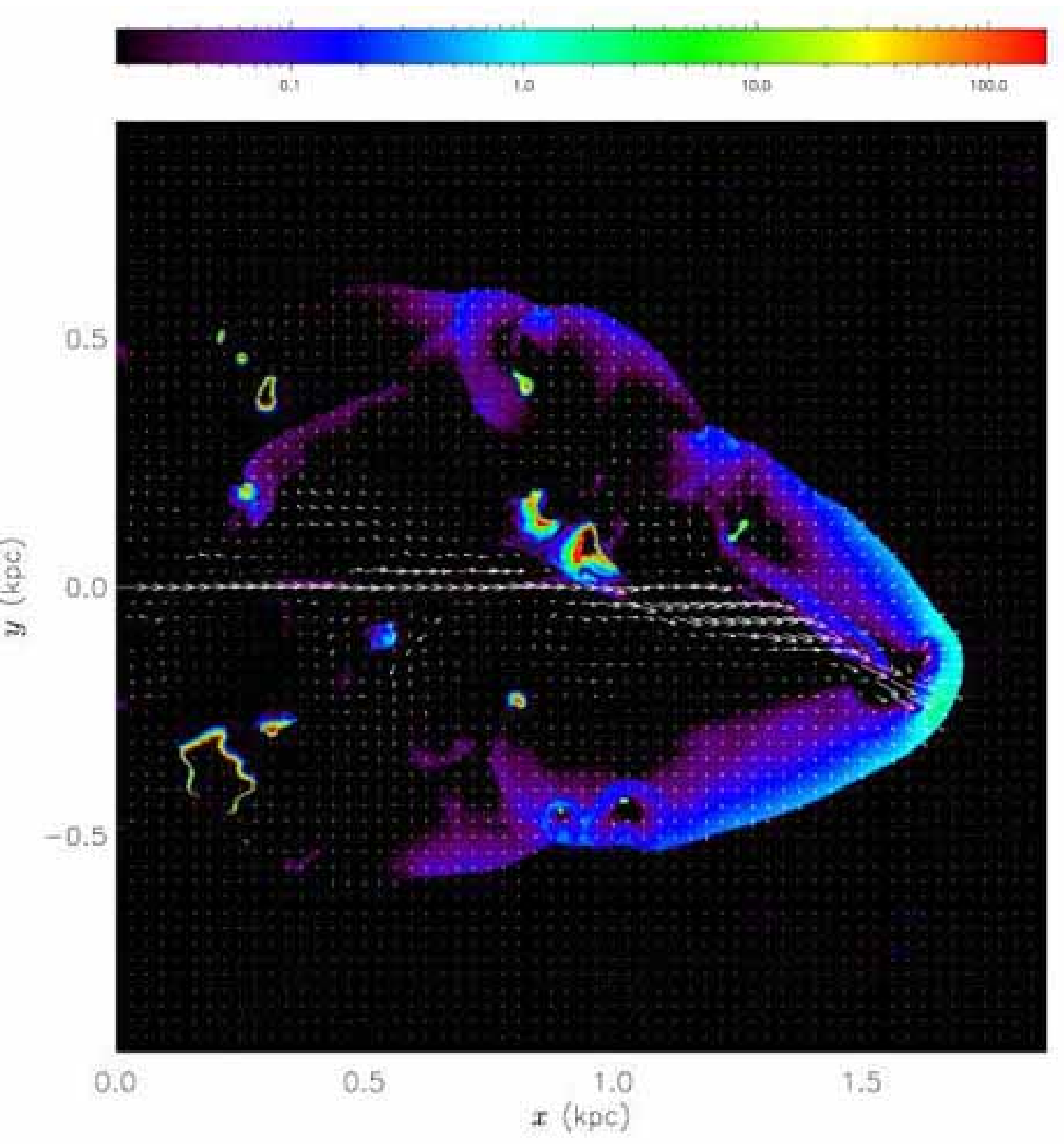}
\\
\includegraphics[width=9cm]{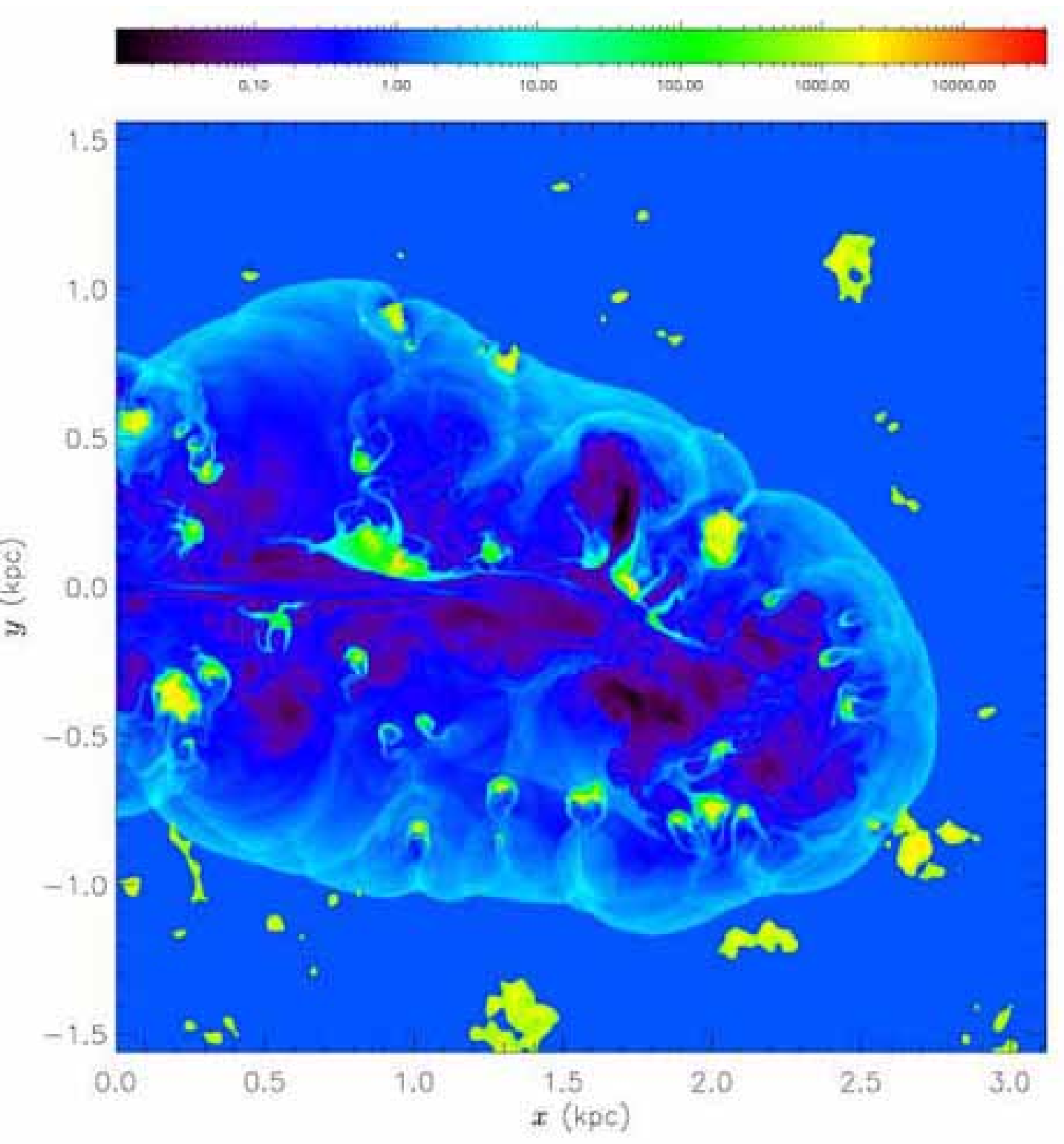}
&\includegraphics[width=9cm]{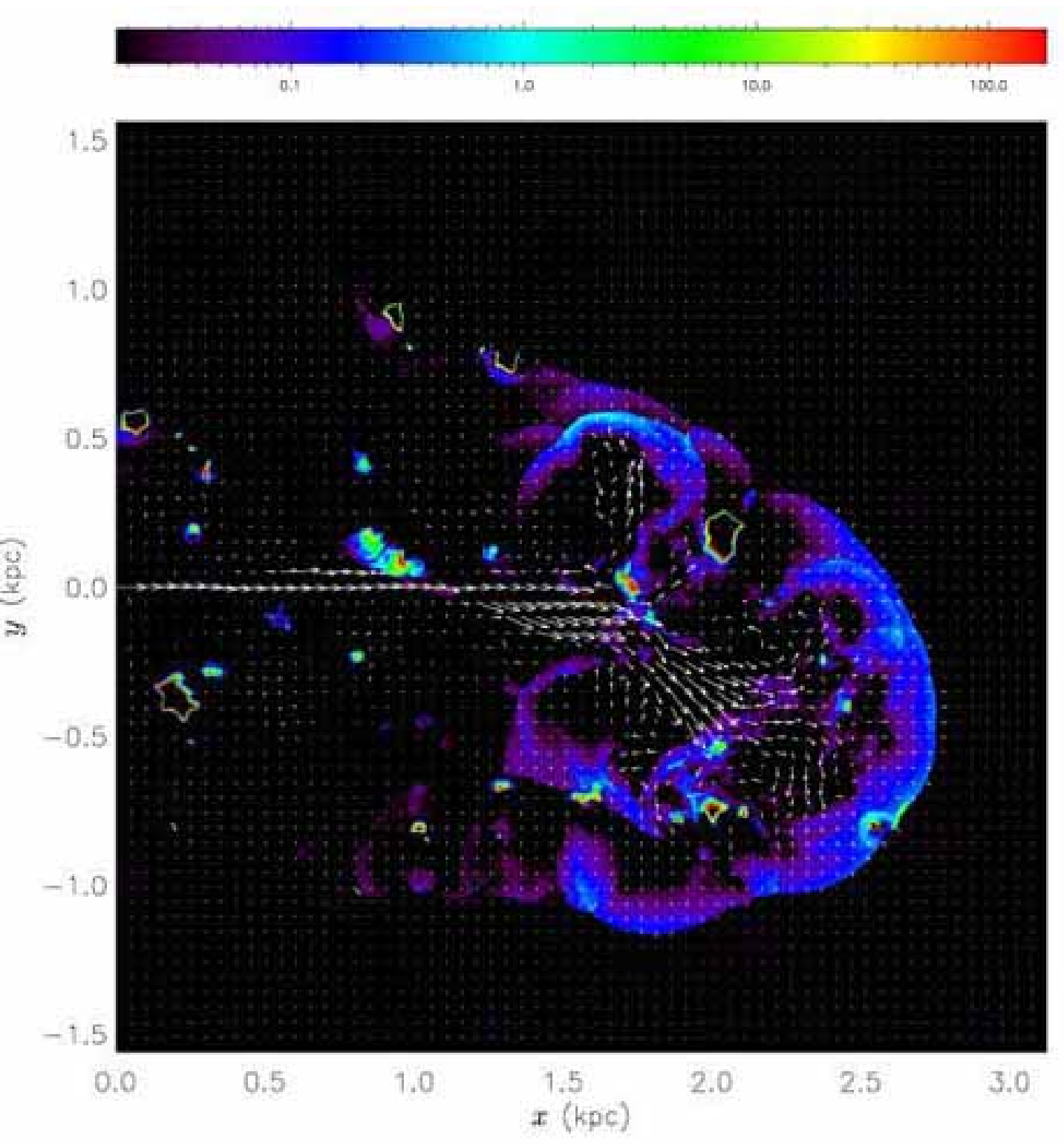}
}
\end{array}
$
\caption{
Subregions of model {\tt A1}
showing the area disturbed by the jet
at the times $t=5t_0$ (upper row) and $10t_0$ (lower row).
The left panels show density $\rho/\rho_\mathrm{x}$
and the right panels show velocity vectors
superimposed on maps of the volumetric rate of cooling.
The cooling is measured in units of
$\mathrm{erg}\ \mathrm{cm}^{-3}\ \mathrm{s}^{-1}$,
but the upper end of the scale is deliberately saturated
to keep the bow shock visible.
The actual maximum rate of radiative cooling
in the cloud shocks
is $1.8\times10^3\ \mathrm{erg}\ \mathrm{cm}^{-3}\ \mathrm{s}^{-1}$
when $t=5t_0$.
Reflected and refracted shocks are visible
at distances of order $10\ \pc$ around some of the clouds.
}
\label{fig.slab02b.1000.dens.cut}
\label{fig.slab02b.1000.cool.cut}
\label{fig.slab02b.2000.dens.cut}
\label{fig.slab02b.2000.cool.cut}
\end{center}
\end{minipage}\end{figure*}

The left panel of Figure~\ref{fig.slab02b.1000.dens}
shows the jet interacting with clouds
during an early stage of the simulation,
$t=5t_0$.
The bow shock and therefore the influence of the jet
have only propagated about $1.7\ \kpc$ forward
and $0.5\ \kpc$ laterally,
and most of the clouds remain in their initial condition.
The upper panels of Figure~\ref{fig.slab02b.1000.dens.cut}
show details of the subregion immediately surrounding the jet
at this time of the simulation.
Within the high-pressure region bounded by the bow shock,
all clouds are shocked.
The cloud shocks are the intensely cooling regions on the cloud perimenters
visible in the upper-right panel of Figure~\ref{fig.slab02b.1000.cool.cut}.
The cloud density is sufficiently high
that the shocks have penetrated only a few pc at this stage
--- a small depth compared to a cloud thickness.
The other noteworthy cooling component visible in
the cooling map (Figure~\ref{fig.slab02b.1000.cool.cut})
is the X-ray emission from the bow shock.
This is more intense near the apex of the bow shock than at its sides.

The lower panels of Figure~\ref{fig.slab02b.2000.dens.cut}
show the density and cooling
at a later time, $t=10t_0$,
when the jet is more advanced and the clouds are more eroded.
Some clouds that lay in the path of the jet at time $5t_0$
(located at $(0.85,0.15)\,\kpc$ and $(0.95,0.10)\,\kpc$)
are now completely shocked and destroyed.
Other clouds have survived closer to the nucleus
but with less damage because they are out of the jet's path,
e.g. those at $(0.3,0.4)\,\kpc$ and $(0.3,0.25)\,\kpc$.
The jet survives an oblique interaction with one group of eroded clouds
near $(0.9,0.15)\,\kpc$,
but its width increases to the right of this point.
A second cloud collision near $(1.7,0.0)\,\kpc$
deflects the jet into the negative $y$ direction,
and causes its decollimation
into a turbulent bubble of hot plasma
accumulating at the apex of the bow shock.
The head of the jet is unsteady
and changes direction completely within a few $t_0$.

The density distribution at time $t=19t_0$,
near the finish of the simulation,
appears in the right panel of Figure~\ref{fig.slab02b.3800.dens}.
Clouds are visible in all stages of shock compression.
The two major cloud collisions,
around $(1.00,0.05)\,\kpc$
and $(1.75,0.00)\,\kpc$,
now leave the jet well collimated and relatively straight.
The jet loses collimation at a third collision,
around $(2.5,^-\!\!0.2)\,\kpc$.
The turbulent bubbles around the head of the jet have grown,
and spurt between the clouds in several different directions.
As was the case at $t=5t_0$ and $10t_0$,
the clouds situated off to the sides
have survived much better than clouds
that lay close to the direct path of the jet.

In summary, whether a cloud behaves radiatively and contracts,
or adiabatically and ablates
is affected by its size, location and density.
Small clouds are more effectively adiabatic and less durable than large clouds.
This may be partly because their exposed surfaces
are large compared to their internal volume,
and partly because they are geometrically narrower
and on average less dense than large clouds.
Some of the smallest swell and develop ablative tails
almost as soon as they enter the cocoon,
e.g. those between
$(1.3,^-\!\!0.7)\,\kpc$ and
$(2.1,^-\!\!0.7)\,\kpc$
when $t=10t_0$, in Figure~\ref{fig.slab02b.2000.dens.cut}.
A larger cloud near $(0.2,^-\!\!0.5)\,\kpc$
was within the cocoon for most of the simulation
and yet it was barely affected,
even until $t=19t_0$ (right panel, Figure~\ref{fig.slab02b.3800.dens}).
Qualitatively, each cloud appears to suffer one of three possible fates.
Small clouds are destroyed quickly
regardless of their position.
Large clouds near the path of the jet
affect the jet through reflection shocks
(even if the jet doesn't strike directly)
but they inflate and ablate away in the long term.
Large bystander clouds situated well off the jet axis
endure almost indefinitely

Interaction with clouds alters the propagation
and shape of the bow shock.
The bow shock is severely retarded when it strikes a cloud
and refracts through the gaps between clouds.
Thus the bow shock surface is conspicuously indented
(``dimpled'') for as long as it remains amidst the cloud field.
We would expect this effect to manifest in 3D simulations also,
although the extra degree of freedom in 3D
is also likely to bring about some differences.

The jet generally remains identifiable
and collimated
after deflection in two or at most three cloud interactions.
These interactions need not be direct:
it is sufficient for the jet to interact with
the reverse shocks that reflect off a cloud
lying within several times $r_\mathrm{j}$ off the jet's path.
These backwashing shocks spread continually in the turbulent cocoon;
they appear to act like a ``cushion'' against collision between jet and cloud.
More than three consecutive cloud interactions
disrupts the jet.

\subsection{Low density, overpressured jets}

\subsubsection{Jet stability properties, model {\tt B0}:}
\label{results.eruptyy}

Our main series of simulations depicts a more powerful jet
($F_E=10^{46}~\mathrm{erg}~\mathrm{s}^{-1}$)
with a much lower density contrast
($\eta=1.16\times10^{-3}$)
and a large overpressure ($\xi=100$).
The initial internal Mach number of the jet is $M=12.72$ at the nucleus.
The jet radius is $r_\mathrm{j}=10\ \pc$,  the square grid is $4\ \kpc$ wide and the resolution is
$2000\times2000$ cells.
The background gas again has a temperature $T_\mathrm{x}=10^7~\mathrm{K}$
but the density is raised to $n_\mathrm{x}=1\ \mathrm{cm}^{-3}$.
These parameters are suggested by the observations of GPS sources
\citep[e.g. in the summary by ][]{conway02a}.

\begin{figure}
\begin{center}
\ifthenelse{\isundefined{\rainbow}}{
}{
}
$\begin{array}{cc}
\ifthenelse{\isundefined{\rainbow}}{
\includegraphics[width=6cm,trim=1.5cm 0.5cm 1.5cm 0.0cm]{f07b.eps}
\\\includegraphics[width=6cm, trim=1.5cm 0.5cm 1.5cm 0.0cm]{f07c.eps}
\\\includegraphics[width=6cm, trim=1.5cm 0.5cm 1.5cm 0.0cm]{f07av.eps}
}{
\includegraphics[width=6cm,trim=1.5cm 0.5cm 1.5cm 0.0cm]{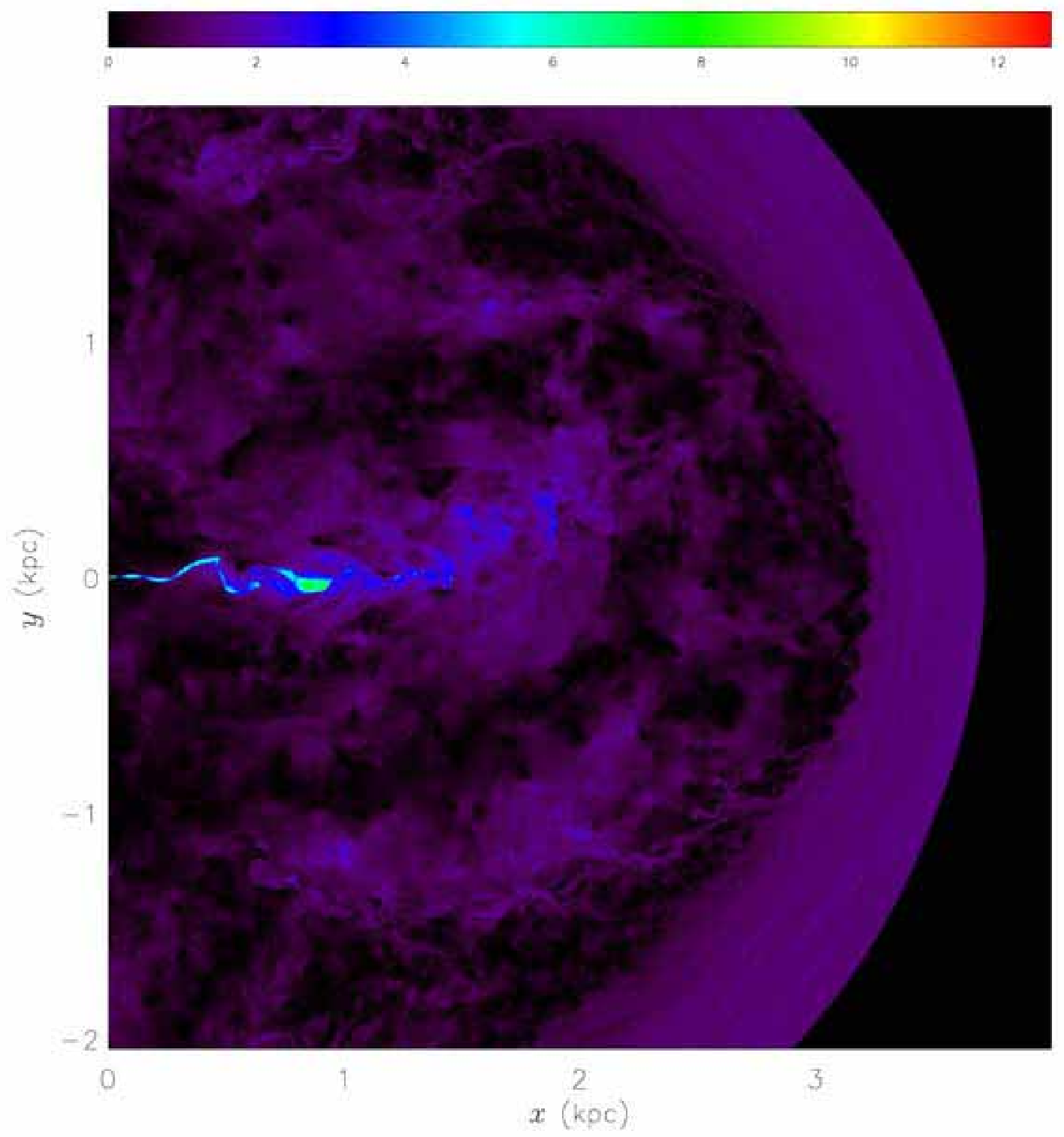}
\\\includegraphics[width=6cm, trim=1.5cm 0.5cm 1.5cm 0.0cm]{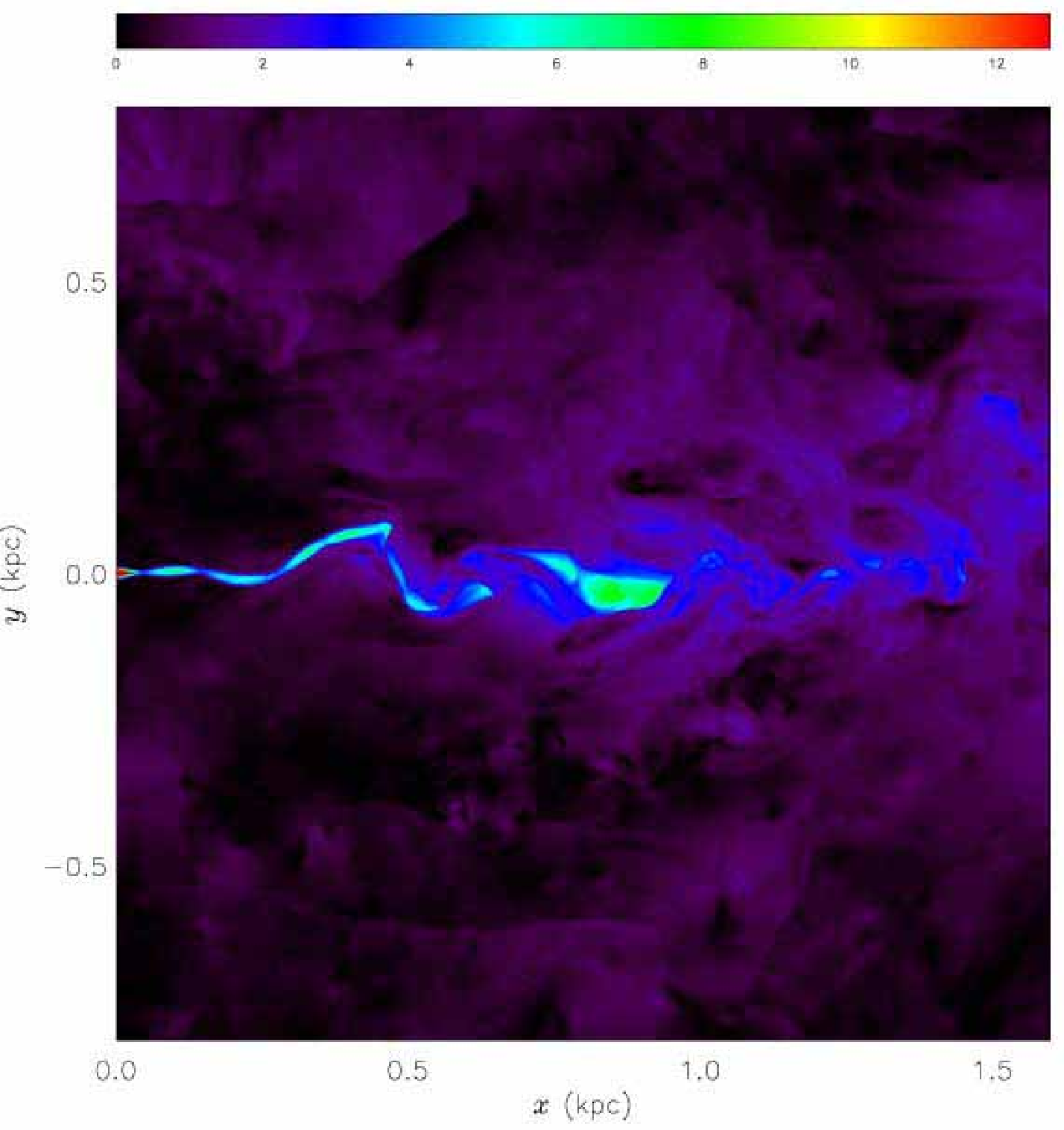}
\\\includegraphics[width=6cm, trim=1.5cm 0.5cm 1.5cm 0.0cm]{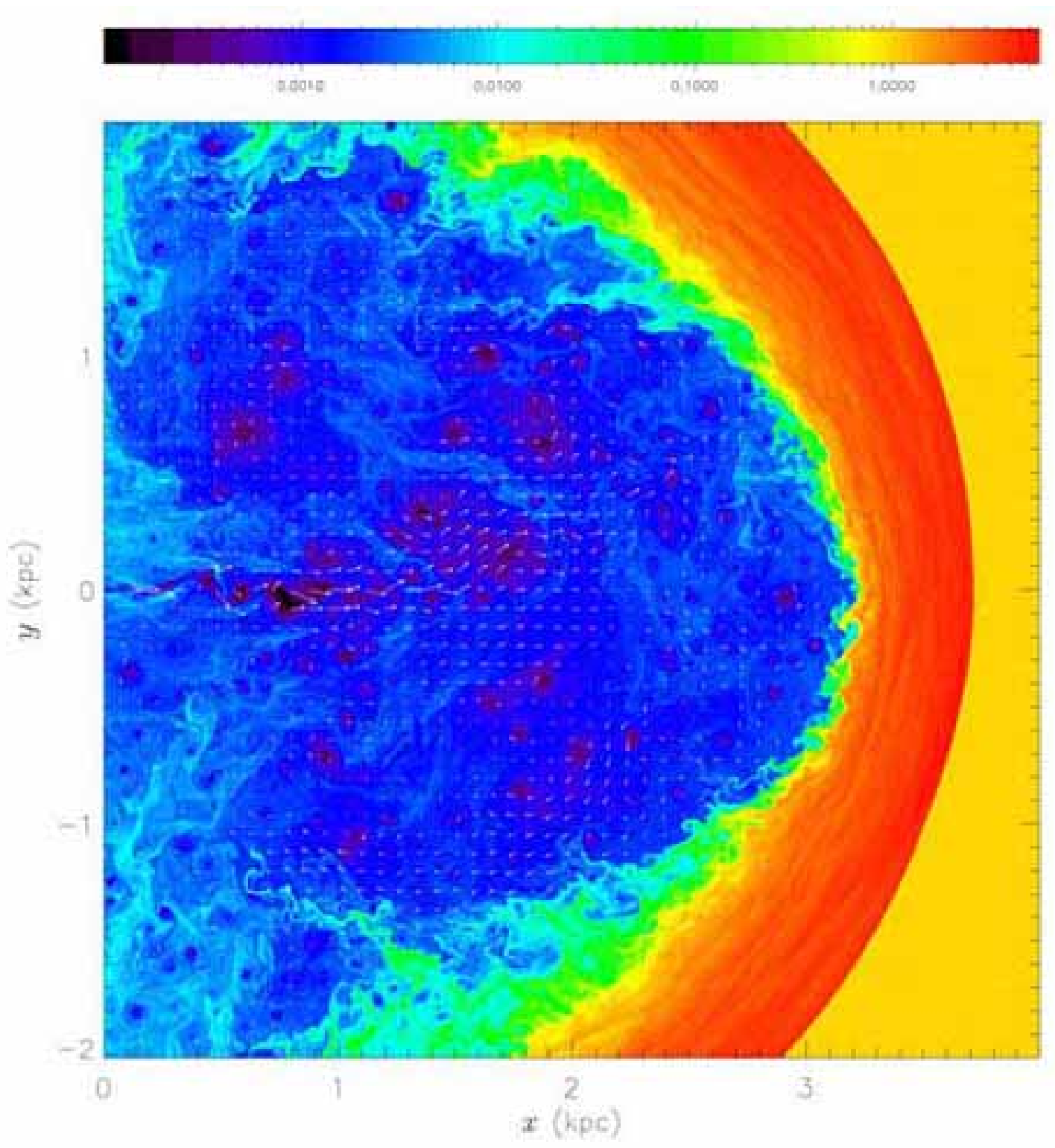}
}
\end{array}$
\caption{
Snapshots of the cloudless test model {\tt B0} at time $t=48t_0$.
The upper panels show the local Mach number:
the full computational grid,
and a subregion around the jet.
The lower panel maps density,
$\rho/\rho_\mathrm{x}$, overlain with velocity vectors.
}
\label{fig.eruptyy.0320.dens}
\label{fig.eruptyy.0320.mach}
\end{center}
\end{figure}

One of the aims of this work
is to distinguish the effects of an inhomogeneous medium
from the phenomena that are intrinsic to the jet itself.
Therefore we have performed a control simulation
containing only the jet and an initially uniform intercloud medium.
The main large-scale features at the end of this simulation
(see Figure~\ref{fig.eruptyy.0320.dens})
are:
an outer shell of highly overpressured gas processed through the bow shock,
enclosing a turbulent cocoon where jet plasma mixes with external gas.
The length of the jet is small compared to
the diameter of the bubble surrounding it.
The jet is highly unstable:
it stays recognisably straight throughout its first $0.1\ \kpc$
but at greater distances from the origin
it flaps transversely with increasing amplitude,
losing collimation at positions beyond $\sim0.5\ \kpc$ from the origin.
This highlights a feature of two--dimensional simulations
that may not be present in three dimensions.
The instability in this jet is driven by 2D turbulence in the cocoon
that results from the mixing of gas at the edge,
although the instability is not as rapid as in
the ensuing cases with cloudy media.
The turbulent driving is probably stronger here
because of the persistence of vortices
associated with the well--known inverse cascade in two dimensions. The instability evident here is seeded by a very small amount of numerical noise in the initial data.

\citet{hardee1988} performed a linear stability analysis
of the instabilities of two-dimensional slab jets.
Two qualitative kinds of instabilities occur.
An unstable  pinching mode
is responsible for diamond shocks
at quasi-periodic locations along the length of the jet.
Sinusoidal instabilities are responsible for
transverse oscillations,
which disrupt and ultimately decollimate the jet.
According to the analytical and numerical results of \citet{hardee1988},
the maximally growing fundamental mode has wavelength
\begin{equation}
\lambda^*\approx
{{0.70}\over{\alpha_\mathrm{sj}}}
{{2\pi M r_\mathrm{j}}\over{0.66+\eta^{1/2}}} \, ,
\label{hardee.wave}
\end{equation}
an $e$-folding length scale of
\begin{equation}
l^*_e \approx
{{1.65}\over{\alpha_\mathrm{sj}}}
M r_\mathrm{j}
\ ,
\label{hardee.growth}
\end{equation}
and angular frequency
\begin{equation}
\omega^*\approx 1.5\ \alpha_\mathrm{sj} a_\mathrm{e} / r_\mathrm{j}
\ ,
\label{hardee.omega}
\end{equation}
where
$a_\mathrm{e}$
is the adiabatic sound speed in the cocoon
immediately exterior to the jet surface,
the parameter $\alpha_\mathrm{sj}\rightarrow 1$ for $M\gg1$,
and
$\eta$, in this case, is the density contrast between the jet
and its immediate surroundings within the cocoon.

In simulation {\tt B0},
the internal Mach number within the jet is reduced to typically $M\approx6$,
downstream of the first diamond shock
(see Figure~\ref{fig.eruptyy.0320.mach}).
The cocoon gas immediately surrounding the jet
has a density typically $\sim4\rho_\mathrm{j}$
and adiabatic sound speed $a_\mathrm{e}\sim800v_0$,
where $v_0\approx365\ \mathrm{km} \> \mathrm{s}^{-1}$
is the isothermal sound speed of
the undisturbed medium outside the bow shock.
Upon substitution into equations~(\ref{hardee.wave}) and (\ref{hardee.growth}),
these values imply exponential growth of sinusoidal instabilities
on scales $l\sim100\ \mathrm{pc}$
and with wavelengths $\lambda\sim200\ \mathrm{pc}$.
This is consistent with the growing oscillations that are
apparent in individual frames of our simulations,
e.g. in Figure~\ref{fig.eruptyy.0320.mach}.
From equation (\ref{hardee.omega}),
the implied period of the instability is
$T^*<0.052\ t_0$.
The simulated jet thrashes back and forth transversely
within this typical time.

The sinusoidal transverse instability
revealed in this homogeneous ISM simulation
are similar in wavelength and growth time
to instabilities manifest in the inhomogeneous simulations discussed below.
However,
the appearance of this instability in this ``clean'' simulation shows that
the instability can be properly regarded as
a Kelvin-Helmholtz instability of the jet itself
--- albeit, in the inhomogeneous cases,
stimulated by the turbulence in the cloudy medium.
Note that growth and wavelength of the transverse KH instability
at first sight seems to be suggestive of a transonic jet.
However, the comparison with the \cite{hardee1988}
expressions for maximally growing wavelength and growth time
shows that our simultions are consistent with
the expected behavior of \emph{supersonic} jets with the quoted Mach numbers.

\subsubsection{Small cloud filling factor: model {\tt B1}}

\begin{figure}
\begin{center}
$\begin{array}{cc}
\ifthenelse{\isundefined{\rainbow}}{
\includegraphics[width=7.3cm]{f08av.eps}
\\\includegraphics[width=7.3cm]{f08b.eps}
\\\includegraphics[width=7.3cm]{f08c.eps}
}{
\includegraphics[width=7.3cm]{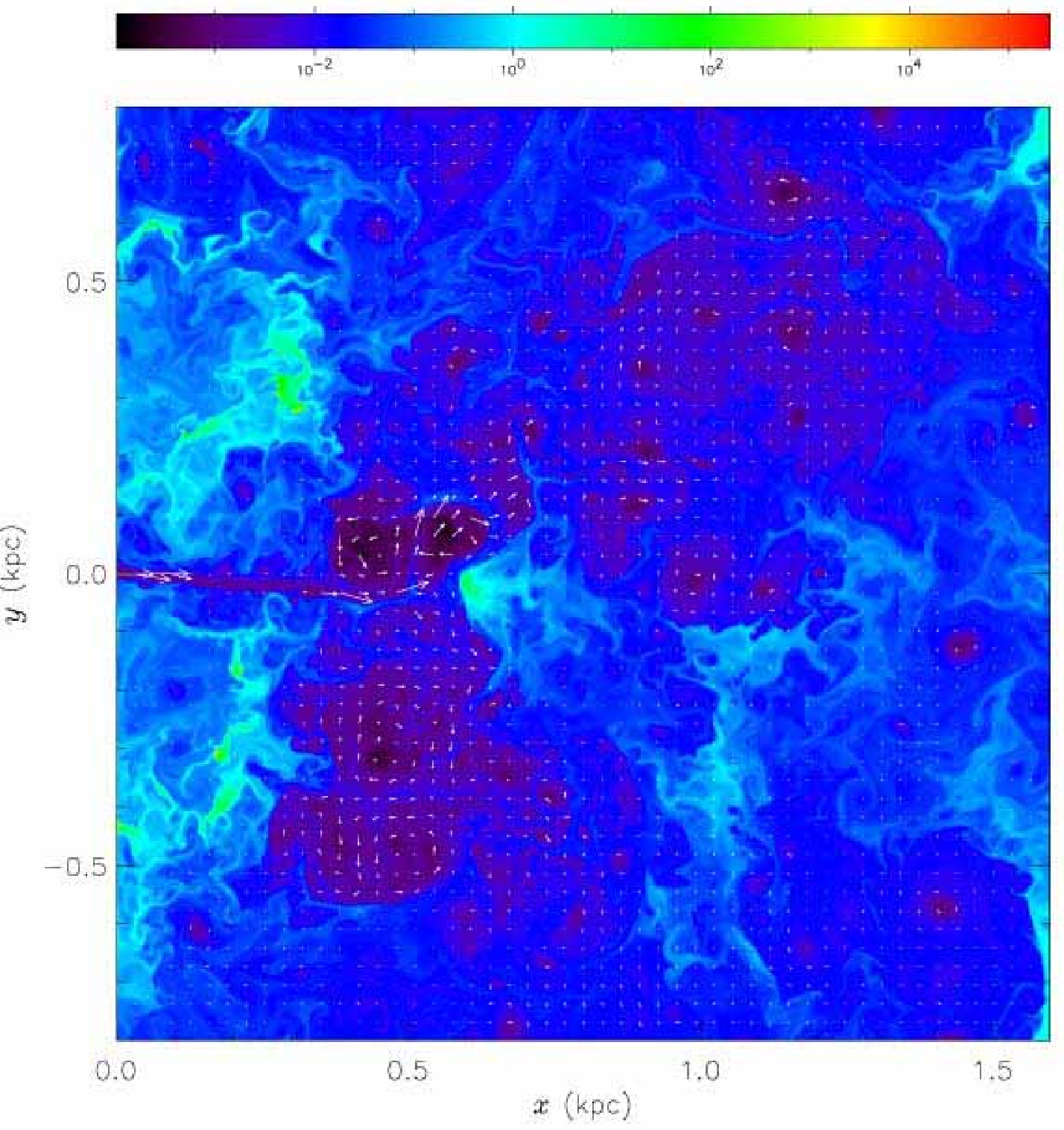}
\\\includegraphics[width=7.3cm]{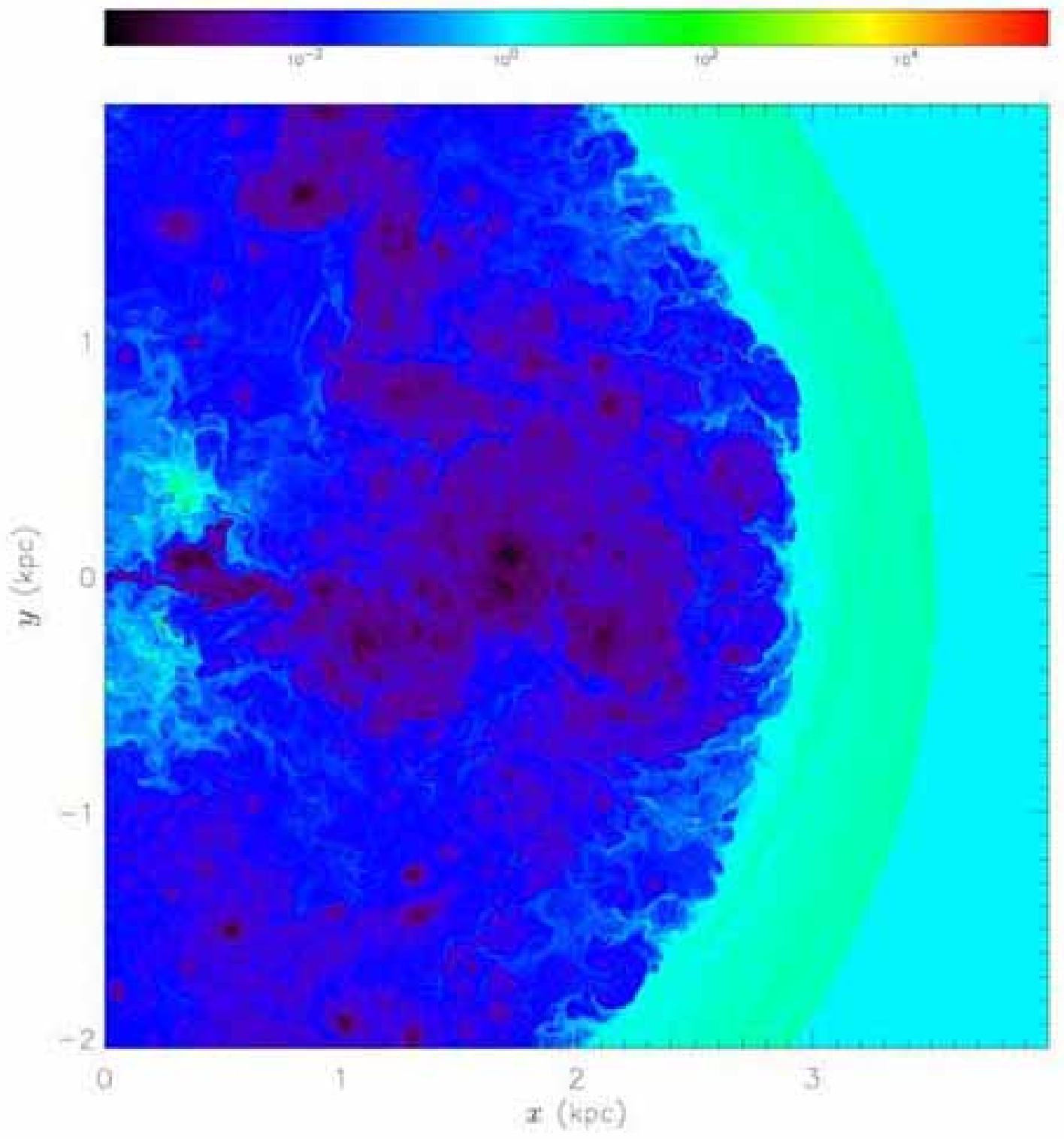}
\\\includegraphics[width=7.3cm]{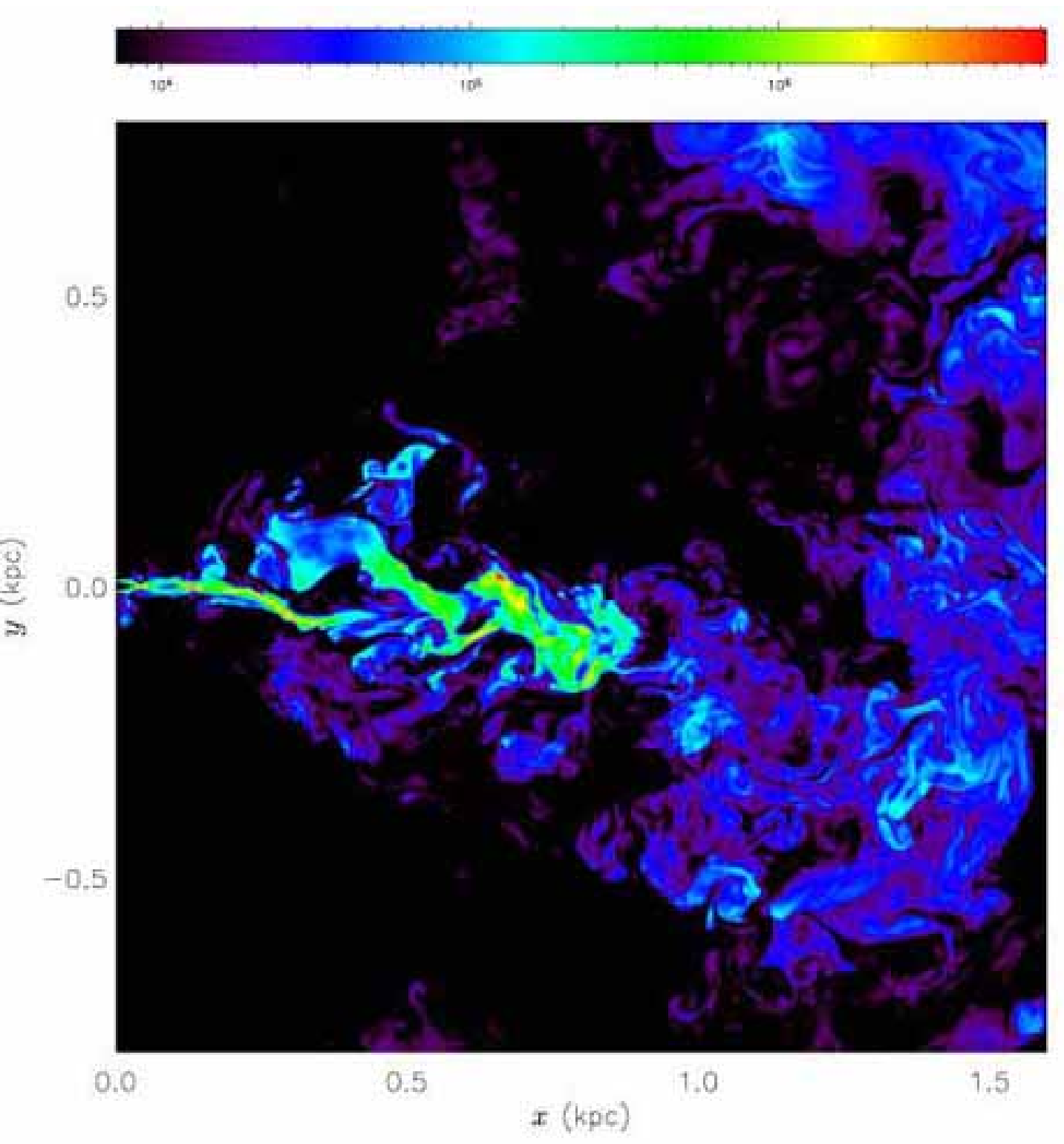}
}
\end{array}$
\caption{
Snapshots of model {\tt B1}.
Top-left: 
$\rho/\rho_\mathrm{x}$
and velocity field
in a subregion at time $t=24t_0$.
Top-right:
$t=48t_0$, $\rho/\rho_\mathrm{x}$ in the full grid.
Bottom-left:
$t=48t_0$,
simulated radio emissivity in a subregion
(the lowest background intensity is $1/1024$ times the peak).
}
\label{fig.erupt03.0160.cut.dens}
\label{fig.erupt03.0320.dens}
\label{fig.erupt03.0320.cut.glow}
\end{center}
\end{figure}

This is our simulation with the fewest clouds:
their filling factor is $f_\mathrm{c}=4.1\times10^{-3}$
($\sigma_\mathrm{min}=2.6$)
within a region spanning radially between $0.1\kpc$ and $0.5\kpc$
from the nucleus.
All of the clouds are quickly overtaken by the expanding bow shock.
As in model {\tt A1},
each cloud is shocked by the overpressure of the cocoon
(up to $5\times10^4$ times greater the external pressure),
and contracts as it cools radiatively.
Some clouds are accelerated away from their initial positions
by the thrust of the jet.
For some other clouds,
radiative cooling eventually proves inadequate;
the tendency to contract is overcome,
and the cloud's fate is expansion and ablation.
Halfway through this simulation at time $t=24t_0$,
(top panel, Figure~\ref{fig.erupt03.0160.cut.dens}),
shows the destruction of one such cloud in the path of the jet,
situated at $(0.6,0.0)\,\kpc$
but ablating away across about $0.2\ \kpc$ to the right.
Meanwhile,
a few extremely dense and compact clouds
persist in apparent safety off to the sides of the jet,
within a few hundred $\pc$ of the nucleus.
They appear to survive indefinitely and
appear scarcely changed by the time the simulation ends,
as seen in the middle panel of Figure~\ref{fig.erupt03.0320.dens}.
In comparison with Figure~\ref{fig.eruptyy.0320.dens}
we see much more dense gas (derived from the ablated clouds)
intermixed with jet plasma in the cocoon.

The last panel in Figure~\ref{fig.erupt03.0320.cut.glow}
represents the appearance of the jet in synchrotron emission
at late stages of the simulation.
Here we have used the approximation
\citep[e.g.][]{saxton2001,saxton2002a}
that the emissivity
$j_\nu\propto \varphi p^{(3+\alpha)/2}$,
where $p$ is the pressure,
$\varphi$ is the relative concentration of mass originating from the jet
(a scalar tracer that is passively advected in our code) and $\alpha$ is the spectral index that we take to be $0.6$.
Several diamond shocks are visible
in the well collimated section of the jet closest to the nucleus.
The brighter, barred ``fishbone'' shaped pattern
seen at distances $x>0.4\ \kpc$
is a transient and typical manifestation of
the transverse instability of the jet
(as in \S~\ref{results.eruptyy}).

\subsubsection{Intermediate cloud filling factor: model {\tt B2}}

\begin{figure}
\begin{center}
$\begin{array}{cc}
\ifthenelse{\isundefined{\rainbow}}{
  \includegraphics[width=7.3cm]{f09av.eps}
\\\includegraphics[width=7.3cm]{f09b.eps}
\\\includegraphics[width=7.3cm]{f09c.eps}
}{
  \includegraphics[width=7.3cm]{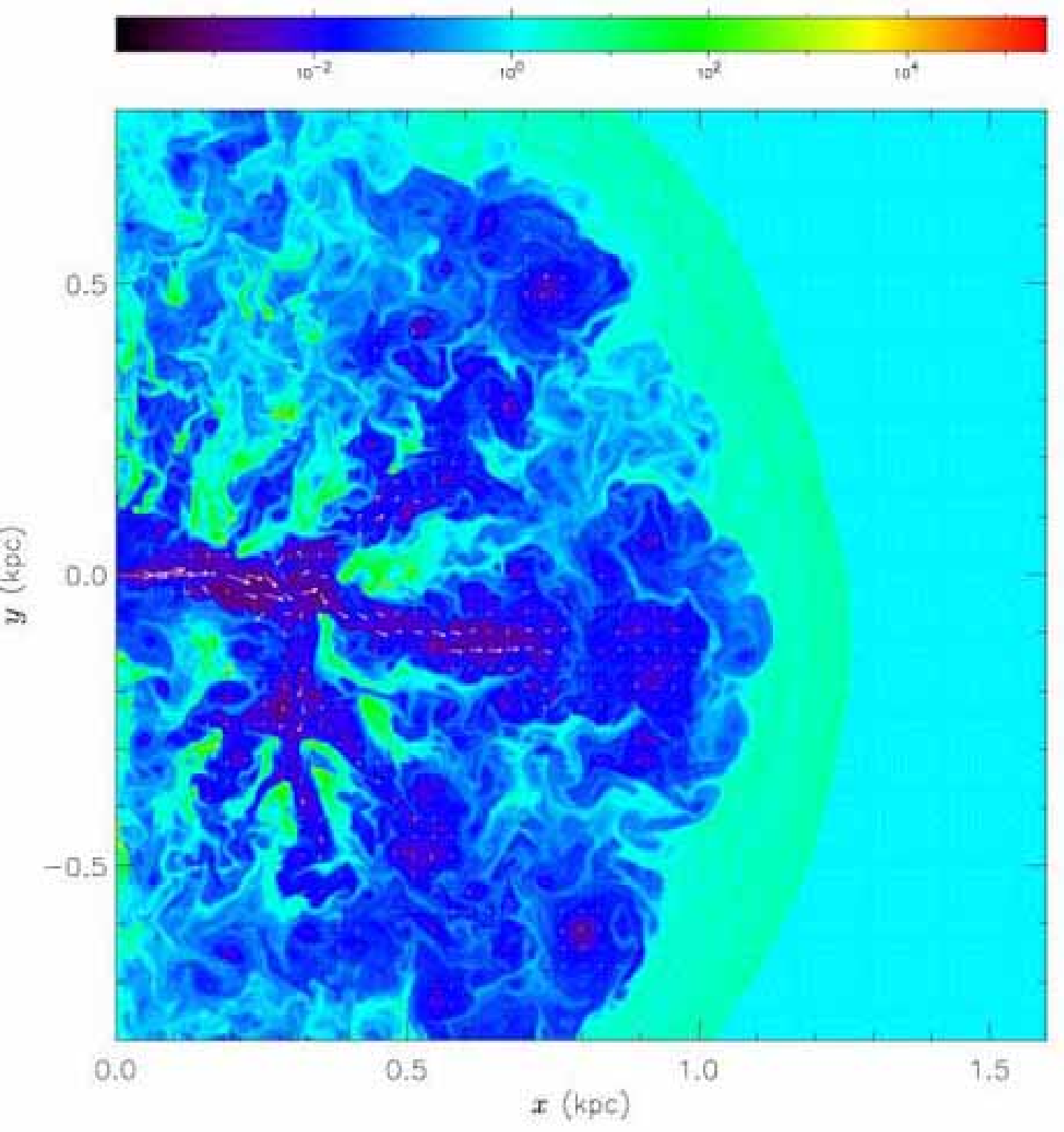}
\\\includegraphics[width=7.3cm]{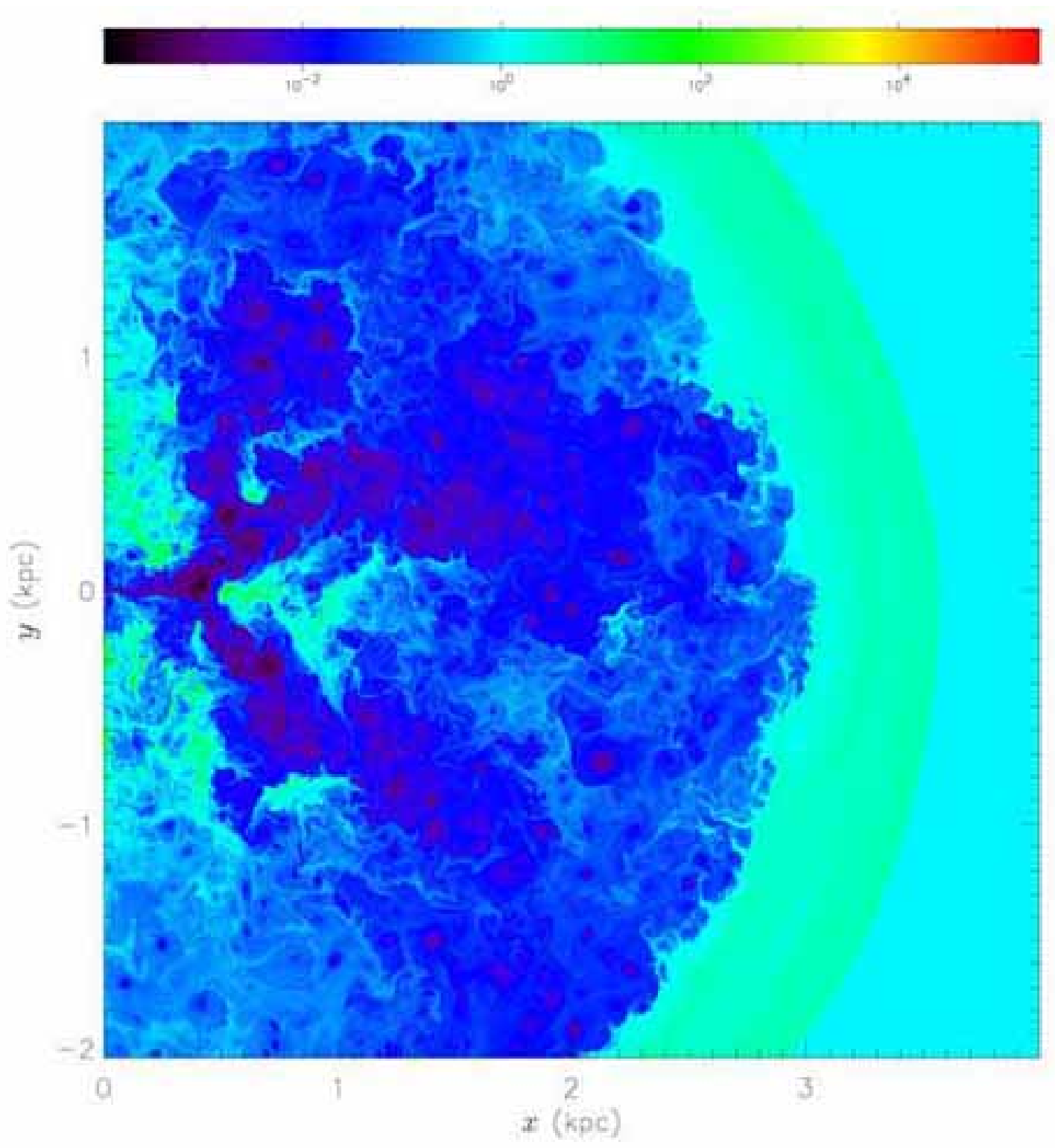}
\\\includegraphics[width=7.3cm]{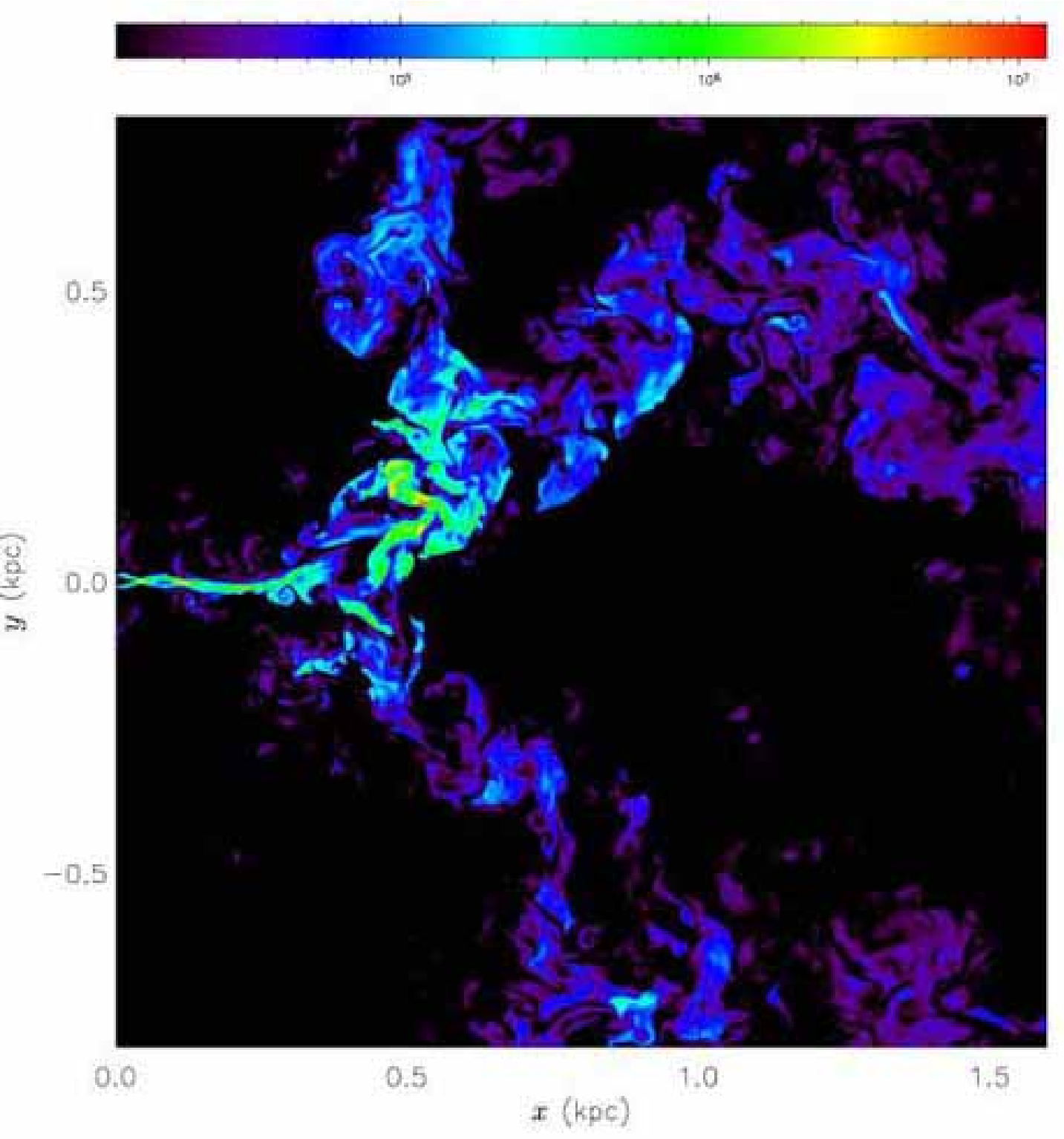}
}
\end{array}$
\end{center}
\caption{
Snapshots of model {\tt B2}.
Top-left:
$\rho/\rho_{\rm x}$ and flow velocities
near the jet when $t=12t_0$.
Top-right:
$t=48t_0$,
$\rho/\rho_{\rm x}$ throughout the full grid.
Bottom-left:
$t=48t_0$,
simulated radio emissivity around the jet
(with a dynamic range of 1024).
}
\label{fig.erupt06.0080.dens}
\label{fig.erupt06.0320.dens}
\label{fig.erupt06.0320.cut.glow}
\end{figure}

In this case the filling factor of clouds is higher with
$f_\mathrm{c}=3.0 \times 10^{-2}$,
corresponding to $\sigma_{\mathrm{min}}=1.9$.
In the first few $t_0$ of simulation time
the jet is completely disrupted within $< 0.1\ \kpc$ of the nucleus
as a result of the KH instability
induced by waves reflected off the innermost clouds.
By
$t\sim12t_0$
(see top panel, Figure~\ref{fig.erupt06.0080.dens}),
the jet has ablated the nearest clouds
and cleared an unobstructed region
out to $\sim0.1\ \kpc$ on either side.
Within this region the jet remains well collimated
during the rest of the simulation.
The clouds that define the sides of this sleeve
are shocked but they survive until the last frame
(middle panel, Figure~\ref{fig.erupt06.0320.dens}).
The flow of plasma from the head of the jet
splits into a fork on either side
of a cloud group
that sits nearly on the jet axis
at about $x\approx0.5\ \kpc$.
Throughout the simulation
these clouds retain dense, efficiently cooling cores
but they continually lose dense gas
entraining into the stream of jet plasma passing to either side.
Near the end of the simulation
($t=48t_0$, Figure~\ref{fig.erupt06.0320.dens})
these obstructive clouds
are trailed by dense ablated gas
extending throughout a triangular area
up to $0.5\ \kpc$ further out in the positive $x$ direction.

The radio emissivity
(last panel, Figure~\ref{fig.erupt06.0320.cut.glow})
has some similarities to the appearance of model {\tt B1},
with a bright jet remaining well collimated out to around $0.5$~kpc.
However the turbulent plumes of jet-derived plasma
beyond the jet's head are more disrupted
by the dense obstructions of the cloud field,
Radio bright plasma divides in several directions,
as Figure~\ref{fig.erupt06.0320.cut.glow} shows.
The brightest areas of the jet and post-jet flows
are typically concentrated closer to the nucleus
than in the less cloudy model {\tt B1}.

\subsubsection{Large filling factor: model {\tt B3}}

\begin{figure*}\begin{minipage}{180mm}
\begin{center}
$\begin{array}{cc}
\ifthenelse{\isundefined{\rainbow}}{
\includegraphics[width=9cm]{f10a.eps}
&
\includegraphics[width=9cm]{f10b.eps}
\\
\includegraphics[width=9cm]{f10c.eps}
&
\includegraphics[width=9cm]{f10d.eps}
}{
\includegraphics[width=9cm]{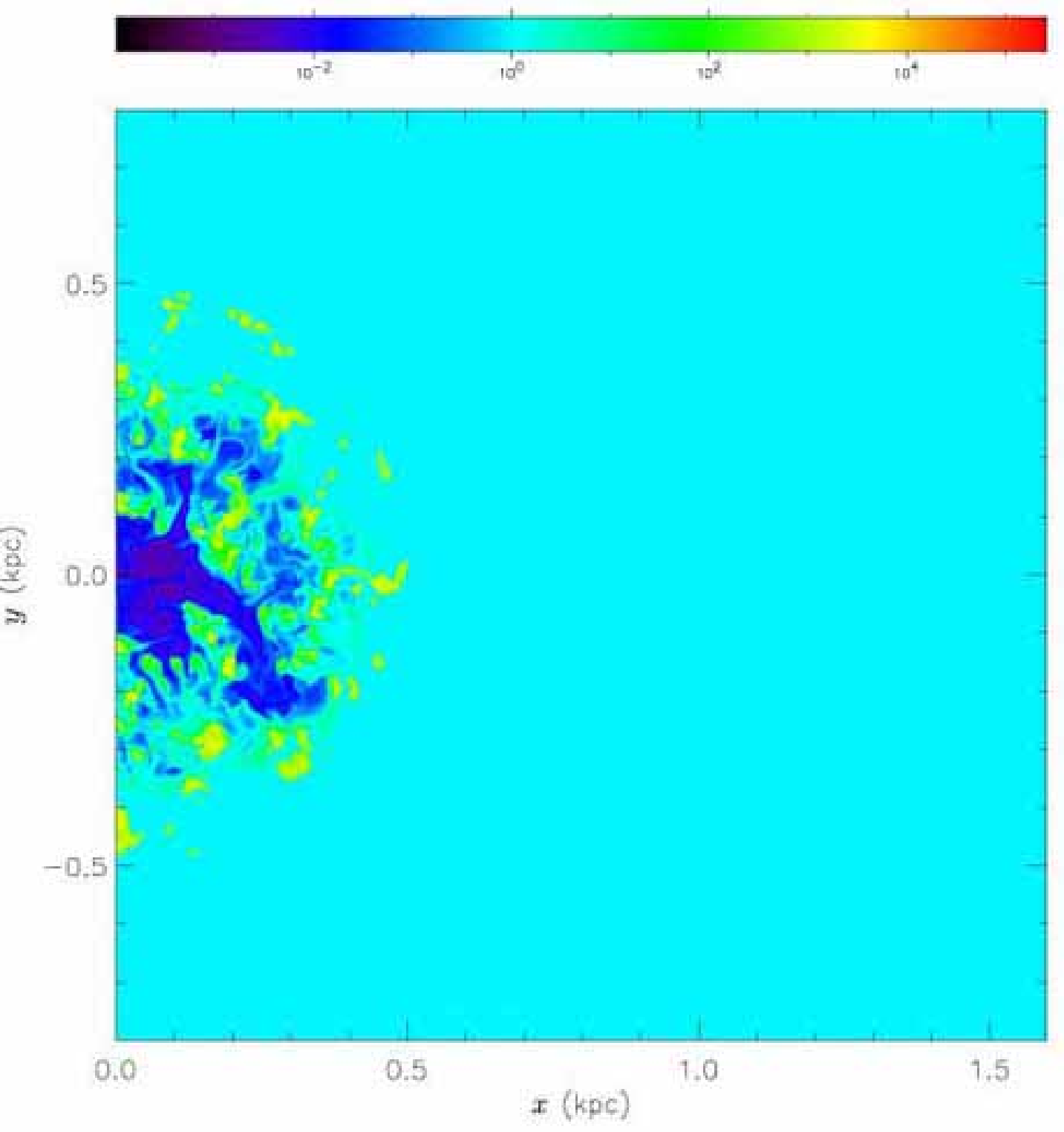}
&
\includegraphics[width=9cm]{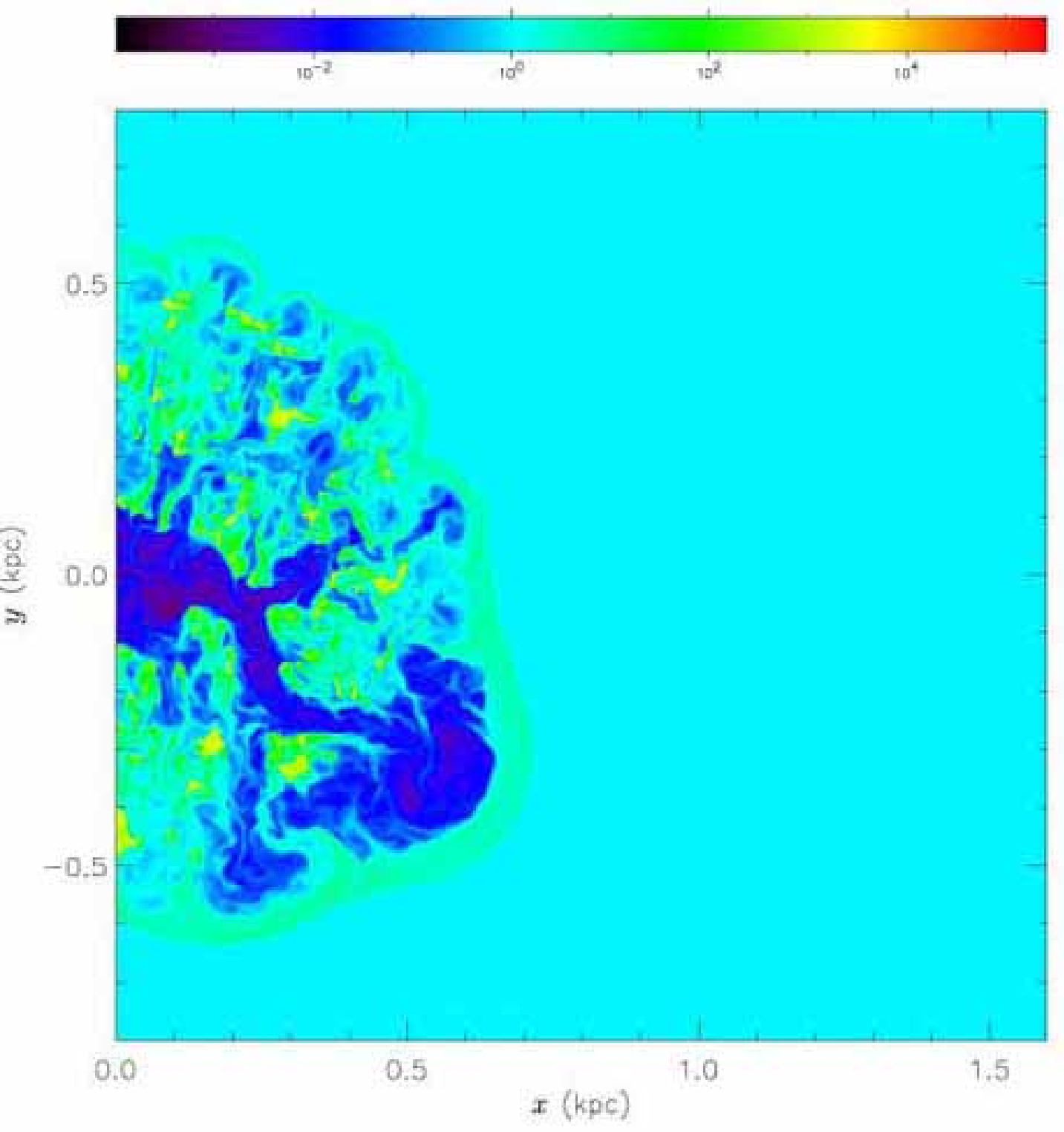}
\\
\includegraphics[width=9cm]{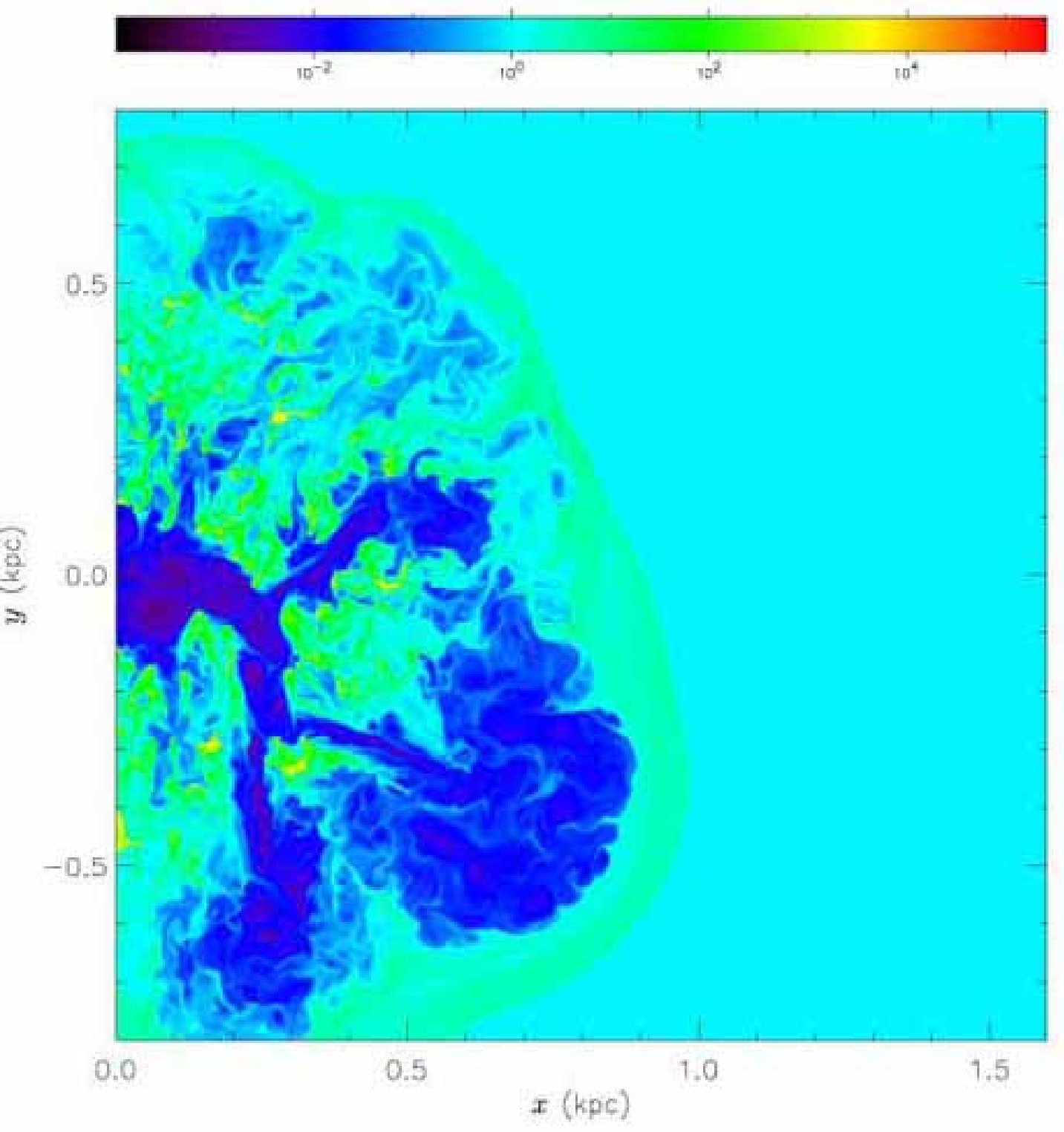}
&
\includegraphics[width=9cm]{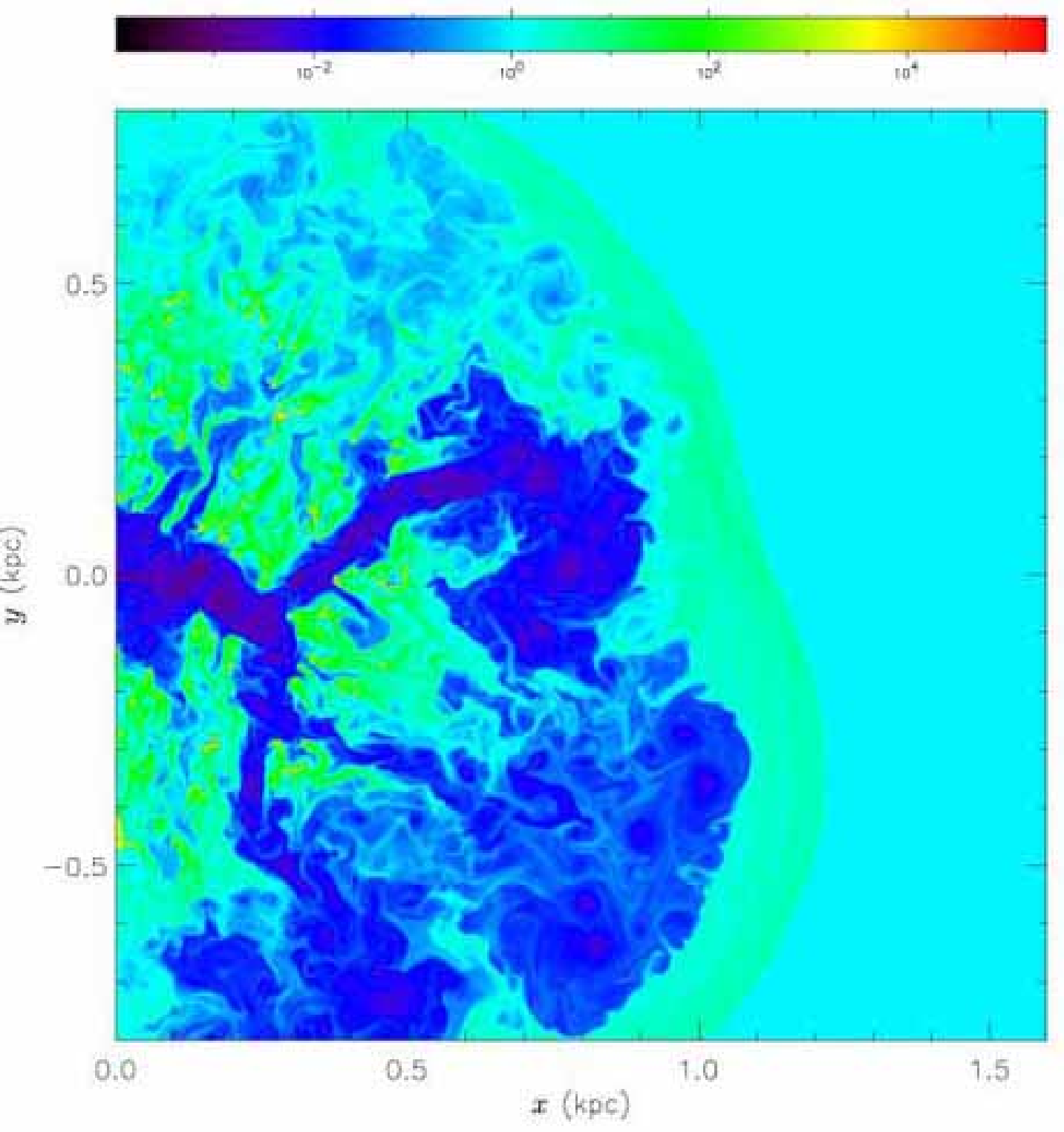}
}
\end{array}$
\caption{
Logarithmically scaled density images ($\rho/\rho_\mathrm{x}$)
of a central subregion of simulation {\tt B3}
during the initial erosion of channels of jet plasma,
and the earliest eruptions of radio bubbles
beyond the surface of the cloudy zone.
The time steps are
$t=3t_0, 6t_0, 9t_0, 12t_0$
in the
top-left, top-right, bottom-left and bottom-right
frames respectively.
}
\label{fig.erupt05.channels}
\end{center}
\end{minipage}\end{figure*}

\begin{figure*}\begin{minipage}{180mm}
\begin{center}
$\begin{array}{cc}
\ifthenelse{\isundefined{\rainbow}}{
\includegraphics[width=9cm]{f11a.eps}
}{
\includegraphics[width=9cm]{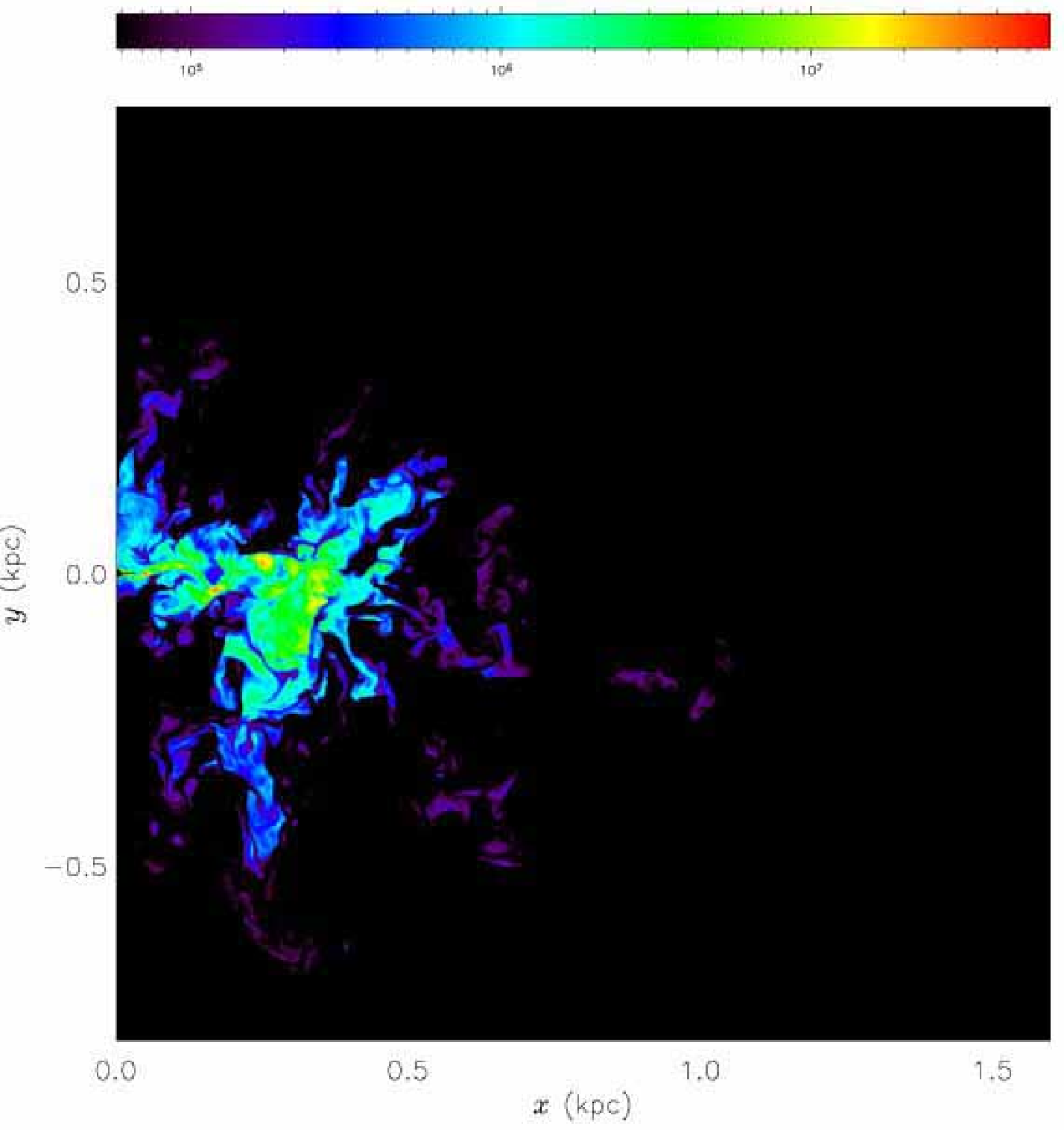}
}
&
\ifthenelse{\isundefined{\rainbow}}{
\includegraphics[width=9cm]{f11b.eps}
}{
\includegraphics[width=9cm]{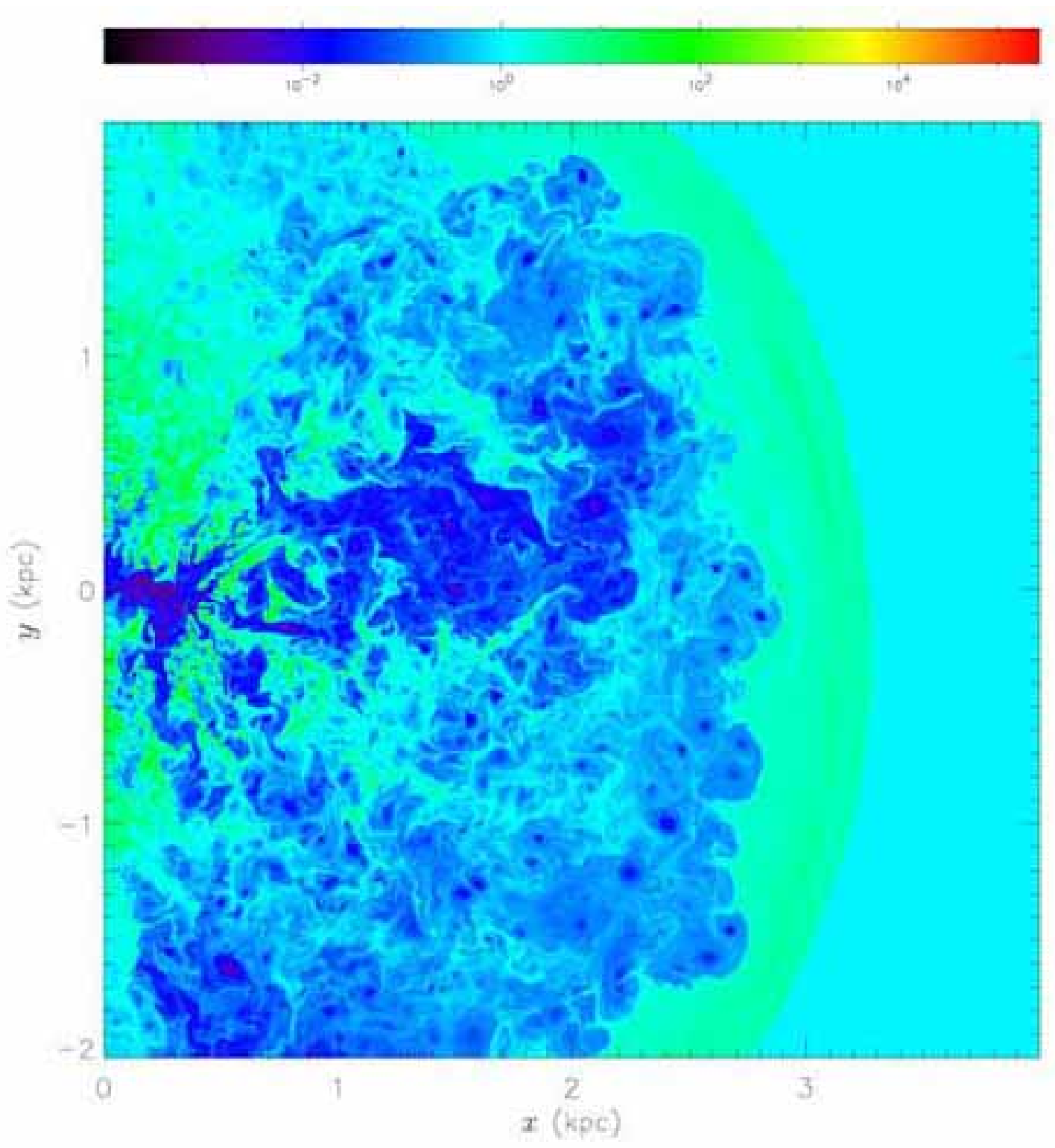}
}
\end{array}$
\end{center}
\caption{
The final condition of model {\tt B3} ($t=48t_0$),
comparable to the lower panels of Figures~\ref{fig.erupt03.0320.cut.glow}
and \ref{fig.erupt06.0320.cut.glow}.
The left panel is a simulated radio emissivity image
of the jet's vicinity,
with dynamic range $1024$ as in previous figures.
The right panel represents
$\rho/\rho_\mathrm{x}$
throughout the full grid.
}
\label{fig.erupt05.0320.cut.glow}
\label{fig.erupt05.0320.dens}
\end{minipage}\end{figure*}

This is our simulation with largest filling factor of clouds,
$f_\mathrm{c}=0.12$
and $\sigma_{\mathrm{min}}=1.2$.
As a result of the high coverage of clouds,
the jet is disrupted when it is less than $0.1\ \kpc$ from its origin.
Jet plasma streams through the available gaps
between the persistent clouds
(see the density distribution
in right panel of Figure~\ref{fig.erupt05.0320.dens}).
These low-density, radio-bright channels
evolve as the cloud distribution changes.
A channel may close when nearby clouds
expand ablatively or
are blown into positions where they create a new obstruction.
Turbulent eddies of jet plasma
accumulate in pockets within the cloud field,
most conspicuously in places where a channel has closed.
The density image, Figure~\ref{fig.erupt05.channels},
illustrates the initial development of intercloud channels.

The left panel of Figure~\ref{fig.erupt05.0320.cut.glow}
shows simulated radio emissivity
in the inner regions of the galaxy
at the end of the simulation.
The radio-bright material is concentrated near the nucleus
within an irregular, branched region that is not directly related
to the alignment of the initial jet.
This particular configuration of channels is ephemeral,
occurring after the opening and closure of previous channels
as the cloudfield evolves.

Eventually, one or more channels
pushes into the open region outside the cloudy zone.
A spurt of turbulent jet plasma accumulates at that point.
In later stages of our simulations,
these eruptions eventually become large
compared to the cloudy zone,
and merge with adjacent eruptions
to form a diffuse, turbulent, approximately isotropic outer cocoon
composed of a mixture of jet plasma, ablated cloud gas
and intercloud gas processed through the bow shock.

\subsubsection{Effect of cooling: model {\tt B3a}}

\begin{figure*}\begin{minipage}{180mm}
\begin{center}
$\begin{array}{cc}
\ifthenelse{\isundefined{\rainbow}}{
\includegraphics[width=9cm]{f12a.eps}
&\includegraphics[width=9cm]{f12b.eps}
}{
\includegraphics[width=9cm]{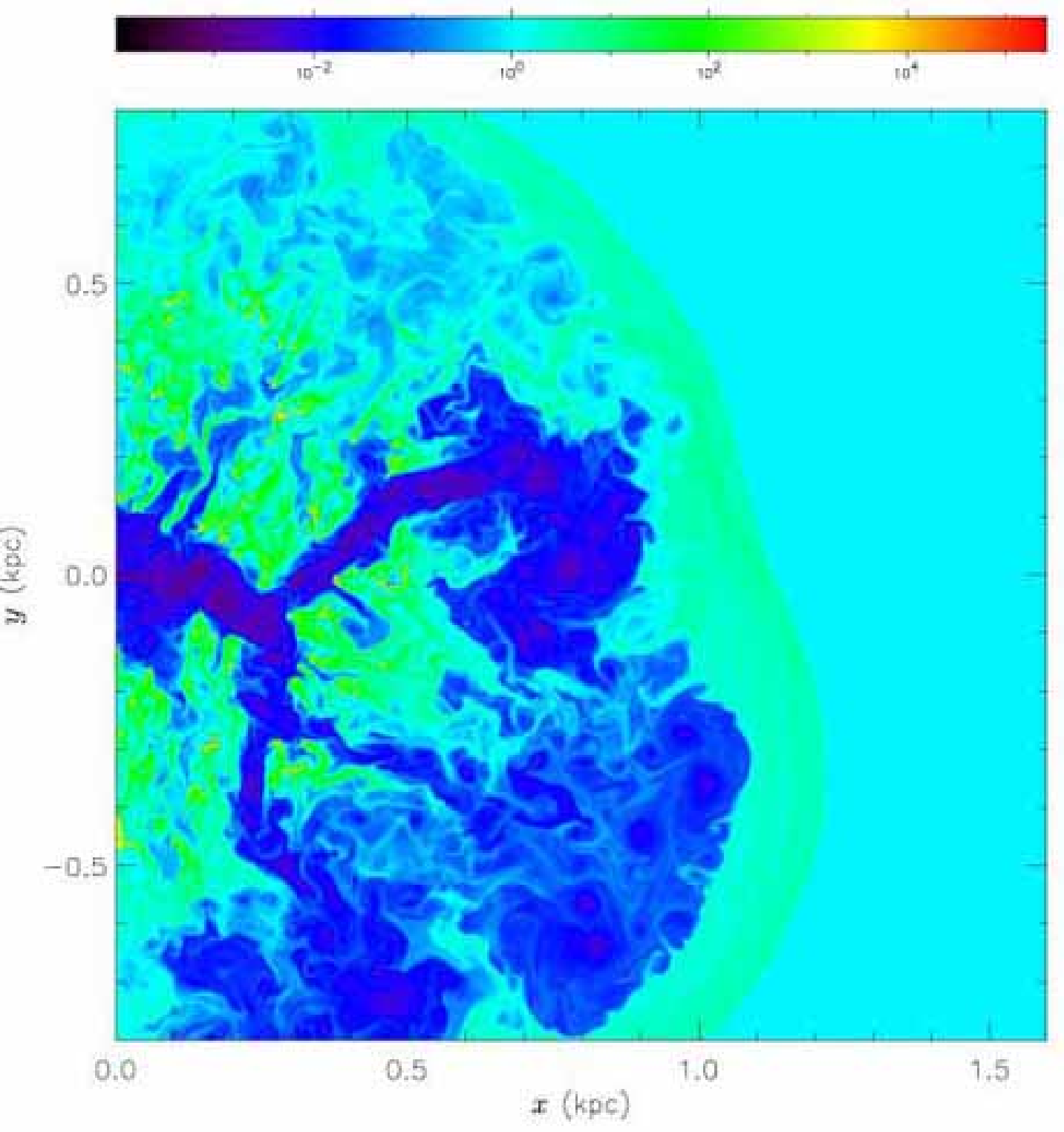}
&\includegraphics[width=9cm]{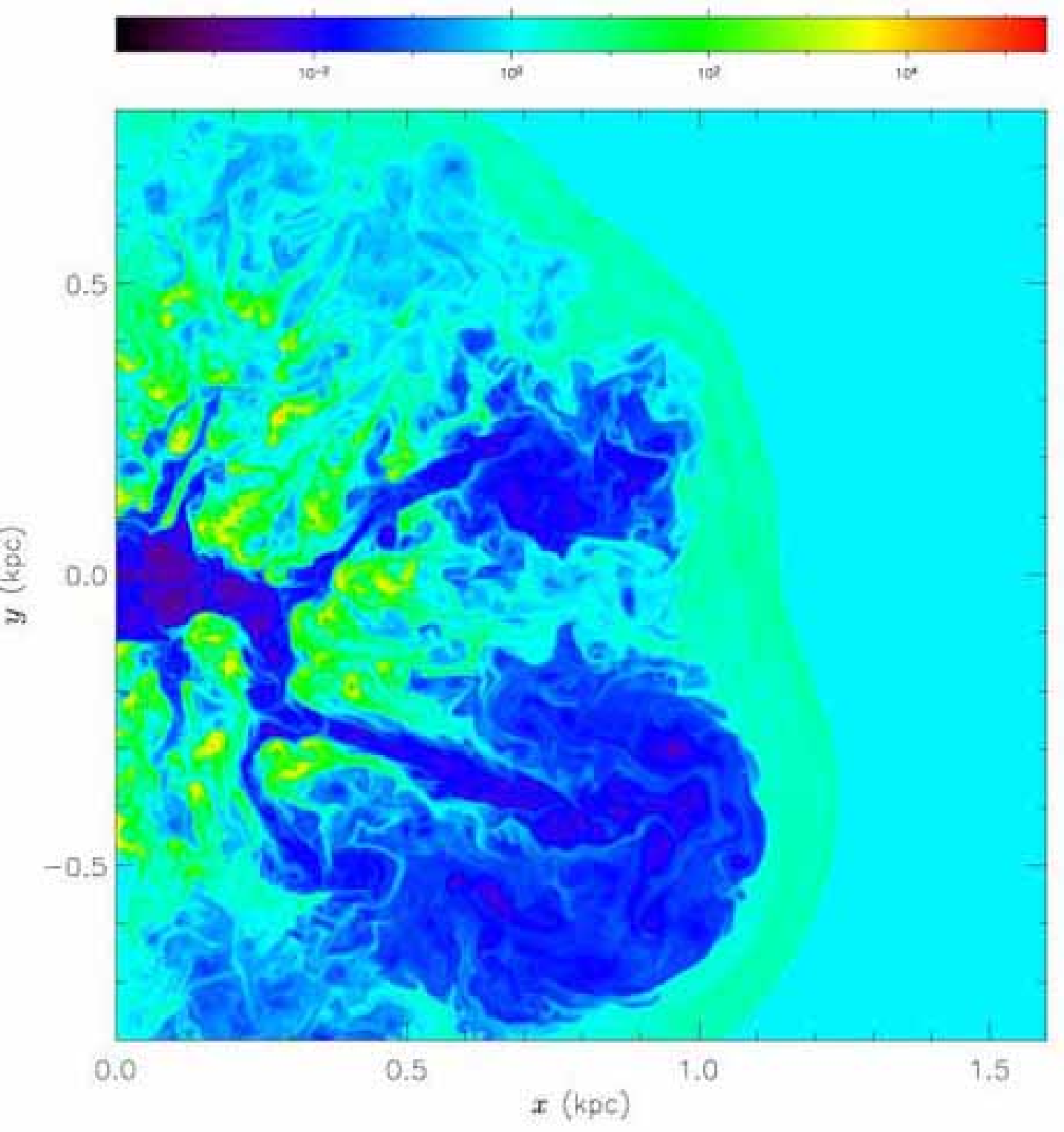}
}
\end{array}$
\caption{
Demonstration of the effects of radiative cooling
upon cloud survival.
Density frames ($\rho/\rho_{\rm x}$) are compared at time $t=12t_0$,
with the left panel showing simulation {\tt B3}
and the right panel shows simulation {\tt B3a},
which is the equivalent simulation with radiative cooling switched off.
In the absence of radiative cooling,
clouds expand and ablate rapidly
as soon as they pass through the bow shock.
}
\label{fig.erupt00}
\end{center}
\end{minipage}\end{figure*}

We ran a further simulation with the same cloud field as in
{\tt B3}
but with radiative cooling switched off
(see Figure~\ref{fig.erupt00}).
In the absence of radiative cooling,
the compression of clouds is insignificant.
The outflow of jet plasma rapidly turns the clouds
into ablating, flocculent banana shapes.
Large clouds and small clouds ablate efficiently.
The survival time of individual clouds
is shorter than in the simulation including the effects of radiative cooling.
Dense material is more readily ablated from the clouds,
and this may block channels of jet plasma
more readily or at an earlier time,
but otherwise the radio morphology is qualitatively similar
to that of {\tt B3}.

\subsection{Evolution of the outer part of the radio source}

\begin{figure*}\begin{minipage}{180mm}
\begin{center}
$\begin{array}{cc}
\includegraphics[width=8cm]{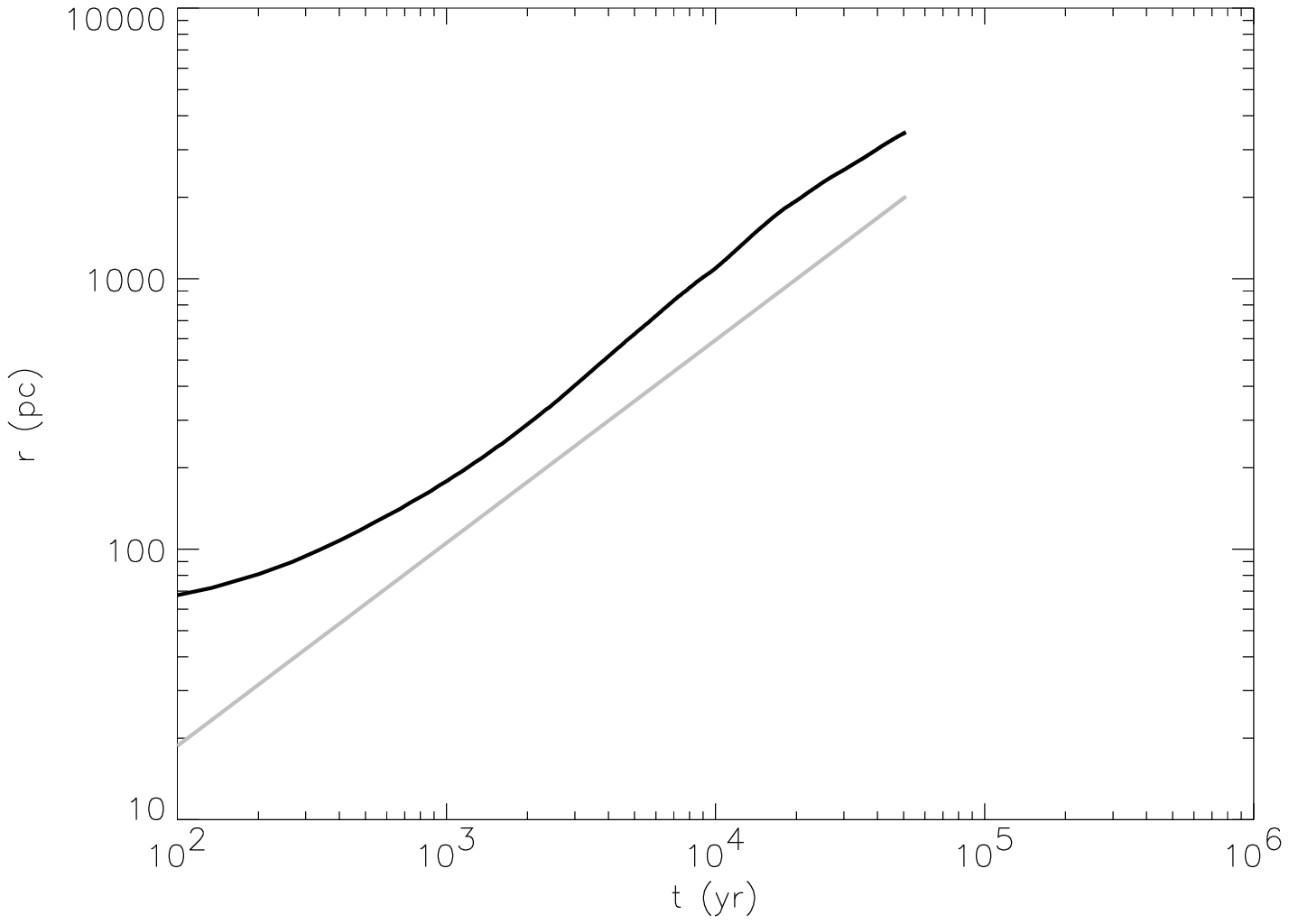}
&\includegraphics[width=8cm]{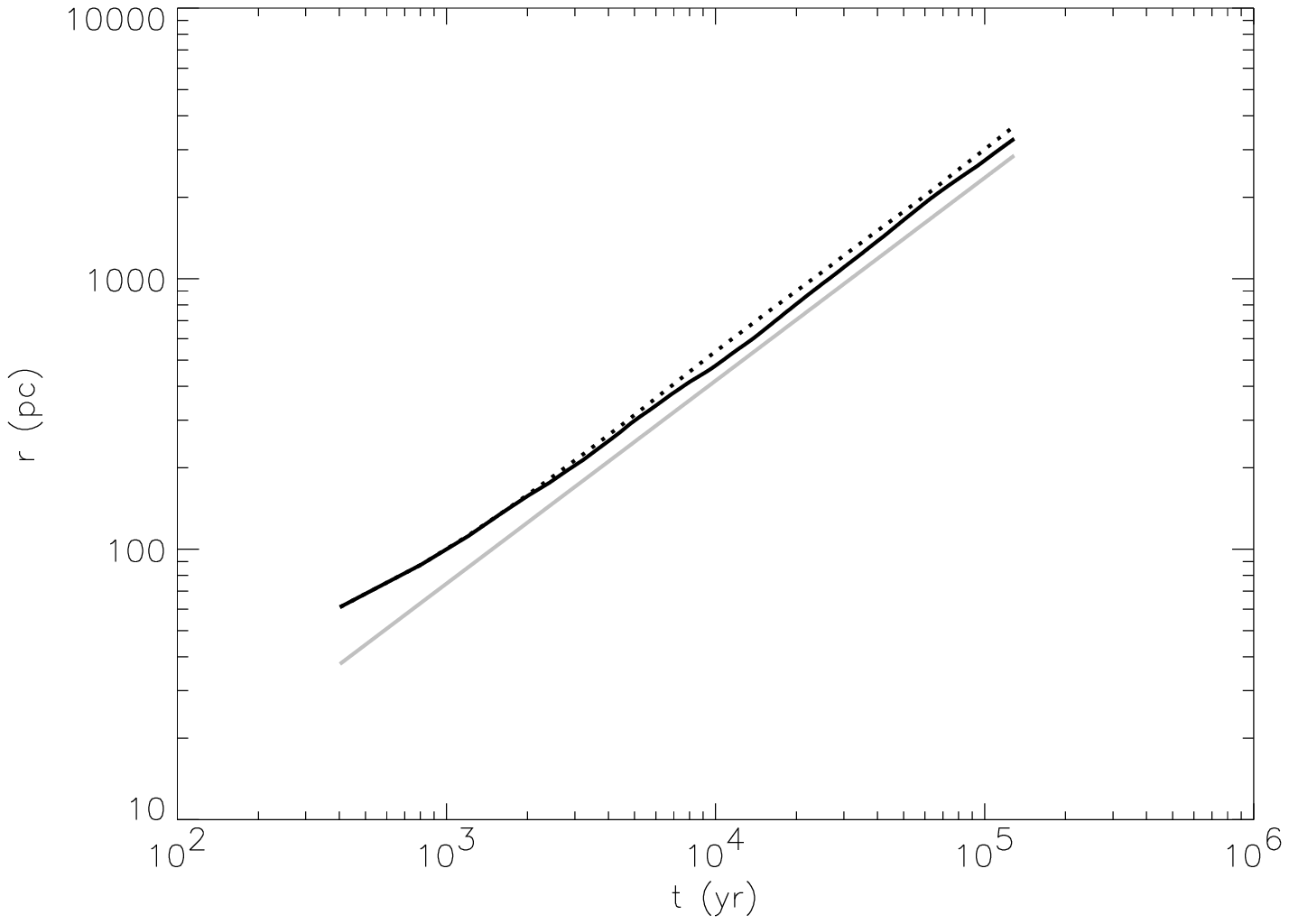}
\end{array}$
\end{center}
\caption{
Evolution of the bow shock's mean radius
compared with the radial growth of an equivalent energy-driven bubble model.
The left panel shows results for the simulation {\tt A1};
the right panel shows results for model {\tt B3}.
In each case the grey line is the ideal model,
and the black line shows simulation results.
The dotted line in the right panel shows
the cloudless control model {\tt B0}.
\label{f.2d_bubble}
}
\end{minipage}\end{figure*}

In each case, as a consequence of the disruption of the jet,
its momentum is distributed throughout the cocoon in an isotropic way.
Hence, the long term evolution is similar to that of an energy-driven bubble.
Figure~\ref{f.2d_bubble}
presents plots of the average radius of the furthermost extent of the radio plasma versus time,
for represenative simulations.
The plots also show the corresponding analytic solution for
a two dimensional bubble with the same parameters of energy flux per unit length and density
(grey line,
see Appendix~\ref{app.bubble} for a derivation).
It is evident that the analytic slope of the radius/time curves
is replicated very well by the simulations.
The offset may be attributed to the time taken for the isotropic phase to be established.
This may be related to the finite minimum width of the jet at its nozzle,
the size of the innermost nuclear region that is initially cloudless,
and the intrinsic length scale of the jet's instability
(\S~\ref{results.eruptyy}).

We expect that a similar power-law expansion will occur in three dimensions,
although this of course is yet to be confirmed.
However, given that jet momentum and power
can be redistributed isotropically in three dimensions,
it is reasonable to expect that the long term evolution of the outer bubble
will be described by a three--dimensional energy--driven bubble.

\section{Comparison of model {\tt B1} with 3C48}
\label{s:3C48}

Whilst these simulations have been two-dimensional,
some of their features should persist
in three--dimensional simulations with similar parameters.
In particular the trends with filling factor will probably persist,
to some extent, so that comparisons with existing observationsla are of interest, even at this stage. 
We therefore make a tentative comparison of one of these simulations (model {\tt B1}) with the CSS quasar 3C48.  
We can also draw some general conclusions from these simulations,
concerning the likely overall morphology of young radio sources.

\begin{figure*}\begin{minipage}{180mm}
\begin{center}
$\begin{array}{ccc}
\ifthenelse{\isundefined{\rainbow}}{
\includegraphics[height=8cm]{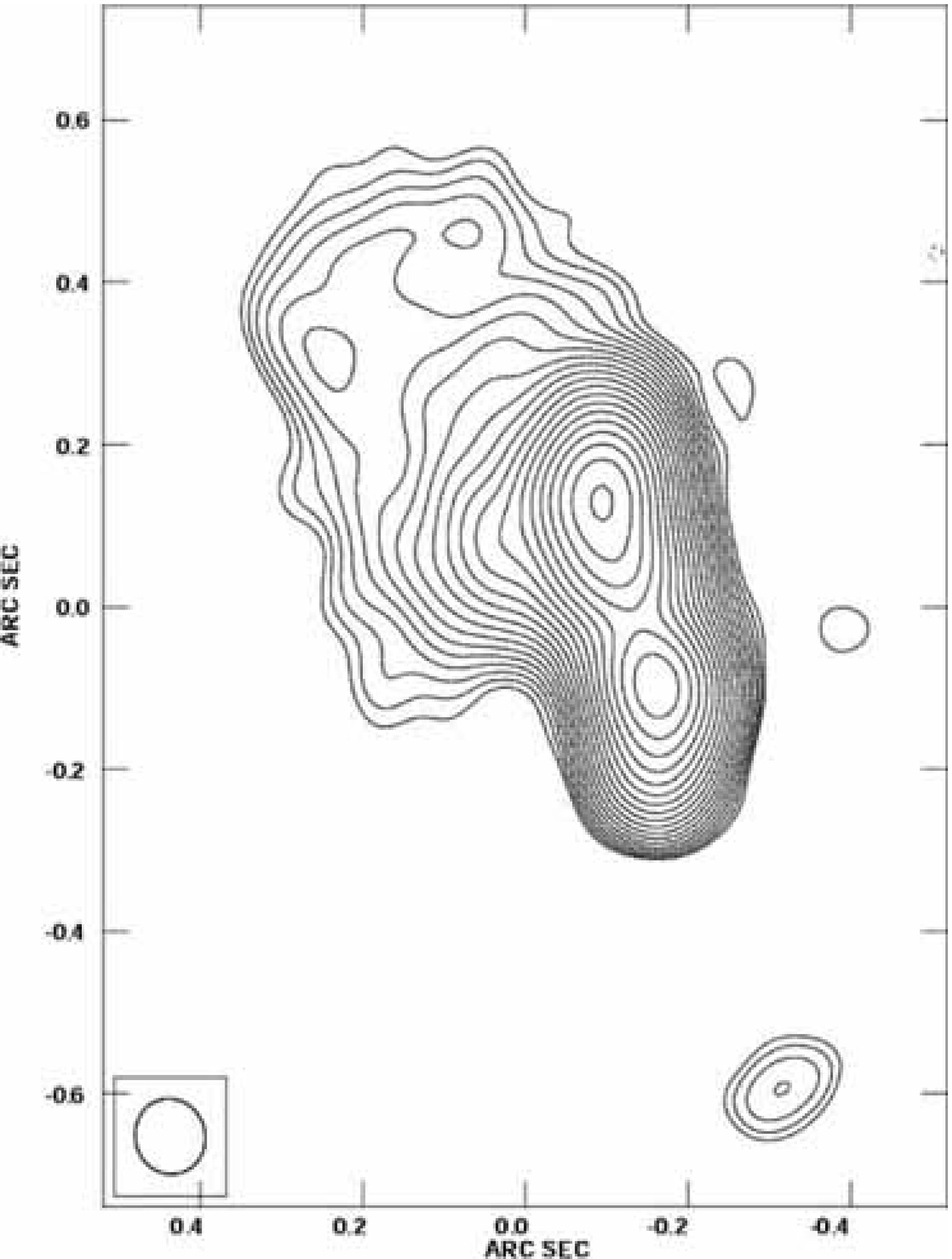}
&\includegraphics[height=8cm]{f14b.eps}
&\includegraphics[height=8cm]{f14cc.eps}
}{
\includegraphics[height=8cm]{f14a.eps}
&\includegraphics[height=8cm]{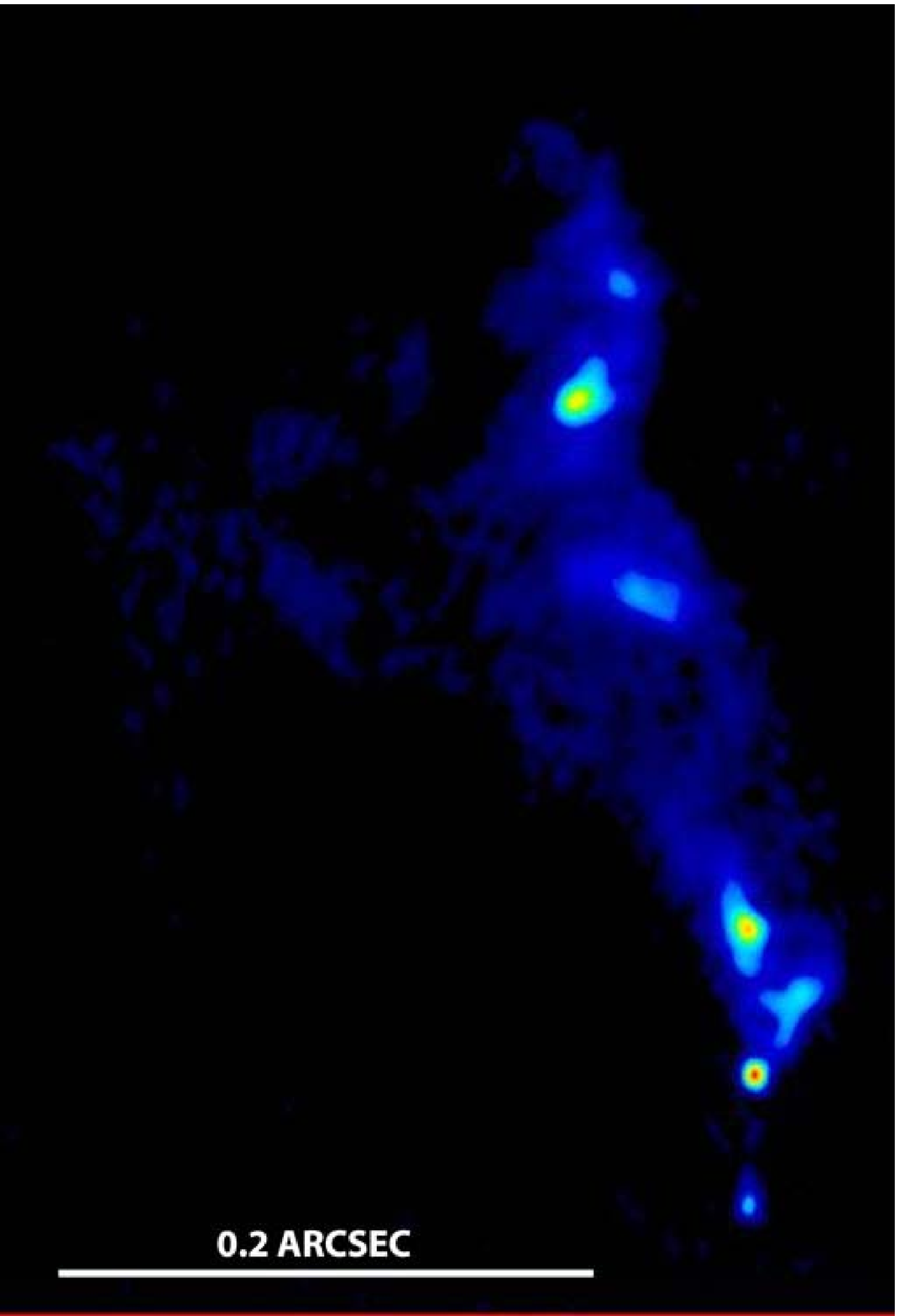}
&\includegraphics[height=8cm]{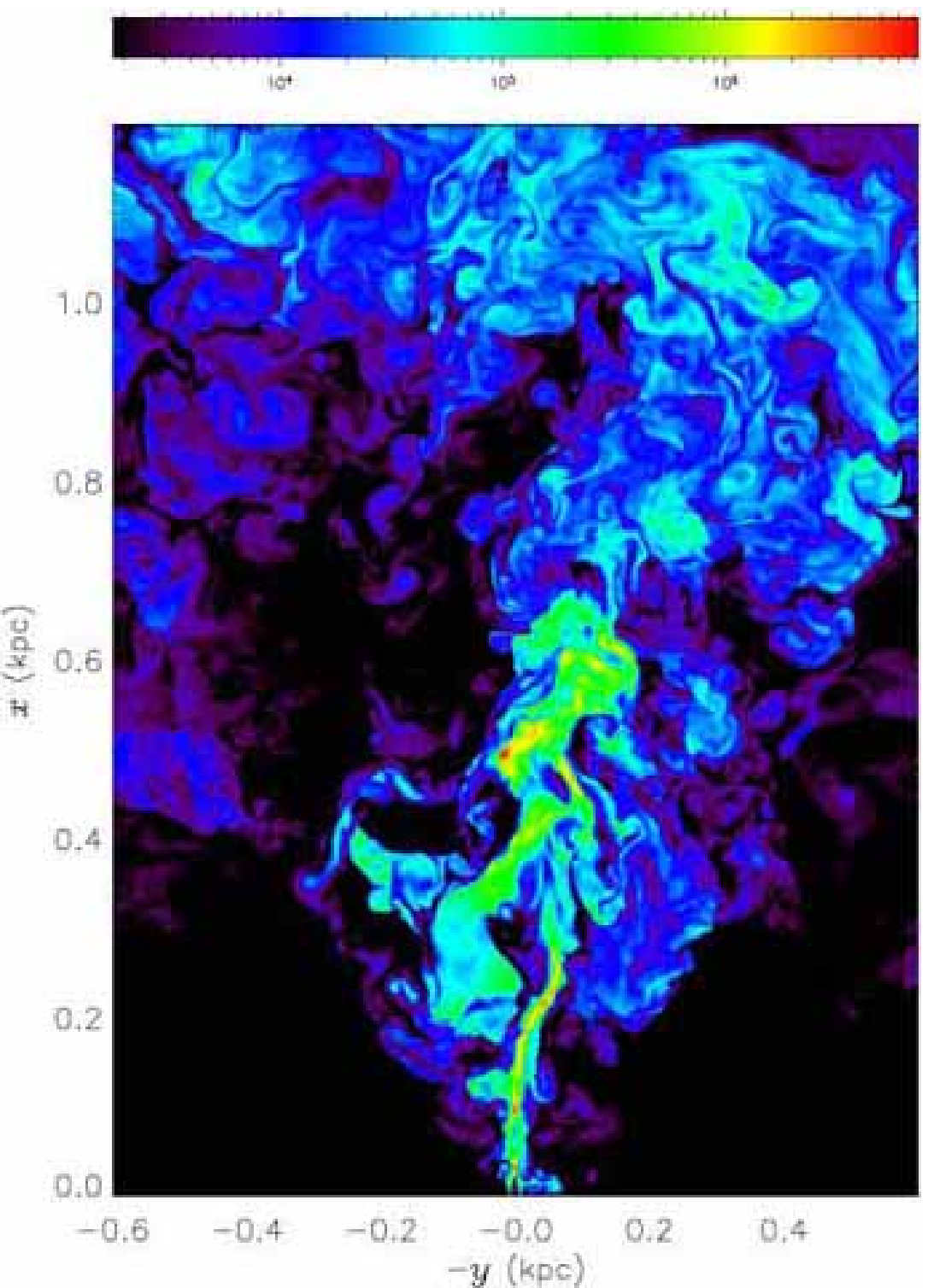}
}
\end{array}$
\end{center}
\caption{
The radio features of 3C48 compared with simulation {\tt B1}.
The left panel is a 22~GHz (K-band) VLA A-array image
illustrating the large--scale (arcsecond scale) structure of 3C48
\citep{ojha04a}.
This image shows diffuse emission on a larger scale to that of the jet,
consistent with the prediction of the simulation.
The middle panel is a 1.6~GHz VLBI image
featuring the jet and bright knots
(Wilkinson et al. 1991).
At the redshift of 3C48 ($z=0.3670$)
an angular scale of 0.1~arcsec corresponds to 0.514~kpc
(assuming $H_0 = 70\ {\rm km}\ {\rm s}^{-1}\ {\rm Mpc}^{-1}$,
$\Omega_m = 0.27$
and
$\Omega_\Lambda = 0.73$).
The right panel shows a snapshot of radio emissivity
from the {\tt B1} simulation, at time $t=48t_0$.
Left and middle images 
are reproduced with the kind permission of R.~Ojha and P.~Wilkinson
respectively.}
\label{f.3C48.1}
\label{f.3C48.2}
\end{minipage}\end{figure*}

Simulation {\tt B1} has the lowest filling factor of dense clouds
and impedes the jet the least.
The jet is still disrupted but the flow associated with it
continues with reduced collimation and with its forward momentum spread over a much larger region.
In the disrupted section,
transverse oscillations are produced, with shocks created
at points where the jet bends significantly,
leading to local enhancements in emissivity.
A large scale bubble is produced outside the cloudy region
but the jet also retains some coherency in the intermediate region.
This particular simulation bears a strong qualitative resemblance
to 3C48.
In Figure~\ref{f.3C48.1}
a snapshot of the simulated radio emissivity at a particular time
is compared with the 1.6~GHz VLBI image of 3C48 \citep{wilkinson91a}.
In two dimensions,
the alternating direction of the jet results from
the transverse instability that we have already discussed
(\S\ref{results.eruptyy}).
In three dimensions,
the corresponding instability would be a helical $m=1$ mode.
In both cases the alternating direction of the supersonic flow produces shocks.
We see evidence for these,
in both the snapshot of the emissivity from the 2D simulation and the image of the jet.

This particular simulation predicts that
there will be a diffuse halo of emission external to the jet region
of the source.
This was, in fact, detected in larger scale images of 3C~48.
One such image presented in \citet{wilkinson91a},
shows diffuse emission on a 1~arcsecond scale in 3C48
extending well beyond the compact jet structure evident in the VLBI image.
More recent higher quality images have been obtained by \citet{ojha04a}
and one of their contour maps is reproduced in Figure~\ref{f.3C48.2}.
These images clearly reveal the extended structure of 3C48.
\citet{wilkinson91a}
had already concluded that the morphology of 3C48
is likely the result of jet--cloud interactions
and cited evidence for asymmetric emission line gas
northwest of the core
as well as the presence of dust revealed by the high IRAS flux.
They interpreted the radio knot close to the core as the result of
a jet--cloud interaction.
Whilst this is possible,
our simulation shows that the sort of jet disruption observed in 3C48
can occur without direct jet--cloud encounters.
The general turbulence in the jet cocoon can excite jet instabilities,
producing shocks in the supersonic flow
either as a result of the jet being bent through
an angle greater than the Mach angle, or though pinching--mode instabilities.
(Both of these occur in the simulations.)
An explanation for the first bright knot involving cocoon turbulence may therefore be more favorable
since direct jet--cloud encounters are usually highly disruptive and
\cite{wilkinson91a}
appealed to a glancing deflection in order to explain the radio structure.
Given that the sort of jet--ISM interaction revealed in our simulations,
is occurring, we expect that the emission line gas
would show evidence for shock excitation
as a result of radiative shocks induced by the overpressured radio cocoon.
\cite{chatzichristou01a}
has suggested that the emission line spectrum of 3C48
may show evidence for a mixture of AGN and shock excitation.

\citet{wilkinson91a}
also noted out that 3C48, classified as a CSS quasar,
is much less extended than other similar power radio--loud quasars.
They speculated that the confinement results from
the emission line gas and dust inferred to be present in the ISM.
Qualitatively, we agree with this,
except to point out that the dispersion of the momentum flux
(by a small number of clouds)
prevents a radio source from expanding rapidly.
As we have shown,
the source becomes an expanding bubble
with a much slower rate of expansion than a jet with the same kinetic power.
Hence the gas and dust is not required to physically confine the radio plasma.
The slower expansion can be effected by a small filling factor of dense clouds
that redistribute the jet momentum isotropically.

We also note a qualitative resemblance between the simulation {\tt B3} and M87. The channeling of the flow in {\tt B3} produces a hierarchy of structures resembling the hierarchy of radio structures in M87 apparent in the low frequency radio image \citep{owen00a}. However, we have not emphasized this with a detailed comparison since the scale of the source and the simulation are quite different. Nevertheless, the production of the structure observed in simulation {\tt B3} depends mainly on the filling factor of the clouds and this mechanism should also work on larger scales. We propose to confirm this in future with a fully relativistic 3D simulation.

Other sources with similar morphology to 3C48
and which may be being disrupted in a similar way include
the TeV $\gamma$--ray blazars, MKN~501 and MKN~421.
The most recent VLBI images of these sources
\citep{giovannini00a,piner99a}
show similar structures in the form of disrupted jets
and large scale plumes suggesting that
these sources are disrupted on the sub--parsec scale by a cloudy medium.
Again the scale of the simulation {\tt B1}
is such that it does not {\em strictly} apply.
However, the physical principle of these jets being disrupted
by a small filling factor of dense clouds
is probably still valid with interesting repercussions for
the production of the $\gamma$--ray emission \citep{bicknell04a}.

\section{DISCUSSION}

We have presented a series of two--dimensional simulations of slab jets
propagating through inhomogeneous media.
Despite the limitations of the restrictions to two dimensions, and to non--relativistic flow,
these simulations provide useful insight into the characteristics of
jets in radio galaxies which
have an inhomogeneous interstellar medium.
These include GPS and CSS radio sources,
high redshift radio galaxies and galaxies at the center of cooling flows.
In all of these galaxies there is ample evidence for dense gas clouds
that may interact with the radio source producing at least some of the features
evident in our simulations.
In the case of GPS and CSS sources the inhomogeneous medium
may be the result of a fairly recent merger;
in the case of high redshift radio galaxies an inhomogeneous medium
is likely to be the result of the galaxy formation process at those epochs,
as large galaxies are assembled from sub-galaxian fragments.

As well as the limitations to two dimensions and non--relativistic flow,
there is another feature of our simulations that requires some comment
and that is the ablation of cloud material that is evident in most cases.
There is likely to be a large contribution of numerical mass diffusion
due to the very high density contrasts between jet and cloud gas.
Although this has some effect on the evolution,
the dominant processes in the case of a low filling factor
are waves emanating from the clouds impacted by the high pressure cocoon and,
in the high filling factor simulations,
direct obstruction of the flow plus wave effects.

Despite the above--mentioned caveats,
we have been able to deduce these general features from our simulations:
\begin{enumerate}
\item {\bf Jet instability.}
When propagating through the type of inhomogeneous interstellar medium
that we have imposed, the jets become unstable.
However, this instability does not always result from
direct jet--cloud collisions
but occurs indirectly from the waves
generated by the impact of the high--pressure jet cocoon on the dense clouds.
This leads to a general level of turbulence in the medium
that induces unstable Kelvin--Helmholtz oscillations in the jet.
An unstable jet was also produced in the homogeneous simulation ({\tt B0})
and this was related to turbulence caused by
the entrainment of ISM into the cocoon
as well as the higher level of large scale turbulence in two dimensions.
However, the simulations with an inhomogeneous medium
became unstable in a shorter distance
and we attribute this to the effect of the clouds.

\item {\bf Production of halo/bubble.}
The instability of the jet is associated with
the spreading of the jet's momentum
and a turbulent diffusion of the jet material into a halo or bubble.
These are characteristics that we expect to be present in three dimensions.
The radius-time curve for the outer part of the halo
is well--described by the analytic expression for a two--dimensional bubble.
In three dimensions a energy driven bubble model should also be appropriate.
This is useful since it gives us a way of estimating the energy flux fed into
a large scale radio source from the dimensions of the outer halo
and the density of the surrounding interstellar medium.
The production of a bubble means that
the overall increase of size of the radio source with time is diminished
with respect to a source that propagates unimpeded.
However, this restriction on the size of the source
is not due to direct physical confinement of the radio plasma by denser gas.
Rather, it is the result of the isotropic spread of the jet momentum
by the process of jet instability described above.

\item {\bf The effect of filling factor.}
The main parameter that we have varied in these simulations
is the filling factor of the dense clouds
and we can see that this makes a considerable difference to
the morphology of the radio source.
With only a small filling factor of clouds,
the jet is readily disrupted,
but retains some directed momentum
and produces the 3C48--type morphology described in \S~\ref{s:3C48}.
As the filling factor increases,
the jet propagates via a number of different channels
which open and close leading to an hierarchical structure of radio plasma
that we noted is similar to the large--scale structure of M87.

\item {\bf The effect of jet density ratio.}
We also varied the jet density ratio parameter, $\eta$,
but only in one simulation.
The simulation {\tt A1} had $\eta = 0.38$
whereas in the other simulations $\eta \sim 10^{-3}$.
The main noticeable effect is that the heavier jet
is affected less by instabilities and direct jet--cloud collisions
than the lighter jets.
This points to an important difference between models with
$\eta \gtapprox 0.01$
and models with $\eta \ltapprox 0.001$.
Of course, such low density ratios are important
to provide the relevant power for radio galaxy jets.
However, reassuringly for physics but difficult for modelers,
such low values of $\eta$ push models into the relativistic regime,
as we have discussed in \S~\ref{s:jet_pars}.
There are two important points to be made here.
The first is that many jet simulations
usually involve a modest value of $\eta \ga 0.01$
and these may be misleading.
Second, the most realistic simulations will be those that directly
include relativistic physics
and which also incorporate the mixing of relativistic and thermal plasma.
\end{enumerate}

This paper complements other work on the physics of direct jet--cloud collisions 
(e.g. \citet{fragile04a}, \citet{wiita02a}) in which the focus has been on the effect of a jet on a single cloud. 
Direct jet--cloud collisions  are clearly important 
in the context of sources such as Minkowksi's object \citep{vanbreugel85a}. 
In this paper we have clearly been more concerned with the overall radio source morphology, 
the evolution of an ensemble of clouds, the effects of indirect jet--cloud collisions 
(i.e. the effect of the high pressured cocoon on dense clouds) and trends with cloud  filling factor. 
However, we note an interesting connection between \citet{wiita02a} and this work: 
We have both noted that less jet disruption occurs in the case of slow dense jets. 
We also note that  the morphology of radiative cloud shocks deduced by \citet{fragile04a} 
is also reproduced in some of the individual clouds in our simulations. 

Our simulations have clearly all been two--dimensional. 
What are the expected differences between two--dimensional and three--dimensional work? 
There are at least two (1) In two dimensions, vortices survive for a lot longer as a result of 
the inverse cascade of turbulent energy. These vortices play a significant role in disrupting the jet and clouds. 
We expect much less disruption from vortical motions in three dimensions.
(2) The number of independent directions that the jet can take in three dimensions is greater 
so that we expect that the clouds will impede the jet less severely 
and that escape from the cloudy region surrounding the nucleus will be easier. 
These issues as well as the inclusion of special relativistic effects will be addressed in future papers.

\section*{ACKNOWLEDGMENTS}
Our research has been supported by ARC Large Grant A699050341
and allocations of time by the ANU Supercomputer Facility.
Professor Michael Dopita provided financial support
from his ARC Federation Fellowship.
We are very grateful to P.N.~Wilkinson and R.~Ojha
for their permission to use their respective images of 3C48.
We thank the anonymous referee for the several suggestions
improving the format of the paper.

\appendix

\section{GRID SUBDIVISION FOR MULTIPROCESSOR COMPUTATION}
\label{app.mpi}

We confront two practical obstacles
to the calculation of high resolution hydrodynamics simulations
in higher dimensions:
the duration of the run in real-time
(which inhibits the exploration of a wide parameter space),
and the finite computer memory
(which directly limits the number of grid cells that can be computed).
We circumvent these limitations
by subdividing the spatial grid
into $N = 2^n$ subgrids (with $n$ an integer)
and assigning each to a program running in parallel on a separate processor.
The processors communicate between themselves using
functions of the {\tt MPI} (Message Passing Interface)
library\footnote{\tt http://www-unix.mcs.anl.gov/mpi/}.

\begin{figure*}\begin{minipage}{180mm}
\begin{center}
\ifthenelse{\isundefined{\rainbow}}{
\includegraphics[width=15cm]{fB1.eps}
}{
\includegraphics[width=15cm]{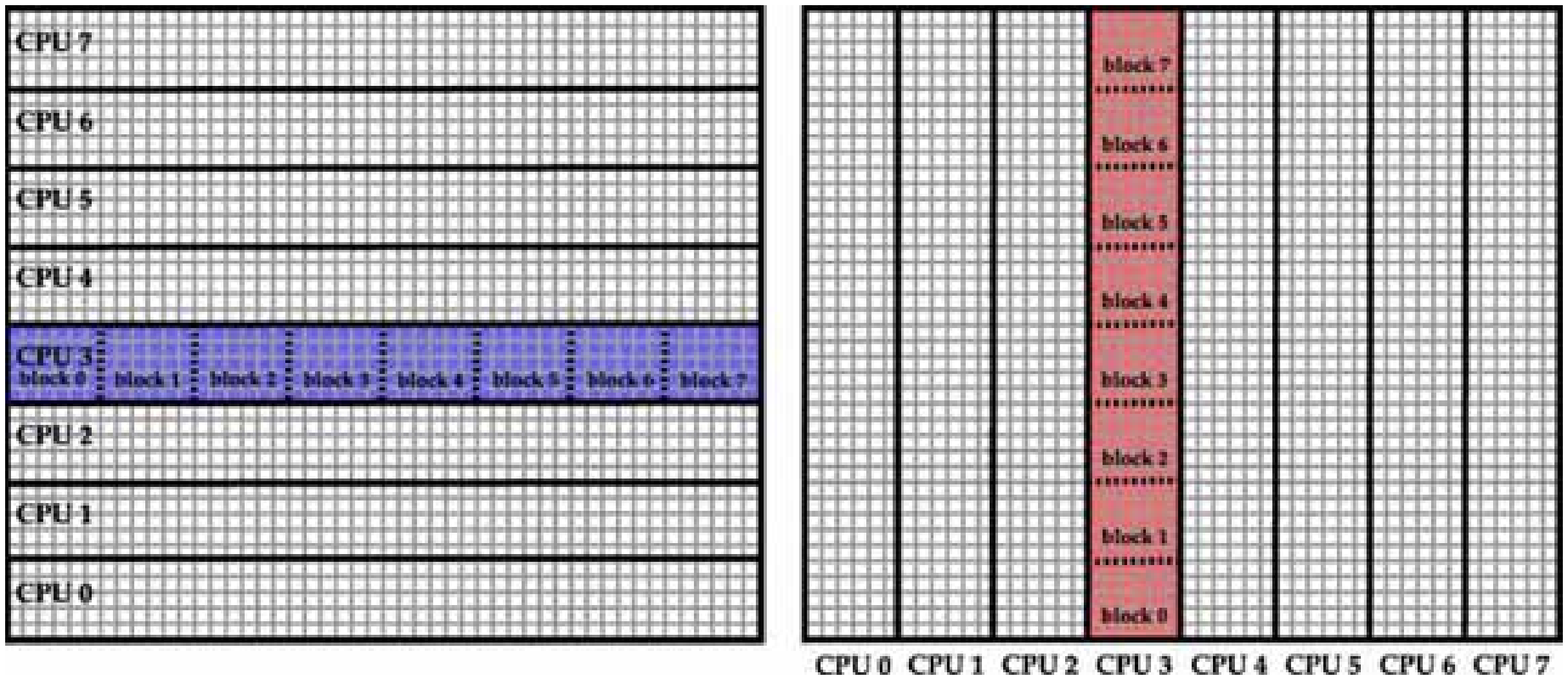}
}
\caption{
Distribution of the long and the wide subgrid partitions
in a simulation run with 8 processors.
For illustrative clarification,
the long subgrid assigned to processor 3 is highlighted in the left panel,
and
the wide subgrid assigned to processor 3 is highlighted in the right panel.
Each processor's memory needs only to retain
one long subgrid array and one wide subgrid array,
plus scalar global variables.
Collective maps of the global computational grid
need only be assembled from output data files after the simulation is run.
}
\label{fig.mpi.1}
\end{center}
\end{minipage}\end{figure*}

If the global grid has dimensions of
$I\times J$ cells in the $x$ and $y$ directions,
then we define
a ``long subgrid'' as an area with $I \times (J/N)$ cells
(see the left panel of Figure~\ref{fig.mpi.1}),
and a ``wide subgrid'' as an area of $(I/N) \times J$ cells
(see the right panel of Figure~\ref{fig.mpi.1}).
Within the memory at each processor we maintain representations of
one long subgrid or one short subgrid:
see for example the shaded long and wide subgrids held by processor 3
in Figure~\ref{fig.mpi.1}.
The total number of cells in a long subgrid
is the same as that of a wide subgrid.
In fact for each hydrodynamic variable on each processor,
we identify the long and wide subgrid arrays
with the same section of memory,
which is feasible and reliable since
the program uses only one of the subgrid representations at any time
(as we shall detail below).

We then conceptually subdivide each long or wide subgrid
into $N$ blocks separated by transverse cuts
(e.g. the dotted line divisions in processor 3's subgrids
depicted in Figure~\ref{fig.mpi.1}).
For each processor $p$ the block numbered $p$
is the only area of overlap between
the long and wide subgrids that it is assigned.

\begin{figure*}\begin{minipage}{180mm}
\begin{center}
\ifthenelse{\isundefined{\rainbow}}{
\includegraphics[width=15cm]{fB2.eps}
}{
\includegraphics[width=15cm]{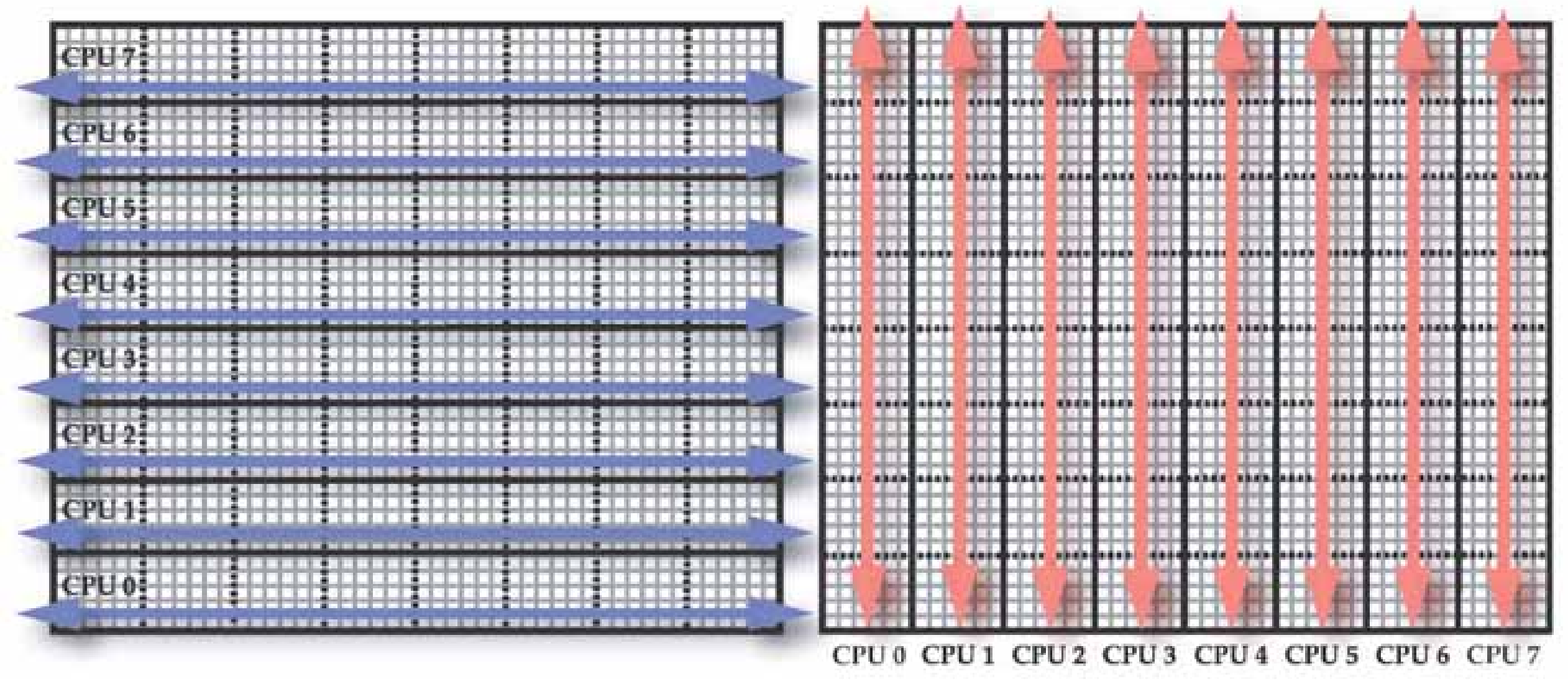}
}
\caption{
Hydrodynamic sweeps in the $x$ direction are only performed
in the long subgrid arrays (left panel),
and sweeps in the $y$ direction are only performed
in the wide subgrid arrays (right panel).
Thus every sweep operation occurs
along the largest possible array dimension
and involves only the arrays kept in the memory of the responsible processor.
Communication between processors is only required
to exchange updated block arrays of the hydrodynamic variables
after each sweep is finished.
}
\label{fig.mpi.2}
\end{center}
\end{minipage}\end{figure*}

At the start of the simulation each processor
is assigned long and wide subgrids according to its identity number.
These regions are filled according to the initial conditions
which are defined in terms of global grid coordinates.
The first hydrodynamic sweep
\citep{colella1984}
is in the $x$ direction,
and is performed by each processor within its long subgrid arrays
(see left panel of Figure~\ref{fig.mpi.2}).
Thus the $x$--sweeps do not encounter any of the subgrid interfaces,
obviating the need for any communication of flux information
between processors at this stage.
The one-dimensional array vectors associated with each sweeep
are the maximum possible length,
which makes optimal use of processor cache.

After the $x$ sweep is complete,
the evolved density, velocity, pressure and other zone data
are communicated between the processors.
By use of the {\tt MPI\_ALLTOALL} function,
the data of block $i$ of the long subgrid of processor $j$
is copied to block $j$ of the wide subgrid of processor $i$,
for all $i$ and $j$.
These are corresponding areas of the global computational grid.
As the communication sequence proceeds,
each processor receives data blocks
and assembles them into complete, ordered arrays
describing the wide subgrid that it is conceptually assigned.
The wide subgrid arrays overwrite the memory areas
that was previously used for long subgrid data.
However the sequence of the {\tt MPI\_ALLTOALL} communication
provides for a complete redistribution of data across the set of processors
without any collective loss of information.

When the communication is complete,
each processor performs a hydrodynamic sweep in the $y$ direction
within its respective wide subgrid arrays.
(See right panel of Figure~\ref{fig.mpi.2}.)
Again, as in the $x$ sweep,
no processor boundaries are encountered during the computation,
and the one-dimensional sweep arrays are long.
At the end of the $y$ sweep,
the processors communicate blocks of their evolved wide grid arrays
in another {\tt MPI\_ALLTOALL} operation,
ensuring that the long subgrid arrays are up to date.
At this point program control operations are performed,
e.g. generating output files recording the hydrodynamic variable arrays
of each processor.

Care is needed to ensure that the processors evolve
their subgrids synchronously in simulation time.
Within each subgrid $g$
the maximum allowable timestep of integration for each $x$ or $y$ sweep
is determined by a Courant condition:
$\delta t_g$ must be less than the characteristic timescales
of fluid motion or sound waves crossing a cell,
e.g.
$\delta t_g < 0.6 \min( \delta x / v_{x,i}, \delta x / c_{s,i})$,
for all cells $i$ with side length $\delta x$.
In order to synchronise the simulation across all the subgrids,
the processors are made to inter-communicate
their locally calculated values of
$\delta t_g$,
and then adopt the global minimum allowable timestep,
$\delta t = \min(\delta t_0, \delta t_1, ... \delta t_{N-1})$.

This straightforward scheme for
multiprocessor {\tt PPM} hydrodynamics simulations
is readily generalised to a three-dimensional grid.
The 3D analogues of long and wide subgrids
are ``flat slabs''
(full size in the $x$ and $y$ directions,
but subdivided into $N$ segments in the $z$ direction)
and ``tall slabs''
(full size in the $x$ and $z$ directions,
but subdivided into $N$ parts in the $y$ direction).
Using 32, 64, 128 or more of the processors of the
Australian National University's
Compaq Alphaserver SC,
we have been able to perform 3D simulations with cubic dimensions
of 256, 512 or more cells per side
(work in preparation).

\section{SELF-SIMILAR SOLUTION FOR INFLATION OF A TWO-DIMENSIONAL BUBBLE}
\label{app.bubble}

\subsection{Dimensional analysis}
\label{app.dimensional}

This appendix provides an analytical solution
that can be used for the comparison of
slab-jet simulations with expected asymptotic behaviour.
The two parameters that determine the evolution of the bubble are
$A$, the energy flux per unit transverse length,
and $\rho_0$ the density of the background medium.
In terms of jet parameters,
\begin{equation}
A = \left( \frac{1}{2} \rho v^2 + \frac{\gamma}{\gamma-1} p \right) \, v D
\end{equation}
where $D$ is the width of the jet.

Consider the parameter $A/\rho_0$.
The dimensions of this parameter are given by:
\begin{equation}
\left[ \frac{A}{\rho_0} \right] = L^4 T^{-3} \Rightarrow
\left[ \frac{A}{\rho_0} \right]^{1/4} = L T^{-3/4}
\end{equation}
Let $R(t)$ be the bubble radius at time $t$.
The dimensions of $R t^{-3/4}$ are also $L T^{-3/4}$.
Hence the self-similar evolution of the bubble will be given by:
\begin{equation}
\frac{R}{t^{3/4}} = C \left[ \frac{A}{\rho_0} \right]^{1/4}
\end{equation}
where $C$ is a numerical constant to be determined.
The radius of the bubble is therefore given as a
function of time by:
\begin{equation}
R = C \, \left[ \frac{A}{\rho_0} \right]^{1/4} \, t^{3/4}
\end{equation}
This derivation is justified and elaborated in the next section,
where the constant $C$ is derived.

\subsection{Temporal evolution of bubble}

\subsubsection{Formulation of equations}

Consider a cylindrical bubble that is being fed by a jet
with an energy density per unit length (transverse to the jet) equal to $A$. We
consider the jet energy to be effectively thermalised so that the pressure derived from the shocked jet plasma drives cylindrical expansion.
Let the energy density of the bubble be $\epsilon$ and
the pressure of the bubble be $p$.

We represent the bubble in two dimensions in the $x-y$ plane
and take a length, $L$ of
bubble in the $z-$direction.
The total energy,
$E= \epsilon \pi R^2 L$ of this length of bubble with
volume $\pi R^2 L$ is governed by the equation:
\begin{equation}
\frac{d}{dt} \left(
\epsilon \pi R^2 L \right) + p \frac{d}{dt} \left( \pi R^2 L \right) = AL
\end{equation}
The second term represents the work done by the pressure during the expansion;
the term on the right represents the rate of energy fed into the bubble.
Note that we are assumig that the dominant form of
energy in this bubble is internal and that kinetic energy
does not contribute significantly.

Taking out the common factor of $L$,
the evolution of the bubble is governed by the equation
\begin{equation}
\frac{d}{dt} \left(
\epsilon \pi R^2 \right) + p \frac{d}{dt} \left( \pi R^2 \right) = A
\label{e:bubble}
\end{equation}
We solve this equation by introducing the equation of state
[$\epsilon = p/(\gamma-1)$]
and by relating the pressure in the bubble to its rate of expansion.

Assuming that the bubble is expanding supersonically
with respect to the background medium,
then its structure consists of a strong shock outside the bubble
and a contact discontinuity separating the
bubble material from the shocked ISM.
In this region,
the pressure of the shocked ISM and the bubble are equal.
Suppose that the strong shock is moving with velocity
$v_{\rm sh}$.
Then the gas behind the shock is moving with speed
$3v_{\rm sh}/4$ so that
\begin{equation}
\dot R = \frac{3}{4} \, v_{\rm sh}
\label{e:rdot}
\end{equation}
Let the density of the background medium be $\rho_0$.
Then the pressure behind the shock is given by
\begin{equation}
p \approx \frac{3}{4} \rho_0 v_{\rm sh}^2
\approx \frac{4}{3} \rho_0 {\dot R}^2
\label{e:pressure}
\end{equation}
Note that we have not specified the $\gamma$ for the bubble
but we have assumed that $\gamma=5/3$ for the background ISM.

Substituting the equation of state into (\ref{e:bubble}),
and expanding the first term,
the bubble equation becomes:
\begin{equation}
\frac{\gamma}{\gamma-1} \pi p \frac{dR^2}{dt}
+ \frac{1}{\gamma-1} \pi R^2 \frac{dp}{dt} = A
\end{equation}
Then,
using equation~(\ref{e:pressure}) for the pressure,
we obtain,
after some simplification:
\begin{equation}
R {\dot R}^3 + \frac{1}{\gamma} R^2 \dot R \stackrel{..}{R} = \frac
{\gamma-1}{\gamma}
\frac{3A}{8 \pi \rho_0}
\label{e:final_equation}
\end{equation}

\subsubsection{Solution}

It would be reasonably straightforward to develop a general solution
of equation~(\ref{e:final_equation}).
However, if we settle for a self-similar solution,
for which:
\begin{equation}
R = a t^\alpha
\end{equation}
where $a$ and $\alpha$ are constants.
Subsitution into equation~(\ref{e:final_equation}) gives:
\begin{equation}
a^4 \alpha^2 \, \left[
\alpha + \frac{\alpha-1}{\gamma}
\right]\, t^{4 \alpha -3} =
\frac{\gamma-1}{\gamma} \, \frac{3 A}{8 \pi \rho_0}
\ .
\end{equation}
Equating the powers of $t$ on both sides:
\begin{equation}
\alpha = \frac{3}{4}
\end{equation}
as expected.
Equating the constants:
\begin{equation}
a =
\left[
\frac{3}{8} \frac{\gamma-1}{\alpha^2 \left[ (\gamma+1) \alpha -1
\right]}
\frac{A}{\pi \rho_0} \right]^{1/4}
\ .
\end{equation}
The form of the dependence on the parameter $A/\rho_0$
was anticipated in
\S\ref{app.dimensional}.
For $\gamma = 5/3$,
as used in our simulations:
\begin{equation}
a = \left[ \frac{4 A}{9 \pi \rho_0} \right]^{1/4}
\end{equation}
and the solution for $R$ is:
\begin{equation}
R = \left[ \frac{4 A}{9 \pi \rho_0} \right]^{1/4} \, t^{3/4}
\ .
\end{equation}

\subsection{Radius of shock wave}

In comparing the analytic solution with simulations
it is easiest to locate the position of the discontunity in the pressure.
This is the location of the outer shock wave,
whose velocity is derived from
equation~(\ref{e:rdot}),
i.e.
\begin{equation}
v_{\rm sh} = \frac{4}{3} \, \dot R
\end{equation}

Hence the radius of the pressure discontinuity is given by
\begin{equation}
R_{\rm p} = \frac{4}{3} a t^\alpha
= \frac{4}{3} \left[ \frac{4 A}{9 \pi \rho_0} \right]^{1/4} \,
t^{3/4}
\end{equation}
for $\gamma = 5/3$.


\bsp

\label{lastpage}


\begin{thebibliography}{}

\bibitem[\protect\citeauthoryear{Begelman}{Begelman}{1996}]{begelman96a}
Begelman, M.~C. 1996, in Cygnus A: Study of a Radio Galaxy, ed. C.~L. Carilli
  \& D.~A. Harris (Cambridge: University Press), 209
\bibitem[\protect\citeauthoryear{Bicknell, Dopita, \& O'Dea}{Bicknell
  et~al.}{1997}]{bicknell97a}
Bicknell, G.~V., Dopita, M.~A., \& O'Dea, C.~P. 1997, ApJ,{ 485}, 112
\bibitem[\protect\citeauthoryear{Bicknell et~al.}{Bicknell
  et~al.}{1998}]{bicknell98a}
Bicknell, G.~V., Dopita, M.~A., Tsvetanov, Z.~I., \& Sutherland, R.~S. 1998,
  ApJ,{ 495}, 680
\bibitem[\protect\citeauthoryear{{Bicknell} et~al.}{{Bicknell}
  et~al.}{2005}]{bicknell04a}
{Bicknell}, G.~V., {Safouris}, V., {Saripalli}, L., {Saxton}, C.~J.,
  {Subrahmanyan}, R., {Sutherland}, R.~S., {Midgley}, S., \& {Wagner}, S.~J.
  2005, NewAR,{ in press}, 000
\bibitem[\protect\citeauthoryear{{Bicknell}, {Saxton}, \&
  {Sutherland}}{{Bicknell} et~al.}{2003}]{bicknell2003a}
{Bicknell}, G.~V., {Saxton}, C.~J., \& {Sutherland}, R.~S. 2003, PASA,{ 20},
  102
\bibitem[\protect\citeauthoryear{{Bicknell}}{{Bicknell}}{1994}]{bicknell1994a}
{Bicknell}, G.~V. 1994, ApJ,{ 422}, 542
\bibitem[\protect\citeauthoryear{Blondin \& Lukfin}{Blondin \&
  Lukfin}{1993}]{blondin93a}
Blondin, J.~M. \& Lukfin, E.~A. 1993, ApJS,{ 88}, 589
\bibitem[\protect\citeauthoryear{{Chatzichristou}}{{Chatzichristou}}{2001}]{ch%
atzichristou01a}
{Chatzichristou}, E.~T. 2001, in IAU Symposium 162
\bibitem[\protect\citeauthoryear{Colella \& Woodward}{Colella \&
  Woodward}{1984}]{colella1984}
Colella, P. \& Woodward, P.~R. 1984, J. Comput. Phys.,{ 54}, 174
\bibitem[\protect\citeauthoryear{{Conway}}{{Conway}}{2002a}]{conway02a}
{Conway}, J.~E. 2002a, New Astronomy Review,{ 46}, 263
\bibitem[\protect\citeauthoryear{{Conway}}{{Conway}}{2002b}]{conway2002}
{Conway}, J.~E. 2002b, New Astronomy Review,{ 46}, 263
\bibitem[\protect\citeauthoryear{{de Vries} et~al.}{{de Vries}
  et~al.}{1999}]{devries1999a}
{de Vries}, W.~H., {O'Dea}, C.~P., {Baum}, S.~A., \& {Barthel}, P.~D. 1999,
  ApJ,{ 526}, 27
\bibitem[\protect\citeauthoryear{{de Vries} et~al.}{{de Vries}
  et~al.}{1997}]{devries1997}
{de Vries}, W.~H., {et~al.} 1997, ApJS,{ 110}, 191
\bibitem[\protect\citeauthoryear{{Drake} et~al.}{{Drake}
  et~al.}{2003}]{drake2003}
{Drake}, C.~L., {McGregor}, P.~J., {Bicknell}, G.~V., \& {Dopita}, M.~A. 2003,
  Publications of the Astronomical Society of Australia,{ 20}, 57
\bibitem[\protect\citeauthoryear{{Fragile} et~al.}{{Fragile}
  et~al.}{2004}]{fragile04a}
{Fragile}, P.~C., {Murray}, S.~D., {Anninos}, P., \& {van Breugel}, W. 2004,
  March), ApJ,{ 604}, 74
\bibitem[\protect\citeauthoryear{{Giovannini} et~al.}{{Giovannini}
  et~al.}{2000}]{giovannini00a}
{Giovannini}, G., {Cotton}, W.D., {Feretti}, L., {Lara}, L., \& {Venturi}, T.
  2000, Advances in Space Research,{ 26}, 693
\bibitem[\protect\citeauthoryear{{Hardee} \& {Norman}}{{Hardee} \&
  {Norman}}{1988}]{hardee1988}
{Hardee}, P.~E. \& {Norman}, M.~L. 1988, ApJ,{ 334}, 70
\bibitem[\protect\citeauthoryear{{Kawakatu}, {Umemura}, \& {Mori}}{{Kawakatu}
  et~al.}{2003}]{kawakatu03a}
{Kawakatu}, N., {Umemura}, M., \& {Mori}, M. 2003, ApJL,{ 583}, 85
\bibitem[\protect\citeauthoryear{{Komissarov} \& {Falle}}{{Komissarov} \&
  {Falle}}{1996}]{komissarov96a}
{Komissarov}, S.~S. \& {Falle}, S.~A.~E.~G. 1996, in Astronomical Society of
  the Pacific Conference Series 173--179
\bibitem[\protect\citeauthoryear{{Labiano} et~al.}{{Labiano}
  et~al.}{2003}]{labiano03a}
{Labiano}, A., {et~al.} 2003, Publications of the Astronomical Society of
  Australia,{ 20}, 28
\bibitem[\protect\citeauthoryear{Lind et~al.}{Lind et~al.}{1989}]{lind89}
Lind, K.~R., Payne, D.~G., Meier, D.~L., \& Blandford, R.~D. 1989, ApJ,{ 344},
  89
\bibitem[\protect\citeauthoryear{{Magorrian} et~al.}{{Magorrian}
  et~al.}{1998}]{magorrian98a}
{Magorrian}, J., {et~al.} 1998, AJ,{ 115}, 2285
\bibitem[\protect\citeauthoryear{{Murgia} et~al.}{{Murgia}
  et~al.}{1999}]{murgia1999a}
{Murgia}, M., {Fanti}, C., {Fanti}, R., {Gregorini}, L., {Klein}, U., {Mack},
  K.-H., \& {Vigotti}, M. 1999, A\&A,{ 345}, 769
\bibitem[\protect\citeauthoryear{{Murgia} et~al.}{{Murgia}
  et~al.}{2002}]{murgia2002a}
{Murgia}, M., {Fanti}, C., {Fanti}, R., {Gregorini}, L., {Klein}, U., {Mack},
  K.-H., \& {Vigotti}, M. 2002, New Astronomy Review,{ 46}, 307
\bibitem[\protect\citeauthoryear{{Murgia}}{{Murgia}}{2003}]{murgia2003a}
{Murgia}, M. 2003, Publications of the Astronomical Society of Australia,{ 20},
  19
\bibitem[\protect\citeauthoryear{{O'Dea} et~al.}{{O'Dea}
  et~al.}{2003}]{odea2003a}
{O'Dea}, C.~P., {et~al.} 2003, Publications of the Astronomical Society of
  Australia,{ 20}, 88
\bibitem[\protect\citeauthoryear{{O'Dea} et~al.}{{O'Dea}
  et~al.}{2002}]{odea2002}
{O'Dea}, C.~P., {et~al.} 2002, AJ,{ 123}, 2333
\bibitem[\protect\citeauthoryear{{Ojha}}{{Ojha}}{2004}]{ojha04a}
{Ojha}, R.~{et al.} 2004, in preparation,{ 000}, 000
\bibitem[\protect\citeauthoryear{{Owen}, {Eilek}, \& {Kassim}}{{Owen}
  et~al.}{2000}]{owen00a}
{Owen}, F.~N., {Eilek}, J.~A., \& {Kassim}, N.~E. 2000, ApJ,{ 543}, 611
\bibitem[\protect\citeauthoryear{{Parma} et~al.}{{Parma}
  et~al.}{1999}]{murgia1999b}
{Parma}, P., {Murgia}, M., {Morganti}, R., {Capetti}, A., {de Ruiter}, H.~R.,
  \& {Fanti}, R. 1999, A\&A,{ 344}, 7
\bibitem[\protect\citeauthoryear{{Piner} et~al.}{{Piner}
  et~al.}{1999}]{piner99a}
{Piner}, B.~G., {Unwin}, S.~C., {Wehrle}, A.~E., {Edwards}, P.~G., {Fey},
  A.~L., \& {Kingham}, K.~A. 1999, ApJ,{ 525}, 176--190
\bibitem[\protect\citeauthoryear{{Rosen} et~al.}{{Rosen}
  et~al.}{1999}]{rosen99a}
{Rosen}, A., {Hughes}, P.~A., {Duncan}, G.~C., \& {Hardee}, P.~E. 1999, ApJ,{
  516}, 729
\bibitem[\protect\citeauthoryear{{Saxton}, {Bicknell}, \&
  {Sutherland}}{{Saxton} et~al.}{2002}]{saxton2002b}
{Saxton}, C.~J., {Bicknell}, G.~V., \& {Sutherland}, R.~S. 2002, ApJ,{ 579},
  176
\bibitem[\protect\citeauthoryear{{Saxton} et~al.}{{Saxton}
  et~al.}{2002}]{saxton2002a}
{Saxton}, C.~J., {Sutherland}, R.~S., {Bicknell}, G.~V., {Blanchet}, G.~F., \&
  {Wagner}, S.~J. 2002, A\&A,{ 393}, 765
\bibitem[\protect\citeauthoryear{{Saxton}, {Sutherland}, \&
  {Bicknell}}{{Saxton} et~al.}{2001}]{saxton2001}
{Saxton}, C.~J., {Sutherland}, R.~S., \& {Bicknell}, G.~V. 2001, ApJ,{ 563},
  103
\bibitem[\protect\citeauthoryear{{Silk} \& {Rees}}{{Silk} \&
  {Rees}}{1998}]{silk1998a}
{Silk}, J. \& {Rees}, M.~J. 1998, A\&A,{ 331}, L1
\bibitem[\protect\citeauthoryear{{Sutherland}, {Bicknell}, \&
  {Dopita}}{{Sutherland} et~al.}{2003}]{sutherland2003b}
{Sutherland}, R.~S., {Bicknell}, G.~V., \& {Dopita}, M.~A. 2003, ApJ,{ 591},
  238
\bibitem[\protect\citeauthoryear{{Sutherland}, {Bisset}, \&
  {Bicknell}}{{Sutherland} et~al.}{2003}]{sutherland2003a}
{Sutherland}, R.~S., {Bisset}, D.~K., \& {Bicknell}, G.~V. 2003, ApJS,{ 147},
  187
\bibitem[\protect\citeauthoryear{Sutherland \& Dopita}{Sutherland \&
  Dopita}{1993}]{sutherland93c}
Sutherland, R.~S. \& Dopita, M.~A. 1993, ApJS,{ 88}, 253
\bibitem[\protect\citeauthoryear{{Tremaine} et~al.}{{Tremaine}
  et~al.}{2002}]{tremaine02a}
{Tremaine}, S., {et~al.} 2002, ApJ,{ 574}, 740
\bibitem[\protect\citeauthoryear{{Van~Breugel} et~al.}{{Van~Breugel}
  et~al.}{1985}]{vanbreugel85a}
{Van~Breugel}, W.~J.~M., Filipenko, A.~V., Heckman, T., \& Miley, G.~K. 1985,
  ApJ,{ 293}, 83
\bibitem[\protect\citeauthoryear{{Wiita}, {Wang}, \& {Hooda}}{{Wiita}
  et~al.}{2002}]{wiita02a}
{Wiita}, P.~J., {Wang}, Z., \& {Hooda}, J.~S. 2002, May), New Astronomy
  Review,{ 46}, 439--442
\bibitem[\protect\citeauthoryear{{Wilkinson} et~al.}{{Wilkinson}
  et~al.}{1991}]{wilkinson91a}
{Wilkinson}, P.~N., {Tzioumis}, A.~K., {Benson}, J.~M., {Walker}, R.~C.,
  {Simon}, R.~S., \& {Kahn}, F.~D. 1991, Nature,{ 352}, 313--315
\end{thebibliography}
\end{document}